\documentclass[a4paper,12pt]{article}
\usepackage{jheppub} 
\usepackage[T1]{fontenc} 
\usepackage[latin1]{inputenc}
\usepackage{amsmath}
\usepackage{amsfonts}
\usepackage{amssymb}
\usepackage{makeidx}
\usepackage[toc,page]{appendix}
\usepackage{graphicx}
\usepackage{slashed}
\title{\boldmath Five-body leptonic decays of muon and tau leptons}

\author[a]{A. Flores-Tlalpa,}
\author[b,c]{G. L\'opez Castro,}
\author[b, 1]{P. Roig, \note{Corresponding author.}}

\affiliation[a]{Instituto de F\'{\i}sica, Universidad Nacional Aut\'onoma de M\'exico, Apartado Postal 20-364, 01000 M\'exico D.F., M\'exico}
\affiliation[b]{Departamento de F\'isica, Centro de Investigaci\'on y de Estudios Avanzados del Instituto Polit\'ecnico Nacional, Apartado Postal 14-740, 07000 M\'exico D.F., M\'exico}
\affiliation[c]{Instituto de F\'\i sica Corpuscular, CSIC- Universitat de Val\`encia, Apt. Correus 22085, E-46071 Val\`encia, Spain}

\emailAdd{alain@fisica.unam.mx}
\emailAdd{glopez@fis.cinvestav.mx}
\emailAdd{proig@fis.cinvestav.mx}

\abstract{ We study the five-body decays $\mu^- \to e^- e^+ e^- \nu_{\mu} \bar{\nu}_{e}$ and $\tau^- \to \ell^- \ell'^+\ell'^-\nu_{\tau}\bar{\nu}_{\ell}$  for $\ell, \ell'=e, \mu$ within the Standard Model (SM) and in a general effective field theory description of the weak interactions at low energies. We compute the branching ratios and compare our results with two previous, mutually discrepant, SM calculations. By assuming a general structure for the weak currents we derive the expressions for the energy and angular  distributions of the three charged leptons when the decaying lepton is polarized, which will be useful in precise tests of the weak charged current at Belle II. In these decays, leptonic $\mathbf{T}$-odd correlations in triple products of spin and momenta --which may signal time reversal violation in the leptonic sector-- are suppressed by the tiny neutrino masses. Therefore, a measurement of such $\mathbf{T}$-violating observables would be associated to neutrinoless lepton flavor violating (LFV) decays, where this effect is not extremely suppressed. We also study the backgrounds that the SM five-lepton lepton decays constitute to searches of LFV $L^- \to \ell^- \ell'^+\ell'^-$ decays. Searches at high values of the invariant mass of the $\ell'^+\ell'^-$ pair look the most convenient way to overcome the background.\\ \\ \\ \\ \\
PACS~: 13.35.Bv, 13.35.Dx, 13.66.-a.
\\
Keywords~: Decays of muons, Decays of taus, Leptonic interactions, Michel parameters, T-violation, Lepton Flavor violation.
}

\begin{document}
\maketitle
\flushbottom

\section{Introduction}

Five-body decays of leptons, $L^- \to \ell^-\ell'^+\ell'^-\nu_L\bar{\nu}_{\ell}$ with $L=\tau,\,\mu$ and $\ell,\,\ell'=e,\,\mu$, are allowed processes in the SM, suppressed by a factor $\alpha^2$ with respect to the lowest order three-body $L^-\to \ell^-\nu_L\bar{\nu}_{\ell}$ decays. Their study is interesting because they can provide an important source of background for searches of lepton flavor violating $\mu \to e e^+ e^-$ \cite{Bertl} and $\tau \to \ell \ell'^+\ell'^-$ decays due to undetected neutrinos. Analyses of these decays are also aimed at searches for sterile neutrinos and dark photons \cite{Dib:2011hc, Essig:2013lka}. 

Besides, these processes allow for stringent tests of the weak charged current and, in particular of its Lorentz structure and possible violations of lepton universality (for a recent review on related studies in tau lepton decays see \cite{Pich:2013lsa}). In the closely related radiative $L^- \to \ell^-\nu_{L}\bar{\nu_\ell}\gamma$ decays, the photon carries information about the outgoing lepton polarization and, as a result, two additional Michel-like parameters \cite{Michel:1949qe, Bouchiat:1957zz} can be extracted with precision, in contrast to the non-radiative channel. For the tau lepton case, they shall be measured with comparable precision to the corresponding muon decays \cite{Pocanic:2014mya} using Belle-II data \cite{Denis} thanks to the expected Belle-II statistics and performance \cite{Abe:2010gxa}. These bright experimental prospects demand a corresponding effort on the theory and Monte Carlo \cite{Actis:2010gg} side. In this respect, let us mention the recent papers \cite{Fael:2015gua, Bruser:2015yka} studying radiative $\mu$ and $\tau$ leptonic decays at next-to-leading order and polarized $\tau \to 3 \ell$ decays (focusing on angular correlations of new physics operators), respectively.

In view of these forthcoming good quality data sets, it is timely to attempt to improve the current description of these decays. Among them, we shall include for the first time mass effects of the daughter charged leptons in decays of polarized particles. We will also address the accuracy reached, re-evaluating the branching ratios of different decay channels and comparing with existing (and conflicting) predictions. We first consider the contribution of lepton-flavor conserving charged weak interactions to these processes within an effective field theory framework, so as to test with precision the SM $V-A$ universal structure of the $W$ current through Michel-type parameters both in the spin-independent and spin-dependent cases. 

$\mathbf{T}$-odd correlations in these leptonic decays seem to be absent in the SM and in its most general, lepton-flavor conserving, effective field theory extension here considered. Very light neutrino masses (and thus fixed helicity, to an excellent degree of accuracy) may lead to negligible $\mathbf{T}$-odd spin-momenta correlations in these decays~\footnote{$\mathbf{T}$-odd correlations induced by genuine $\mathbf{T}$ violation would provide indirect signals of $\mathbf{CP}$ violation in the leptonic sector.}; however, no gain is obtained by relaxing the lepton-flavor conserving requirement as it is shown by considering a simple example of a new physics extension with $Z'$ gauge-boson. Therefore, an eventual observation of these $\mathbf{T}$-violating correlations may provide an indirect signal of a non-conventional source of $\mathbf{CP}$ violation. This one, in turn, would contribute to generate a net baryon asymmetry in early stages of the universe \cite{Sakharov:1967dj} through the anomaly-free character of the $B-L$ accidental symmetry of the SM \cite{'tHooft:1976up, Fukugita:1986hr}, with fundamental implications for the enormous matter-antimatter asymmetry of the universe and cosmic evolution.

\begin{figure}\centering
\includegraphics[scale=0.8]{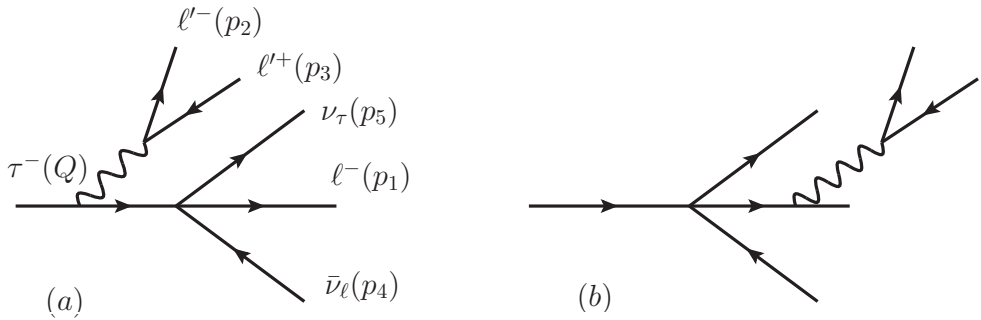}
\caption{\small Feynman diagrams for five-body decays of tau leptons. For identical leptons ($\ell'=\ell$) in the final state, two additional diagrams corresponding to the exchange $p_1\leftrightarrow p_2$ should be considered.}
\label{5leptons}
\end{figure}

Previous studies of the branching ratios for five-body decays of unpolarized leptons within the SM have been reported in \cite{Fishbane, Fetscher, Kersch1, Arbuzov, Dicus-Vega, CLEO, Olga, Djil}. While different calculations of $\mu$ decays agree within numerical uncertainties, the results of two theoretical calculations which consider tau decays \cite{Dicus-Vega, CLEO} differ significantly. Numerical studies of the angular and energy distributions of charged leptons in $\mu$ decays have been reported in \cite{Kersch1} by assuming the most general form of lepton-flavor conserving weak currents including (axial-)vector, (pseudo)scalar and tensor couplings. In the present study, we compute the branching ratios of five-body leptonic decays of tau and muon leptons using an integration method for the phase-space that differs from previous studies. We also provide analytical expressions for the angular and energy distribution of charged leptons in the case of the SM couplings for decays of polarized leptons keeping the whole charged lepton mass dependence but always neglecting neutrino masses. We also compute these distributions in the case of the most general Lorentz structure for the charged weak currents and provide their analytic expressions, in terms of the Michel parameters introduced in Ref.~\cite{Fetscher}, to describe five-body decays in the polarized case keeping finite daughter lepton masses for the first time. Our expressions are aimed to allow  stringent tests of the SM charged weak current by the experimental collaborations. Then, we provide an example of a new physics (LFV) extension which shows how $\mathbf{T}$-odd spin-momenta correlations are absent, as in the case of the charged weak current with the most general Lorentz structure, in the limit of massless neutrinos. The observation of $\mathbf{T}$-violation in these observables would therefore hint at the corresponding neutrinoless LFV processes, which strengths the case for these searches in leptonic muon and tau decays. Finally, we consider in detail the backgrounds that five-lepton lepton decays constitute for searches of the corresponding neutrinoless LFV processes and summarize our conclusions.
The explicit expressions of all contributions to the observables that we have computed are collected in the Appendices.

\section{Five-lepton channels within the SM}

In the SM, the five-body decays of muons and taus can be produced, at lowest order, from a lepton-pair conversion of a virtual photon emitted from the usual leptonic decay~\footnote{Lepton-pair conversion from $Z$ and Higgs bosons replacing the photon are also possible, but are extremely suppressed with respect to electromagnetic conversion, as shown in Figure \ref{5leptons} for the $\tau$ lepton case.}, as shown in Figure \ref{5leptons}. For identical charged leptons in the final state, two other diagrams contribute such that the decay amplitude becomes anti-symmetric under their exchange.

\subsection{Notation and kinematics}

The convention of momenta for the illustrative case of $\tau$ lepton decays, is
\begin{equation}
\tau^-(Q) \to \ell^-(p_1)\;\ell'^-(p_2)\;\ell'^+(p_3)\;\bar{\nu}_{\ell}(p_4)\;\nu_{\tau}(p_5)\ .
\end{equation}
Similarly, the masses of the decaying and final-state charged leptons will be denoted by $M,\ m_1,\ m$ ($m_2=m_3=m$), respectively. In the SM, the charged weak currents have a $V-A$ structure and their strength is encoded in the Fermi constant $G_F$.

For non-identical fermions ($\ell' \not = \ell$), the decay amplitude becomes:
\begin{equation}
{\cal M}_{SM} =\frac{ie^2G_F}{\sqrt{2} \, q^2}  \left({\cal M}_1+{\cal M}_2\right)^{\mu} L_{\mu} \ ,
\end{equation}
with
\begin{eqnarray}
{\cal M}_1^{\mu} &=& \bar{u}(p_5)\gamma_{\alpha}(1-\gamma_5)\left(\frac{i}{\not \!Q-\not\! q-M}\right)\gamma^{\mu}u(Q)\cdot \bar{u}(p_1)\gamma^{\alpha}(1-\gamma_5)v(p_4)\,,  \\
{\cal M}_2^{\mu} &=& \bar{u}(p_5)\gamma^{\alpha}(1-\gamma_5)u(Q) \cdot \bar{u}(p_1)\gamma^{\mu}\left(\frac{i}{\not\! p_1+\not\! q-m_1}\right) \gamma_{\alpha} (1-\gamma_5) v(p_4) \,,
\end{eqnarray}
where $L^{\mu}=\bar{u}(p_2)\gamma^{\mu}v(p_3)$ is the electromagnetic vertex current and $q=p_2+p_3$ is the momentum of the virtual photon ($q^{\mu}L_{\mu}=0$ due to conservation of electromagnetic current). For identical leptons ($\ell'=\ell$), the decay amplitude becomes ${\cal M}={\cal M}_1+{\cal M}_2-({\cal M}_3+{\cal M}_4)$, where the last two terms are obtained from the first two by the exchange $p_1\leftrightarrow p_2$.

Owing to the masslessness of neutrinos, the generic form of the unpolarized squared matrix element can be written as:
\begin{equation}
\overline{|{\cal M}|^2}=\frac{1}{2}\sum_{\rm pols} |{\cal M}|^2 = {\cal T}_{\alpha\beta}p_4^{\alpha}p_5^{\beta}\ .
\label{p4p5}
\end{equation}
This tensor structure is preserved in the case of polarized decaying particle.

\subsection{Branching ratio}

The decay width in the rest frame $Q=(M,\vec{0})$ of the decaying particle is~\footnote{A factor of $1/2$ in the right hand-side needs to be added when dealing with identical particles in the final state.}
\begin{equation} 
\Gamma_5 = \frac{(2\pi)^4}{2M}\int \prod_{i=1}^5 \frac{d^3p_i}{(2\pi)^3 2E_i} \overline{|{\cal M}|^2} \delta^4\left(Q-\sum_{i=1}^5 p_i\right)\,.
\end{equation}
The squared (unpolarized) amplitude $\overline{|{\cal M}|^2}$ of the five-body decay depends upon 8 independent kinematical variables. According to Ref.~\cite{Kumar}, we choose them to be:
\begin{eqnarray}\label{invariants}
s_1 &=& (Q-p_1)^2 \,,\quad  s_2 \, = \, (Q-p_1-p_2)^2 \,,\quad  s_3 \, = \, (Q-p_1-p_2-p_3)^2\,,\nonumber\\
u_1 &=& (Q-p_2)^2 \,,\quad  u_2 \, = \, (Q-p_3)^2\,,\quad  u_3 \, = \, (Q-p_4)^2\,,\nonumber\\
t_2 &=& (Q-p_2-p_3)^2\,,\quad  t_3 \, = \, (Q-p_2-p_3-p_4)^2\,.
\end{eqnarray}
For the reader's convenience, we quote the expression of the relevant scalar products in terms of our set of kinematical invariants given by eq.~(\ref{invariants}).
\begin{eqnarray}
& Q\cdot p_1 \,=\, \frac{\textstyle M^2 + m_1^2 - s_1}{\textstyle 2}\;,\quad Q\cdot p_2 \,=\, \frac{ \textstyle M^2 + m^2 - u_1}{ \textstyle2}\;,\quad Q\cdot p_3 \,=\, \frac{\textstyle M^2 + m^2 - u_2}{\textstyle 2}\,, & \nonumber\\
& \quad Q\cdot p_4 \,=\, \frac{\textstyle M^2 - u_3}{\textstyle 2} \;,\quad p_1\cdot p_2 \,=\, \frac{\textstyle s_2 - s_1 - u_1 + M^2}{\textstyle 2}\;,\quad p_1\cdot p_3 \,=\, \frac{\textstyle s_3 - s_2 - t_2 + u_1}{\textstyle 2}\,,&\nonumber\\
& p_1\cdot p_4 \,=\, \frac{\textstyle t_2 - s_3 - t_3}{\textstyle 2}\;,\quad p_2\cdot p_3 \,=\, \frac{\textstyle t_2 - u_1 - u_2 + M^2}{\textstyle 2}\;, \quad & \nonumber \\ & \quad p_2\cdot p_4 \,=\, \frac{\textstyle M^2 - u_3 - t_2 + t_3 - 2 p_3\cdot p_4}{\textstyle 2}\, .&
\end{eqnarray}
The very long expression for $p_3\cdot p_4$ is quoted in Appendix C, for completeness. We have verified that our expression for $p_3\cdot p_4$ agrees with eqs.~(A.6) to (A.8) in Ref.~\cite{Achasov} with suitable replacements for the squared masses $m_i^2$ ($i=1,..,5$).

Our method for integrating the phase-space to get the branching fraction is different from the one used in Ref.~\cite{Dicus-Vega}. We integrate the phase-space directly over the eight independent kinematical variables mentioned above. In contrast, in Ref.~\cite{Dicus-Vega} the phase-space is first (partially) integrated over the momenta of the two neutrinos using the covariance properties of the 
tensor integral ($P=Q-p_1-p_2-p_3$), namely
\begin{eqnarray}
\frac{I_{\alpha\beta}(P)}{48(2\pi)^5} &=&
\int p_{4\alpha}p_{5\beta} \delta^4(P-p_4-p_5)\frac{d^3p_4}{(2\pi)^3 2E_4}\frac{d^3p_5}{(2\pi)^3 2E_5} \nonumber \\
&=& A\, P^2\, g_{\alpha \beta}\,+ \,B \,P_{\alpha}\,P_{\beta}\ ,
\label{tensor45}
\end{eqnarray}
with $B=2A=1/\left(24(2\pi)^5\right)$. Then, the integration over the remaining variables is carried out. Therefore, the phase-space integration using our special choice of kinematics will allow to obtain an independent verification of previous calculations in \cite{Dicus-Vega}.

The matrix element squared and summed over all lepton polarizations reads 
\begin{equation}\label{eqgral1}
\overline{|{\cal M}|^2} \, = \, e^4 G_F^2 \Big[ \widehat{T}_{11} + \widehat{T}_{22} + \widehat{T}_{1221} + \widehat{T}_{33} + \widehat{T}_{44} - \widehat{T}_{1331} - \widehat{T}_{1441} - \widehat{T}_{2332} - \widehat{T}_{2442} + \widehat{T}_{3443} \Big] \,.
\end{equation}
$\widehat{T}_{ii}$ corresponds to the contribution of $\mathcal{M}_i \mathcal{M}_i^\dagger$, while $\widehat{T}_{ijji}$ to that of $\mathcal{M}_i \mathcal{M}_j^\dagger + \mathcal{M}_i^\dagger \mathcal{M}_j$. The subindexes 1 and 2 stand for Figs.~\ref{5leptons} (a) and (b), respectively,  while diags. 3 and 4 are obtained -in this order- by exchanging identical fermions in diagrams 1a and 1b. Subindexes in eq. (\ref{eqgral1}) stand for the corresponding contributions of different diagrams. Analogous notation will be used throughout. For convenience, we define the reduced amplitudes ($\widehat{t}_{ii},\,\widehat{t}_{ijji}$) taking out common factors
\begin{eqnarray}\label{reducedamplitudes}
\widehat{T}_{11} \, = \, \frac{32 \, \widehat{t}_{11}}{[D_1(p_2) D_2(p_2)]^2} & , \  \ & \widehat{T}_{22}\,=\,\frac{32 \, \widehat{t}_{22}}{[D_1(p_2) D]^2} \,,\\
\widehat{T}_{1221} \, = \, \frac{32 \, \widehat{t}_{1221}}{[D_1(p_2)]^2 D_2(p_2) D} & , & \widehat{T}_{33}\,=\,\frac{32 \, \widehat{t}_{33}}{[D_1(p_1) D_2(p_1)]^2} \,,\nonumber\\
\widehat{T}_{44} \, = \, \frac{32 \, \widehat{t}_{44}}{[D_1(p_1) D]^2} & , & \widehat{T}_{1331} \, = \, \frac{32 \, \widehat{t}_{1331}}{D_1(p_1) D_1(p_2) D_2(p_1) D_2(p_2)} \,,\nonumber\\
\widehat{T}_{1441} \, = \, \frac{32 \, \widehat{t}_{1441}}{D_1(p_1) D_1(p_2) D_2(p_2) D} & , & \widehat{T}_{2332} \, = \, \frac{32 \, \widehat{t}_{2332}}{D_1(p_1) D_1(p_2) D_2(p_1) D} \,,\nonumber\\
\widehat{T}_{2442} \, = \, \frac{32 \, \widehat{t}_{2442}}{D_1(p_1) D_1(p_2) D^2} & , & \widehat{T}_{3443} \, = \, \frac{32 \, \widehat{t}_{3443}}{[D_1(p_1)]^2 D_2(p_1) D} \,.\nonumber
\end{eqnarray}
Explicit expressions for the $\widehat{t}_{ii}$ and $\widehat{t}_{ijji}$ are given in Appendix A. 

In eq.~(\ref{reducedamplitudes}) we have introduced additional variables to make expressions shorter. These are ($i=1,2$)
\begin{eqnarray}\label{additionalvars}
D_1 (p_i) &=& m^2 + p_i\cdot p_3 \,, \\
D_2 (p_i) & = & m^2 + p_i\cdot p_3 - Q\cdot p_i - Q\cdot p_3 \,, \nonumber\\
D & = & m^2 + p_1\cdot p_2 + p_1\cdot p_3 + p_2\cdot p_3 \,. \nonumber
\end{eqnarray}
Note that these definitions are true for both decay modes, without identical particles and with identical particles ($m_1=m$).

\begin{table*}[h!]
\begin{center}
\begin{tabular}{|c||c|c|c|c|}
\hline 
Channel & Ref. \cite{Dicus-Vega} & Ref. \cite{CLEO} & This work & PDG \cite{pdg2014}\\
\hline
BR($\tau^- \to e^-e^+e^- \bar{\nu}_e\nu_{\tau}$)$\times 10^5$ & $4.15\pm0.06$ & $4.457\pm0.006$ & $4.21\pm0.01$ & $2.8\pm1.5$\\
BR($\tau^- \to e^- \mu^+\mu^- \bar{\nu}_e\nu_{\tau}$)$\times 10^7$ & $1.257\pm0.003$ & $1.347\pm0.002$ & $1.247\pm0.001$ & -\\
BR($\tau^- \to \mu^-e^+e^-\bar{\nu}_{\mu}\nu_{\tau}$)$\times 10^5$ & $1.97\pm0.02$ & $2.089\pm0.003$ & $1.984\pm0.004$ & $<3.6$\\
BR($\tau^- \to \mu^-\mu^+\mu^- \bar{\nu}_{\mu}\nu_{\tau}$)$\times 10^7$ & $1.190\pm0.002$ & $1.276\pm0.004$ & $1.183\pm 0.001$ & -\\
BR($\mu^- \to e^-e^+e^-\bar{\nu}_e\nu_{\mu}$)$\times 10^5$ & $3.60\pm0.02$ & $3.605\pm0.005$ & $3.597\pm0.002$ & $3.4\pm0.4$\\
\hline 
\end{tabular} 
\caption{\label{Table1}\small{Branching ratios for the five-body decays of $\tau$ and $\mu$ leptons with quoted error bar obtained from accuracy of numerical integration. Some of the previous calculations are shown, for comparison, in the second and third columns. An additional $\pm 0.17\%$ uncertainty owing to the current precision of the $\tau$ lifetime measurement \cite{pdg2014} should be added to our result. Experimental data are scarce, with large error bars, but still consistent with the SM predictions.}}
\end{center}
\end{table*}

Our results for the branching ratios are shown in Table \ref{Table1} and are compared to previous results and available experimental data. The quoted uncertainties arise from integration as given by VEGAS \cite{Lepage:1977sw}, the numerical code used in our numerical calculations (as well as in Ref.~\cite{Dicus-Vega}). Table \ref{Table1} shows that our results are in good agreement with those of Ref.~\cite{Dicus-Vega}, and substantially differ in some cases from those of Ref. \cite{CLEO}. Only tiny differences might be attributed to the use of different inputs. Unfortunately, we cannot read from those papers the exact values of the employed inputs. We use the values of masses --and particularly the tau mass, $(1776.82\pm0.16$) MeV-- quoted in the 2014 edition of the PDG \cite{pdg2014}, while $M_\tau$ of Ref.~\cite{Dicus-Vega} (and \cite{CLEO}) presumably corresponds to the CLEO 1993 measurement \cite{Ballest:1993ji} ($1777.8\pm1.8$) MeV. Analogously, the current tau lifetime, $(2.903\pm0.005)\cdot10^{-13}$ s, although consistent with the ALEPH 1992 measurement \cite{Decamp:1992zt}, $(2.91\pm0.14)\cdot10^{-13}$ s, is by far more precise. Our updated values may explain the small differences between our results and those of Ref. \cite{Dicus-Vega}, but not with those of Ref. \cite{CLEO}. In particular, focusing in the comparison of our numerical results with those of Ref.~\cite{Dicus-Vega}, we notice larger differences in the modes with two or three muons in the final state. Taking into account that different tau lifetime input results in an overall shift for all modes, and that modes with heavier daughter leptons are more sensitive to the tau mass owing to kinematics, we can attribute the small differences to the used tau mass inputs.

\begin{figure}[h!]
\begin{center}
\vspace*{2.0cm}
\includegraphics[scale=0.65,angle=-90]{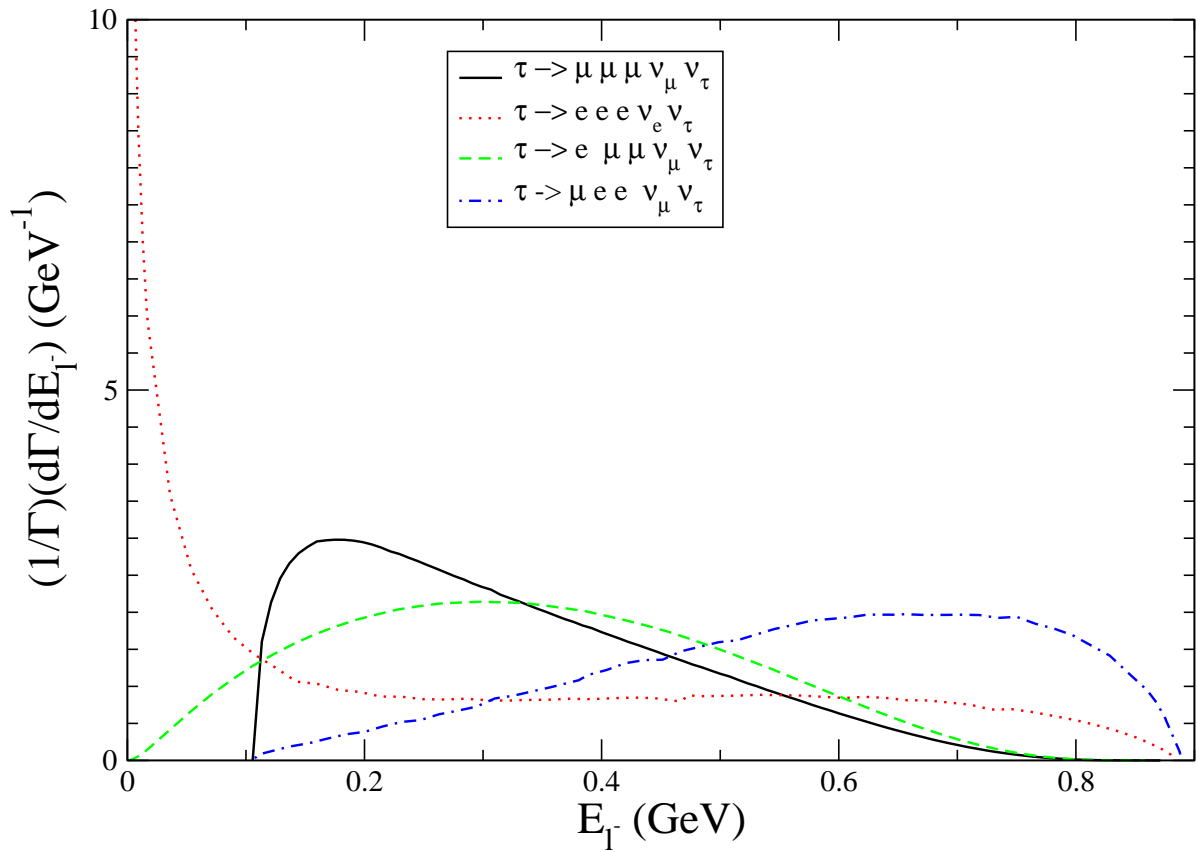}
\caption{\small Differential decay width (normalized to the partial decay width) versus $E_{\ell^-}$ (in modes without identical particles it corresponds to the energy of the different charged lepton) for all five-lepton $\tau$ decay channels.}\label{fig} 
\end{center}
\end{figure}

\begin{figure}[h!]
\begin{center}
\vspace*{1.25cm}
\includegraphics[scale=0.65,angle=-90]{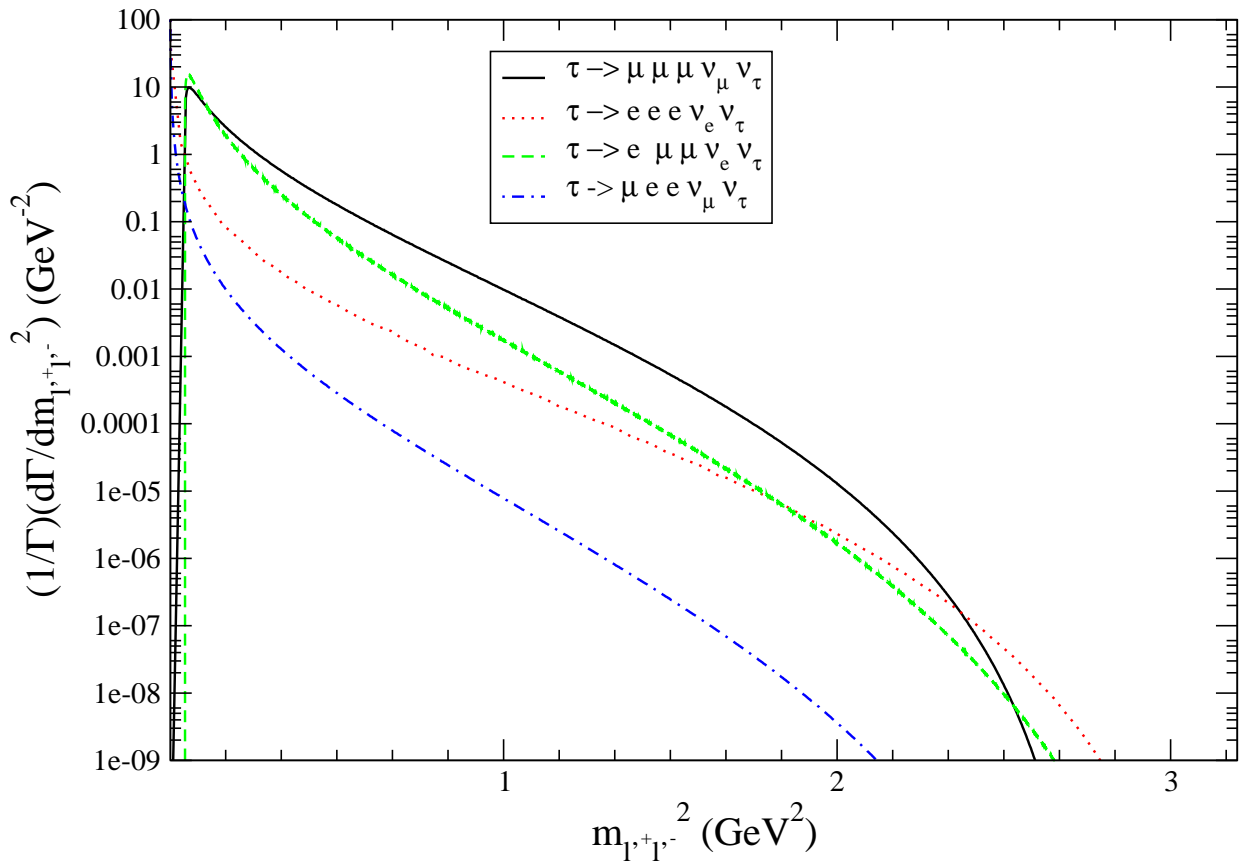}
\caption{ \small Differential decay width (normalized to the partial decay width) versus $q^2=m_{\ell'^+\ell'^-}^2$.}\label{fig2}
\end{center}
\end{figure}

In Fig.~\ref{fig} we plot the normalized differential decay width versus $E_1 = E_{\ell^-}$ for all five-lepton $\tau$ decay channels. As expected, no dynamical structures can be seen and only the kinematical enhancement in the $3e$ channel for low electron energy is noticeable. In Fig.~\ref{fig2} we plot the normalized differential decay width versus $q^2=m_{\ell'^+\ell'^-}^2$ for the considered decays. In this case, all channels tend to peak at low values of $m_{\ell'^+\ell'^-}^2$ because of the low virtuality of the photon near threshold.

Our precise predictions can also be useful to provide an independent test of an anomaly recently reported in tau leptonic decays. In particular, BaBar's measurement of $\tau\to e\gamma\nu_e \nu_\tau$ \cite{Lees:2015gea} differs by 3.5$\sigma$ from the SM prediction \cite{Fael:2015gua}, given at NLO. On the contrary, the agreement on the muon mode is at the 1 $\sigma$ level \cite{Lees:2015gea}. An independent study of these decays with Belle-I data could help to settle this issue. Given our precise prediction for the $\tau\to 3e\nu_e \nu_\tau$ decays and its branching ratio ($\sim4\cdot10^{-5}$), measurable with first generation B-factories BaBar and Belle-I data, another interesting check of the anomaly would come from analysing the former process.

\subsection{Spin-momenta correlation in decays of polarized leptons}

The polarization of the decaying particle is introduced by replacing 
\[
(\not \! Q +M) \rightarrow \frac{1}{2}(\not\! Q+M)(1+\gamma_5\not\! s)
\]
appropriately in the unpolarized squared amplitude. The polarization four-vector $s^\mu$ satisfies the properties $Q\cdot s=0$ and $s^2=-1$. In the rest frame of the decaying particle, 
$s=(0, \vec{\bf s})$.

Considering the polarization of the decaying particle does not change the dependence of the squared amplitude upon the neutrino four-momenta $p_{4,5}$, as compared to the unpolarized case, eq. (\ref{p4p5}). Therefore, the partial integration of the phase-space over neutrino four-momenta is given by Eq. (\ref{tensor45}), and the differential decay rate by:
\begin{equation}\label{distrib}
\frac{d\Gamma_5}{d^3p_1d^3p_2d^3p_3} = \frac{{\cal T}_{\alpha\beta}^s I^{\alpha\beta}(P)}{3\cdot 2^{18} \pi^{10} M E_1E_2E_3} \,,
\end{equation}
where we have added the upper-index $s$ on ${\cal T}_{\alpha\beta}^s$ to make explicit the dependence on the tau polarization.

Thus, after integration over neutrino four-momenta \footnote{As requested by the Belle Collaboration, we have also provided them with expressions for the corresponding spin (in)dependent form factors \textit{without} integrating over neutrino four-momenta. This is intended for direct implementation in the TAUOLA Monte Carlo Generator \cite{Jadach:1990mz, Jadach:1993hs, Shekhovtsova:2012ra, Nugent:2013hxa} for use of the collaboration. We are not including these expressions in this paper, but will send them upon eventual request.}, the differential rate depends only upon the product of charged particle four-momenta ($Q, p_1, p_2, p_3$) and the polarization four-vector $s$. In the rest frame of the decaying particle, the numerator in the right hand side (r.h.s.) of Eq. (\ref{distrib}) can be written as~\footnote{This notation corresponds to Belle's conventions and is useful to display relationships among $G_1$ and $G_2$ in the case without identical particles. Corresponding symmetries are found between $L$ and $G_2$ when there are indistinguishable fermions. $F$  can be checked with earlier computations, i.e. \cite{Dicus-Vega}, for the unpolarized case with the appropriate changes of kinematical invariants. We present, however, our complete expressions to avoid the reader these uncomfortable translations.}:
\begin{equation}\label{polarizedcase}
{\cal T}_{\alpha\beta}^{s} I^{\alpha\beta}(P) = e^4 G_F^2 \Big[ F \, - \,L\,\vec{{\bf p}}_1\cdot \vec{\bf s} \, -\, G_1\,\vec{\bf p}_2\cdot \vec{\bf s}\,-\, G_2\,\vec{\bf p}_3\cdot \vec{\bf s} \Big] \ .
\end{equation}
In the above equation, the coefficients $F$, $L$, $G_1$ and $G_2$ depend only upon dot products of charged leptons momenta.

In order to cast our results in a more compact way we will rewrite the previous result using the most commonly used phase-space integration variables
\begin{equation}\label{polarizedcase2}
\frac{d\Gamma_5}{dx_1 d\Omega_1 dx_2 d\Omega_2 dx_3 d\Omega_3} = \frac{M^2 |\vec{{\bf p}}_1| |\vec{{\bf p}}_2| |\vec{{\bf p}}_3|}{3\cdot 2^{21} \pi^{10}}{\cal T}_{\alpha\beta}^{s} I^{\alpha\beta}(P)\,,
\end{equation}
which also bear a closer relation to the measured observables. In the previous equation, the reduced dimensionless variables $x_i=2E_i/M$ ($i=1,2,3$) were introduced.

In this way, the matrix element squared, summed over final-state lepton polarizations and integrated over the neutrino phase space reads
\begin{eqnarray}\label{eqgral1p}
{\cal T}_{\alpha\beta}^s I^{\alpha\beta}(P) & = & e^4 G_F^2 \Big[ T_{11} + T_{22} + T_{1221} + T_{33} + T_{44} - T_{1331} - T_{1441} - T_{2332} - T_{2442} + T_{3443} \Big] \nonumber\\
 & \equiv & e^4 G_F^2 T_{SM} \,,
\end{eqnarray}
where we made explicit the relative signs due to exchange of identical fermions \footnote{In case all particles are distinguishable, only the first three terms contribute in eq.~(\ref{eqgral1p}).}. For later convenience we will use the reduced $t_{ii}$ and $t_{ijji}$ amplitudes which are defined as $\widehat{t}_{ii}$ and $\widehat{t}_{ijji}$ in eq. (\ref{reducedamplitudes}).

Taking advantage of the decomposition in eqs.~(\ref{polarizedcase}) and (\ref{eqgral1p}) we can split the different contributions to the spin-(in)dependent form factors as
\begin{equation}\label{eqgral3p}
F\,=\, F_{11} + F_{22} + F_{1221} + F_{33} + F_{44} - F_{1331} - F_{1441} - F_{2332} - F_{2442} + F_{3443}\,,
\end{equation}
and analogously for $G_1$, $G_2$ and $L$. We can further introduce the {\it reduced} form factors $f$, $g_1$, $g_2$ and $l$ functions by factoring out the same common coefficients as in eq.~(\ref{reducedamplitudes}). The explicit expressions of these reduced  form factors are given in Appendix B. Trace identities \cite{Sirlin1, Sirlin2} have been used to simplify the calculations.

\section{Effective field theory analysis}\label{EFT}

In the previous section we considered in detail the SM predictions for five-body leptonic decay modes of polarized and unpolarized $\mu$ and $\tau$ leptons. Here, we will generalize this approach within an effective quantum field theory description of the weak charged current at low energies, i.e., much smaller than the electroweak scale. We focus on decays of polarized $\mu, \tau$ leptons as the different spin and charged lepton momenta correlations may be useful to study the effects of New Physics described by new operators and couplings of the effective weak Hamiltonian.

The most general local, derivative-free, lepton-number conserving Lagrangian describing four-lepton interactions consistent with locality and Lorentz symmetry can be written \cite{Michel:1949qe, Bouchiat:1957zz, Fetscher, Kersch1, Kinoshita:1957zz, Scheck:1977yg, Pich:1995vj}
\begin{equation}\label{Lag_gral}
\mathcal{L}\,=\,-\frac{4G_{\ell\ell'}}{\sqrt{2}}\sum_{i,\lambda,\rho} g^i_{\lambda\rho} \left[\overline{\ell'_\lambda} \Gamma^i (\nu_{\ell'})_\xi\right]\left[\overline{(\nu_\ell)_\kappa} \Gamma_i \ell_\rho\right]\,,
\end{equation}
with $i=S,\,V,\,T$; $\Gamma^S=I$, $\Gamma^V=\gamma^\mu$, $\Gamma^T=\sigma^{\mu\nu}/\sqrt{2}$ labelling the Lorentz structure of weak currents and $\lambda,\rho=L,R$ the chiralities of the charged leptons ($\xi$ and $\kappa$, the chiralities of neutrinos, are fixed by $\lambda$ and $\rho$ once a particular Lorentz structure $\Gamma^i$ is chosen). There are 10 independent complex coefficients (4 scalar, 4 vector plus the two tensors-type couplings $g^T_{LR}$ and $g^T_{RL}$) which altogether give rise to 19 independent real couplings once an unphysical global phase is removed~\footnote{The $g^i_{\lambda\rho}$ coefficients, which parametrize beyond the SM effects at low energies in the weak charged current, can be related to the Wilson coefficients of the effective Lagrangian at the electroweak scale \cite{Cirigliano:2009wk} (see \cite{Gonzalez-Alonso:2014bga} for an updated discussion in the context of tau lepton decays). This allows to complement low-energy results with high-energy LEP/LHC searches.}. The global factor $G_{\ell\ell'}$ is fixed by the width of $\ell^- \to \ell'^- \bar{\nu}_{\ell'} \nu_{\ell}$ decay, in such a way that \cite{Fetscher}
\begin{equation}\label{eq_normalization_gral}
1\, = \,\frac{1}{4}\sum_{\lambda,\rho} |g^S_{\lambda\rho}|^2+\sum_{\lambda,\rho} |g^V_{\lambda\rho}|^2 + 3 \big( |g^T_{RL}|^2 + |g^T_{LR}|^2 \big) \, .
\end{equation}
The decay amplitude for five-lepton channels in $\tau$ and $\mu$ decays in the general case will receive contributions from the different terms of the Lagrangian (\ref{Lag_gral}). Those contributions to ${\cal T}_{\alpha\beta}^s I^{\alpha\beta}$ will be denoted as follows
\begin{eqnarray}\label{polarized-gral-case-2}
\mathcal{T}_{\alpha\beta}^{s} I^{\alpha\beta}(P) & = & e^4 |G_{\ell\ell'}|^2 \Big[ |g^S_{LL}|^2 T^S_{LL} + |g^V_{LL}|^2 T^V_{LL} + |g^S_{RL}|^2 T^S_{RL} + |g^V_{RL}|^2 T^V_{RL} + |g^T_{RL}|^2 T^T_{RL} \nonumber\\
 & & + \Re e\Big( g^S_{RL} g^{T\, *}_{RL} \, T^{ST}_{RLRL} + g^S_{LL} g^{V\, *}_{RR} \, T^{SV}_{LLRR} + g^S_{LR} g^{V\, *}_{RL} \, T^{SV}_{LRRL} + g^V_{LR} g^{T\, *}_{RL} \, T^{VT}_{LRRL} \Big) \nonumber\\
 & & + L \leftrightarrow R \Big] \, ,
\end{eqnarray}
in such a way that the SM contribution corresponds to $g^V_{LL}=1$ and all other couplings vanish, i.e. $T^V_{LL} \equiv T_{SM}$. Of course, as in the SM case, each of the new contributions can be written as in eqs. (\ref{polarizedcase}) and (\ref{eqgral1p})-(\ref{eqgral3p}), and their explicit expressions can be given in terms of the reduced form factors $f$, $g_1$, $g_2$ and $l$. Fortunately, it is not necessary to give the expressions of all of them, since  they satisfy the following identities
\begin{eqnarray}
\label{rel-cont-1}
T^V_{LL} & = & 4 T^S_{LL} \doteq T^Q_{LL} \, , \\
T^{VQ}_{RL} & = & 4 T^{SQ}_{RL} = \frac{1}{3} T^{TQ}_{RL} \doteq T^Q_{RL} \, , \\
\label{rel-cont-3}
T^{VB}_{RL} & = & 16 T^{SB}_{RL} = \frac{4}{9} T^{TB}_{RL} = \frac{4}{3} T^{ST}_{RLRL} \doteq T^B_{RL} \, , \\
T^{SV}_{LRRL} & = & T^{SV}_{RLLR} = \frac{1}{6} T^{VT}_{LRRL} = \frac{1}{6} T^{VT}_{RLLR} \doteq \frac{1}{4} T^I_\alpha \, , \\
\label{rel-cont-5}
T^{SV}_{LLRR} & = & T^{SV\, *}_{RRLL} \doteq \frac{1}{2} T^I_\beta \, , \label{rel-cont-51}
\end{eqnarray}
where ($i = S, V, T$)
\begin{equation}\label{rel-cont-6}
T^i_{RL} = T^{iQ}_{RL} + T^{iB}_{RL} \, , 
\end{equation} 
and analogous relations to (\ref{rel-cont-1})-(\ref{rel-cont-3}) and (\ref{rel-cont-6}) hold under $L \leftrightarrow R$. In order to read the results for these contributions eqs. (\ref{polarizedcase}) and (\ref{eqgral1p}) need to be used replacing $G_F^2$ by $|G_{\ell\ell'}|^2$.

In the limit of vanishing neutrino masses (i.e. keeping only left-handed neutrinos) the surviving structures are $T^V_{LL}$, $T^S_{RR}$ and $T^{SV}_{RRLL}$. Consequently, $g^V_{LL} \neq 1$ and $g^S_{RR}\neq 0$ would signal the most-likely first departures of the considered decays from the SM prediction.

Fetscher \textit{et al.} \cite{Fetscher} have introduced a set of eight Michel-like parameters, $Q_{ij}$ ($i,j = L,R$ and $\sum_{ij}Q_{ij}=1$), $B_{LR}$, $B_{RL}$, $I_\alpha$ and $I_\beta$, which fully characterize the three-body decays of polarized muons in the case of the most general interactions defined in (\ref{Lag_gral}). If we use this basis and the notation defined in eqs. (\ref{rel-cont-1})-(\ref{rel-cont-5}), then eq. (\ref{polarized-gral-case-2}) can be written as
\begin{eqnarray} \label{simplest}
\mathcal{T}_{\alpha\beta}^{s} I^{\alpha\beta}(P) & = & e^4 |G_{\ell\ell'}|^2 \Big[ \Big( Q_{LL} T^Q_{LL} + Q_{RL} T^Q_{RL} + B_{RL} T^B_{RL} + L \leftrightarrow R \nonumber \Big)\\
 & & + \Re e\Big( I_\alpha T^I_\alpha + I_\beta T^I_\beta \Big) \Big] \ .
\end{eqnarray}
It is noteworthy that even for the polarized case, and keeping non-vanishing masses of final charged leptons, $\mathcal{T}_{\alpha\beta}^{s} I^{\alpha\beta}$ can be written in terms of the above eight Michel-like parameters. We have checked that, in the limit of massless charged leptons in the final state, the last two terms in the previous expression vanish, in agreement with Eq. (4) in Ref.~\cite{Kersch1}.

In the rest frame of the polarized decaying particle, and after integrating over neutrino four-momenta, the numerator in the r.h.s. of Eq. (\ref{distrib}) has the general form (which inclu\-des $\mathbf{T}$-odd correlations)
\begin{eqnarray}\label{polarized-gral-case}
\mathcal{T}_{\alpha\beta}^{s} I^{\alpha\beta} (P) & = & e^4 |G_{\ell\ell'}|^2 \Big[ F \, - \, L\,\vec{\bf p}_1 \cdot \vec{\bf s} \, -\, G_1\,\vec{\bf p}_2\cdot \vec{\bf s}\,-\, G_2\,\vec{\bf p}_3\cdot \vec{\bf s} \nonumber \\   
&& + H_1\vec{\bf s}\cdot (\vec{\bf p}_1\times \vec{\bf p}_2) + H_2\vec{\bf s}\cdot (\vec{\bf p}_2\times \vec{\bf p}_3) + H_3\vec{\bf s}\cdot (\vec{\bf p}_1\times \vec{\bf p}_3) \Big] \ .
\end{eqnarray}
It is worth to note that the spin-independent coefficient $F$ remains invariant under parity inversion, while the spin-dependent form factors $L$, $G_1$ and $G_2$ ($\mathbf{T}$-even correlation coefficients of spin and momenta) change sign under this operation. Correspondingly, $F$ ($L$, $G_1$, $G_2$) has (have) the same (opposite sign) expression for the contributions $T^V_{LL}$ and $T^V_{RR}$, $T^{SV}_{LLRR}$ and $T^{SV}_{RRLL}$, and so on. Therefore, we only need to provide five contributions, which we choose to be $T^V_{LL}$ (SM case, see Appendix B), $T^{SV}_{LRRL}$, $T^{SV}_{LLRR}$, $T^{S}_{RL}$ and $T^V_{RL}$. It must be noted that the latter two are a more convenient choice than $T^{VQ}_{RL}$ and $T^{VB}_{RL}$ in order to get more compact expressions. However, both sets are related via 
\begin{eqnarray}
\label{transf1}
T^{VB}_{RL} & = & \frac{4}{3} \big( T^V_{RL} - 4 T^S_{RL} \big) \, , \\
\label{transf2}
T^{VQ}_{RL} & = & \frac{1}{3} \big( 16 T^S_{RL} - T^V_{RL} \big) \, .
\end{eqnarray}

A similar decomposition as in Eq. (\ref{simplest}) can be written for the form factors in Eq. (\ref{polarized-gral-case}) \footnote{Or course, the symbol $L \leftrightarrow R$ refers to the chiralities and not to the $\mathbf{T}$-even correlation coefficient $L$.}
\begin{eqnarray}
F & = & \big( Q_{LL} F^{Q}_{LL} + Q_{RL} F^{Q}_{RL} + B_{LR} F^B_{LR} + L \leftrightarrow R \big) + \Re e\big( I_\alpha F_\alpha^I +I_\beta F_\beta^I \big) \, , \\
L & = & \big( Q_{LL} L^{Q}_{LL} + Q_{RL} L^{Q}_{RL} + B_{LR} L^B_{LR} + L \leftrightarrow R \big) + \Re e\big( I_\alpha L_\alpha^I +I_\beta L_\beta^I \big) \, , \\
G_1 & = & \big( Q_{LL} G^{Q}_{1,LL} + Q_{RL} G^{Q}_{1,RL} + B_{LR} G^B_{1,LR} + L \leftrightarrow R \big) + \Re e\big( I_\alpha G_{1,\alpha}^I +I_\beta G_{1,\beta}^I \big) \, , \\
G_2 & = & \big( Q_{LL} G^{Q}_{2,LL} + Q_{RL} G^{Q}_{2,RL} + B_{LR} G^B_{2,LR} + L \leftrightarrow R \big) + \Re e\big( I_\alpha G_{2,\alpha}^I +I_\beta G_{2,\beta}^I \big) \, ,
\end{eqnarray}
and
\begin{equation}
H_1 = H_2 = H_3 = 0
\end{equation}
even in the beyond the SM case.

In Appendix D we provide the expressions for the reduced form factors $f$, $l$, $g_1$ and $g_2$ (see their definition in terms of $F$, $L$, $G_1$ and $G_2$ at the end of Section 2) for the $T^{SV}_{LRRL}$, $T^{SV}_{LLRR}$, $T^{S}_{RL}$ and $T^V_{RL}$ structure. In the case of identical (non-identical) charged leptons, forty (twelve) different reduced form factors are required, in general, to specify each of the five independent Lorentz-chiral amplitudes defined above. As an illustration, we write the explicit expression of $T_{RL}^S$ (analogous expressions hold for the others):
\begin{eqnarray}
T^S_{RL} & = & \sum_{i=1}^4 c_{ii} \left( f^{S, ii}_{RL} - l_{RL}^{S,ii} \, \vec{\bf p}_1\cdot \vec{\bf s} - {g_1}^{S,ii}_{RL}\, \vec{\bf p}_2\cdot \vec{\bf s} - {g_2}^{S,ii}_{RL}\, \vec{\bf p}_3\cdot \vec{\bf s} \right) \\
 & & + \sum_{i<k} \omega_{ik} c_{ikki} \left(f^{S, ikki}_{RL} - l_{RL}^{S,ikki} \, \vec{\bf p}_1\cdot \vec{\bf s} - {g_1}^{S,ikki}_{RL} \, \vec{\bf p}_2\cdot \vec{\bf s} - {g_2}^{S,ikki}_{RL} \, \vec{\bf p}_3\cdot \vec{\bf s} \right) \, , \nonumber
\end{eqnarray} 
where $c_{ii},\ c_{ikki}$ are the factors that relate the amplitudes $\widehat{T}_{ii},\ \widehat{T}_{ikki}$ to the reduced amplitudes $\widehat{t}_{ii},\ \widehat{t}_{ikki}$ in Eq. (\ref{reducedamplitudes}), respectively, and $\omega_{ik}=+1$ for $(i,k) = (1,2), (3,4)$ and $-1$ for $(i,k) = (1,3), (1,4), (2,3), (2,4)$. Once all the independent amplitudes are obtained we can combine them using (\ref{rel-cont-1})-(\ref{rel-cont-6}) and (\ref{transf1})-(\ref{transf2}) and insert them into Eq. (\ref{simplest}) to obtain the final expression for the squared amplitude.

\subsection{Possibility of $\mathbf{T}$-odd correlation terms}

As it was pointed out before, the coefficients $H_i$ (i=1,2,3) of the $\mathbf{T}$-odd correlations may signal violation of time-reversal symmetry and are discussed in the following. $\mathbf{T}$-odd correlations in the form of triple products of spin and momenta can be generated at higher orders due to exchange of virtual photons between charged leptons in five body decays \cite{BLS}. These {\it rescattering} effects that mimic time reversal non-invariance would be suppressed by at least a factor of $O(\alpha/\pi)$ with respect to the leading contributions already considered \cite{BLS}.
  
The contribution of the effective Lagrangian (\ref{Lag_gral}) to the coefficients $H_i$ (i=1,2,3) vanishes in the limit of massless neutrinos, due to the $LL$ and $RR$ structure for the product of weak currents in the amplitude (in the case of the interference between the tensor current and the $V\pm A$ currents the result still holds but we could only check it by \textit{brute force}). These coefficients would arise, in principle, from the imaginary parts of the product of couplings ($\Im m(g^i_{\lambda \rho}g^{j*}_{\sigma\omega}),\ i\not = j$) in the interference terms of the squared amplitude; however they vanish for massless neutrinos, as already mentioned. Thus, interferences between different currents giving rise to $\mathbf{T}$ violation will be extremely suppressed when considering (light) massive neutrinos for the underlying physics giving rise to the Lagrangian in (\ref{Lag_gral}) and we will therefore neglect it.

We would like to emphasize that a measurement of $\mathbf{T}$ violation in these decays may not come from the Lagrangian (\ref{Lag_gral}). Particularly, if mixed chiral structures $LR$ or $RL$ for the product of weak currents are present, they would manifest as non-zero $\mathbf{T}$-odd correlations in the energy and angular distributions. One may think, for instance, about the exchange of additional gauge bosons that couple to {\it flavor-violating} weak currents with both chiralities, as it would be the case for the contribution of a $Z'$ neutral boson \cite{Langacker:2000ju} (see also Sect. V.D of \cite{Langacker:2008yv} and references therein) that couples non-diagonally to leptons as follows
\begin{equation}
{\cal L}=\sum_{i,j}g_{Z'} \overline{\psi}_{i} \gamma^{\mu}(v_{ij}-a_{ij}\gamma_5)\psi_{j} Z'_{\mu} + {\rm h.c.}\,.
\end{equation}
In particular, it is possible that the interference of the corresponding amplitude with the SM $V-A$ term could bring in $\mathbf{T}$-odd terms. However, although allowing for the previous LFV vertices, the process $L^- \to \ell^-\ell'^+\ell'^-\nu_L\bar{\nu}_{\ell}$ is nevertheless lepton flavor-conserving. This implies that Fierz rearrangements in the amplitude will bring the contribution of such interference into those corresponding to the structures already included into the lepton flavor conserving Lagrangian eq. (\ref{Lag_gral}). Consequently, the $\mathbf{T}$ violating $H_i$ form factors will again be zero, as we have checked by explicit computation. Precisely because of this general suppression of $\mathbf{T}$-odd correlations in the SM and some of its extensions (by the tiny neutrino masses), searches for their effects in five-body decays of leptons provide another place to look for (unexpected) indirect manifestations of $\mathbf{CP}$ violation in the lepton sector. On the other hand, their non-observation would be useful to place constraints (although probably mild) on the leptonic $\mathbf{CP}$ phases.

At this point we shall recall that $\mathbf{T}$ violation has been studied in the LFV $\mu^\pm/\tau^\pm \to \ell^+ \ell^- \ell'^\pm$ decays \cite{Okada:1999zk, Kitano:2000fg} \footnote{The SM background to LFV $\tau\to\ell\gamma$ processes given by the $\tau\to\ell\gamma\nu \bar{\nu}$ decays is also computed in \cite{Kitano:2000fg}.}, where these effects are not suppressed in general. Therefore, a non-vanishing measurement of $\mathbf{T}$-odd correlations in leptonic tau decays with three charged leptons would most probably be an indirect signal of LFV neutrinoless processes.

\section{$L\to 3\ell \, 2\nu$ decays as backgrounds for LFV $L\to 3\ell$ searches}

The decays $L^-\to \ell^-\ell'^+\ell'^- \nu_{L}\bar{\nu}_{\ell}$ are some of the main backgrounds in searches for LFV $L^-\to \ell^-\ell'^+\ell'^-$ decays as they may mimic the signal owing to undetected neutrinos. Our thorough analysis of the former processes allows to study how they would mask signals from the latter. In order to do this, we consider the study of  $L^-\to \ell^-\ell'^+\ell'^-$ decays as reported in Ref. \cite{Celis:2014asa} (see Ref.~\cite{Dassinger:2007ru} for an earlier effective-field theory analysis of these processes). There, a general low-energy effective Lagrangian is considered to describe the LFV process and -in addition to the most popular dipole-type amplitude induced by radiative penguin diagram- it also incorporates \footnote{Their operators including two quark fields and their gluonic operators are not relevant for our discussion, as opposed to LFV semileptonic tau decays.} effective four-lepton operators with vector and scalar structures and the different allowed chiralities. Their equation (3.2) \cite{Celis:2014asa} gives the double differential decay distribution in the two independent opposite-sign lepton invariant masses in terms of the coefficients of the relevant operators introduced above. From this, it is straightforward to obtain any observable of interest for our comparison.

 We have first reproduced Figure 5 of Ref.~\cite{Celis:2014asa}, for the same sign di-muon invariant mass spectrum in the $\tau\to 3 \mu$ decay for which the following benchmark points are considered (see definitions in Ref. \cite{Celis:2014asa}):
\begin{itemize}
\item Vector Model: $C_{VLR}=C_{VRL}=0.3$ with all other couplings vanishing.
\item Scalar Model: $C_{SLL}=C_{SRR}=1$ with all other couplings vanishing.
\item Dipole Model: $C_{DL}=C_{DR}=0.1$ with all other couplings vanishing.
\end{itemize}
The scale of LFV is assumed to be $\Lambda=1$ TeV. We have stick exactly to this setting in what follows. We have analyzed the same and opposite sign di-lepton invariant mass distribution and also the differential decay width versus $E_{\ell^-}$ both for the SM $L^-\to \ell^-\ell'^+\ell'^- \nu_{L}\bar{\nu}_{\ell}$ decays and for the $L^-\to \ell^-\ell'^+\ell'^-$ LFV processes. The best discriminating variable for this search of LFV turns out to be the opposite-sign lepton-pair invariant mass distribution. 

The comparison of the current upper limits on $L^-\to \ell^-\ell'^+\ell'^-$ decays to our SM predictions for the five-lepton lepton decays (see Table \ref{Table1}) suggests that the tau decay modes with two or three muons are better suited for these searches than those with two or three electrons. In the case of muon decays the huge background to signal ratio  (i.e. the ratio among the SM process with neutrinos over the corresponding neutrinoless LFV decay) will make detection very challenging.

\begin{table*}[h!]
\begin{center}
\begin{tabular}{|c||c|c|c|}
\hline 
Channel & Current upper limit (UL) \cite{pdg2014, Amhis:2014hma} & S/B (UL) & Expected UL \cite{Hayasaka:2010np}\\
\hline\hline
BR($\tau^- \to e^-e^+e^-$) & $1.4\cdot10^{-8}$ & $\sim3\cdot10^{-4}$ & $\sim10^{-9}$\\
BR($\tau^- \to e^- \mu^+\mu^-$) & $1.6\cdot10^{-8}$ & $\sim0.1$ & $\sim10^{-9}$\\
BR($\tau^- \to \mu^-e^+e^-$) & $1.1\cdot10^{-8}$ & $\sim6\cdot10^{-4}$ & $\sim10^{-9}$\\
BR($\tau^- \to \mu^-\mu^+\mu^-$) & $1.2\cdot10^{-8}$ & $\sim0.1$ & $\sim10^{-9}$\\
BR($\mu^- \to e^-e^+e^-$) & $1.0\cdot10^{-12}$ & $\sim3\cdot10^{-8}$ & $\sim10^{-16}$\\
\hline 
\end{tabular} 
\caption{\label{Table2}\small{Current and expected sensitivities on  LFV $L^-\to \ell^-\ell'^+\ell'^-$ searches. The signal to background ratios (S/B) are estimated from current UL on BR's of LFV decays (signal) and of five-body decays (background).}}
\end{center}
\end{table*}

\begin{figure}[h!]
\begin{center}
\vspace*{1.25cm}
\includegraphics[scale=0.65,angle=-90]{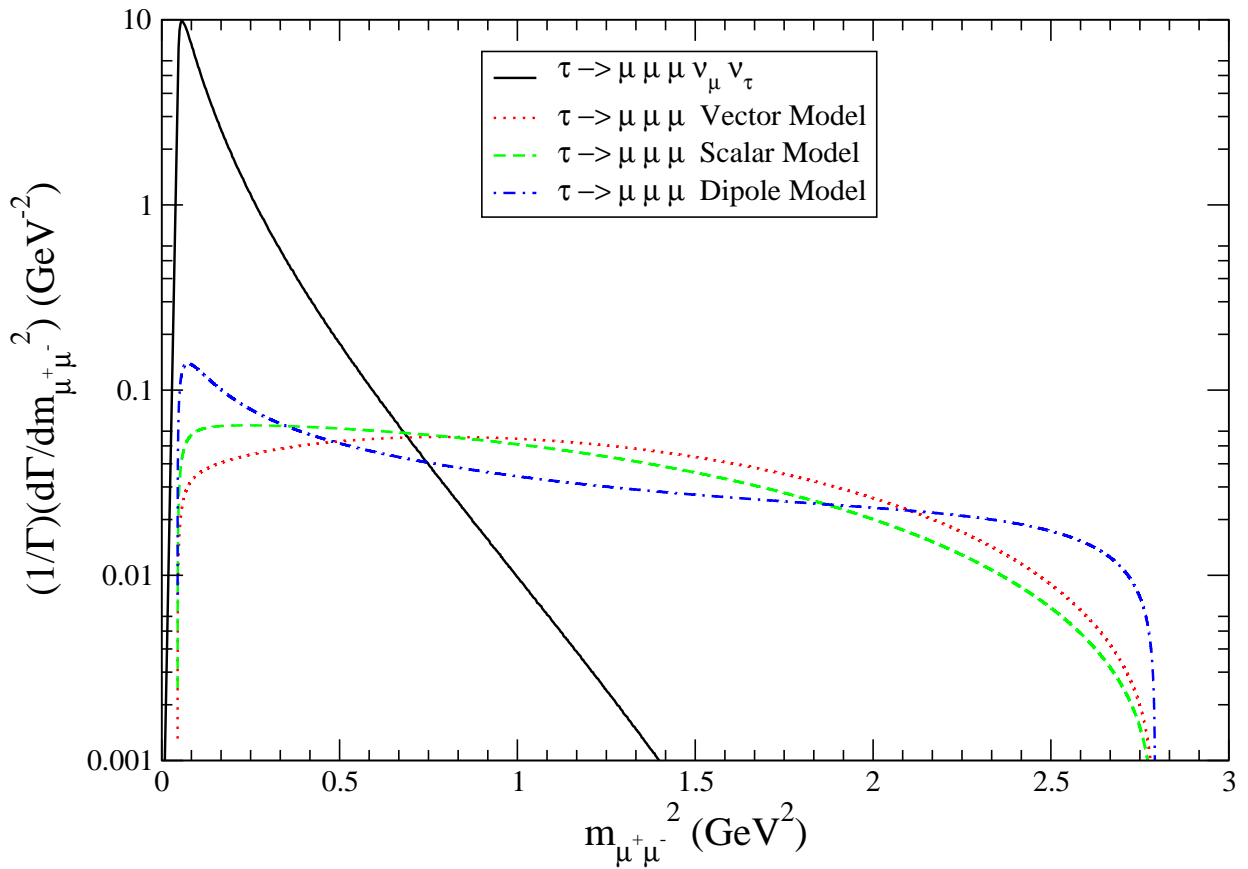}
\caption[]{\label{fig3} \small{Comparison of opposite-sign lepton-pair invariant mass distribution in five-body and LFV decays of tau leptons. Three benchmark scenarios of new physics in $\tau\to3\mu$ are used, according to Ref. \cite{Celis:2014asa}, and $S/B\sim0.1$ (about the current upper limit) is assumed.}}
\end{center}
\end{figure}

In Fig.~\ref{fig3} we confront the SM background to the hypothetical signals of LFV according to the Vector, Scalar and Dipole Models in the case of three muons in the final state. The normalization of the new physics curves is chosen so that the branching ratios for signal over background give $0.1$, which basically corresponds to a $LFV$ signal about the current upper limit $BR(\tau\to3\mu)\leq1.2\cdot10^{-8}$ \cite{Amhis:2014hma}. In this case, a cut for $m_{\mu^+\mu^-}^2\geq0.75$ GeV$^2$ will be most efficient. The similarity of the different new physics models above this cut suggests that it would be hard to disentangle the type of new physics with early data. If the new physics signal is set to the expected \cite{Hayasaka:2010np} upper limit $\sim10^{-9}$, the optimal cut moves to $m_{\mu^+\mu^-}^2\geq1.2$ GeV$^2$. The $\tau\to e \mu\mu$ case is very similar to the three muon channel, with a bit smaller cut: $m_{\mu^+\mu^-}^2\geq0.5$ GeV$^2$ for $S/B\sim 0.1$ and $m_{\mu^+\mu^-}^2\geq0.8$ GeV$^2$ for the envisaged near-future upper limit. The cases with two or three electrons are much harder for detection due to the signal to background ratio $\lesssim 10^{-3}$, which limits the region of the spectrum available for detection, as it can be seen in Figure \ref{fig4} where a cut for $m_{\mu^+\mu^-}^2\geq1.25$ GeV$^2$ shall be needed ($\sim 1.5 $ GeV$^2$ for $\tau\to e \mu\mu$). In addition to a very good statistics, an exquisite control of SM backgrounds will be needed to discover $\mu\to 3e$.

\begin{figure}[h!]
\begin{center}
\vspace*{1.25cm}
\includegraphics[scale=0.65,angle=-90]{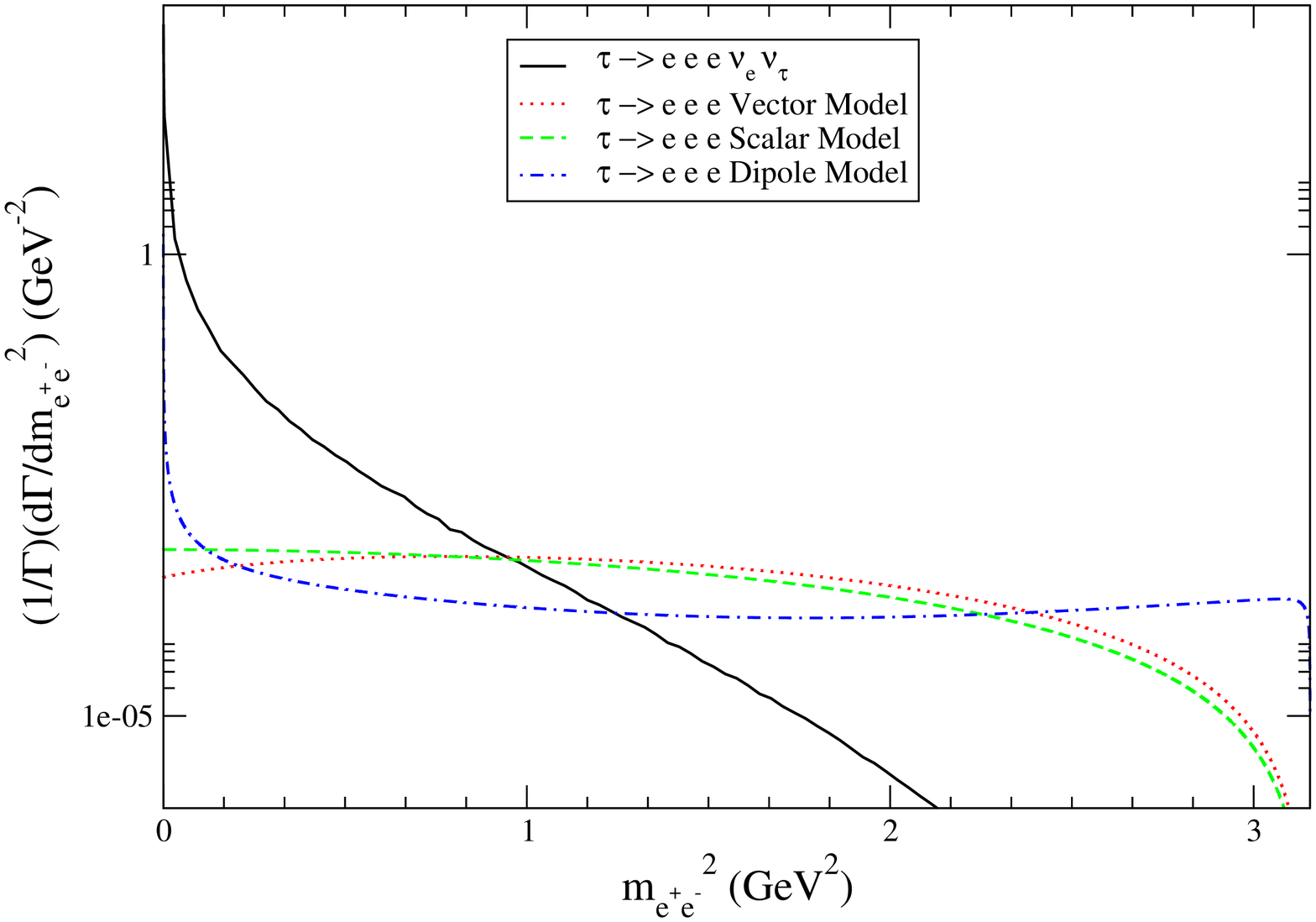}
\caption[]{\label{fig4} \small{SM background confronted to our benchmark scenarios of new physics in $\tau\to3e$ for $S/B\sim0.001$ (about the current upper limit).}}
\end{center}
\end{figure}

\begin{figure}[h!]
\begin{center}
\vspace*{1.25cm}
\includegraphics[scale=0.65,angle=-90]{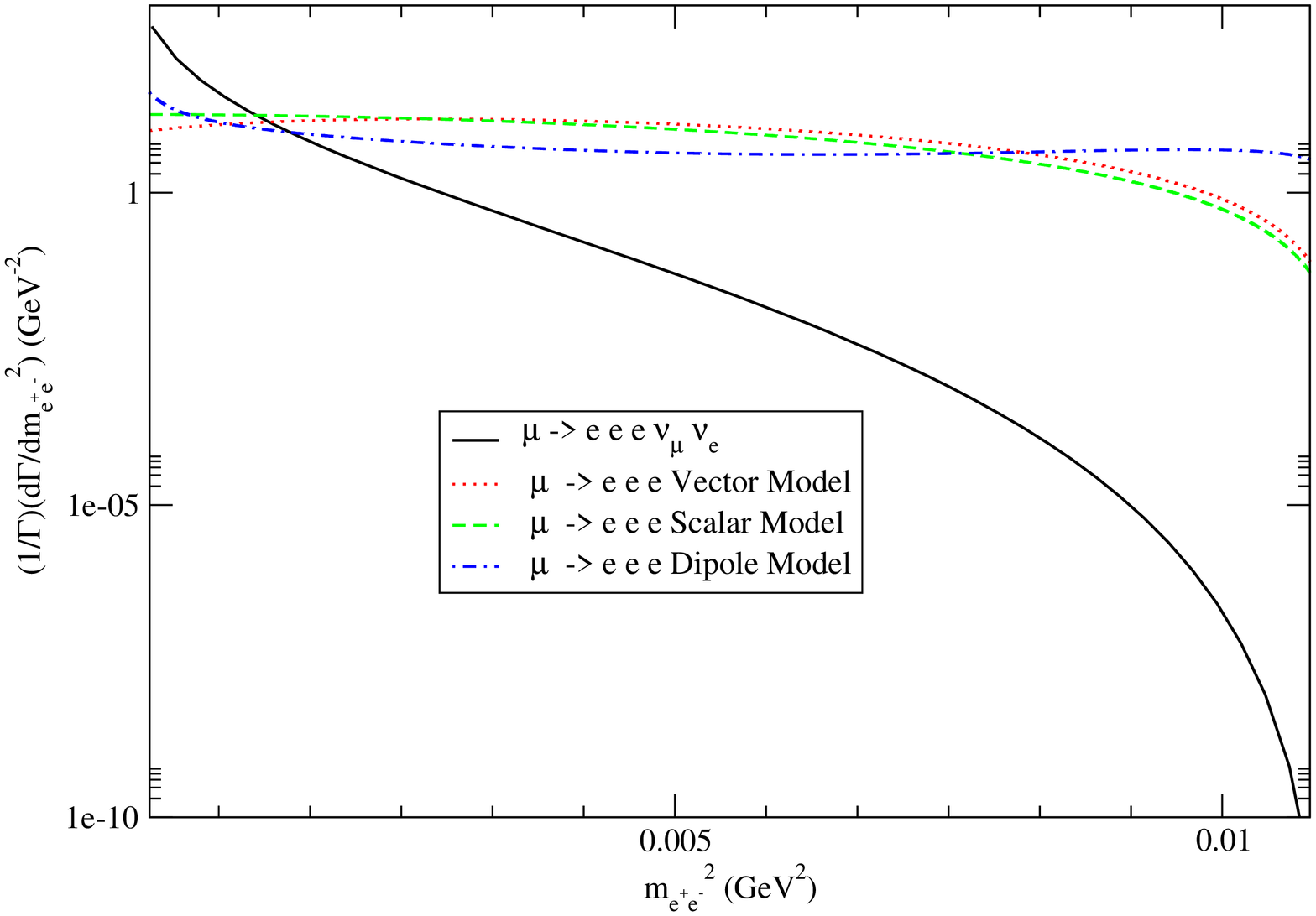}
\caption[]{\label{fig4} \small{SM background confronted to our benchmark scenarios of new physics in $\mu\to3e$ for $S/B\sim0.001$ (about the current upper limit).}}
\end{center}
\end{figure}

\section{Conclusions}

In this paper we have studied some aspects of all possible five-body leptonic channels of $\mu$ and $\tau$ lepton decays. In all our calculations, we have kept the finite masses of the final state charged particles, but neglected neutrino masses. Firstly, we have re-calculated the branching ratios in the SM and have compared our results with two previous and conflicting calculations \cite{Dicus-Vega, CLEO}. Our results were obtained by using a different method for the integration of the five-body phase-space; our yields are in good agreement with the ones of Ref. \cite{Dicus-Vega}.

In the second part, we have calculated the energy and angular distribution of the three-charged leptons, which can be written in terms of $\mathbf{T}$-even and $\mathbf{T}$-odd correlations of the spin $\vec{\bf s}$ of the decaying particle and the momenta $\vec{\bf p}_i\ (i=1,2,3)$ of final state charged leptons. We have derived the expressions for the $\mathbf{T}$-even correlations of the form $\vec{\bf s}\cdot \vec{\bf p}_i$, which are non-zero in the case of the SM and in the framework of the most general low-energy effective Lagrangian which conserves lepton-flavors. In particular, we have provided analytic expressions for the polarized decay probability in terms of the Michel-like parameters introduced by Fetscher \textit{et. al.} \cite{Fetscher} by keeping daughter-lepton mass dependence, as needed by Belle(-II). These results will be useful to test precisely the structure of the weak charged current in five-body leptonic decays of $\mu$ and $\tau$ leptons.
 
 We have also shown that, for massless neutrinos, the coefficients of the $\mathbf{T}$-odd correlations of the form $\vec{\bf s}\cdot (\vec{\bf p}_i\times \vec{\bf p}_j)$ vanish in the SM and in the extension with the most general effective theory of the (current)$\times$(current) form, even if we allow for lepton-flavor violating vertices with extra gauge bosons. Consequently, observation of non-vanishing $\mathbf{T}$-odd correlations either will point to indirect signals of non-conventional mechanisms of $\mathbf{CP}$ violation in the lepton sector (if produced in five-lepton lepton decays) or to the discovery of LFV (through the corresponding neutrinoless processes). These appealing characteristics, in our view, have not been noticed previously in the literature regarding purely leptonic interactions and open a new way to search for effects of $\mathbf{CP}$ violation.

Finally, we have considered in detail the backgrounds that the SM $L^-\to \ell^-\ell'^+\ell'^- \nu_{L}\bar{\nu}_{\ell}$ decays constitute to searches of LFV in the corresponding neutrinoless decays. Tau decays with two or three muons are favoured with a moderate signal to background ratio. Generally, the differential decay distribution versus the invariant mass of the opposite-sign lepton pair is the best observable for maximizing the signal region. This looks similar to the discriminating power of the di-lepton energy spectrum in searches of neutrinoless nuclear double beta decays which would show up as a single line excess over the continuum distribution due to the allowed two neutrinos double beta decays. Forthcoming experiments will be able to perform these searches with reasonable cuts on this variable in the sub-GeV region.

\subsection*{Acknowledgements}
We are grateful to Conacyt for financial support through SNI and under projects 296, 236394,  263916 and 250628. A.F.T. was supported in part by DGAPA-UNAM. The work of G.L.C. has been supported in part by EPLANET project and by the Spanish Government and ERDF funds from the EU Commission [Grants No. FPA2011-23778, FPA2014-53631-C2-1-P, SEV-2014-0398] and by Generalitat Valenciana under Grant No. PROMETEOII/2013/007. We are indebted to Denis Epifanov for his interest in this work and encouraging exchanges and discussions. We are very thankful to Mart\'{\i}n Gonz\'alez-Alonso for pointing us an inconsistency in Eq. (3.2) of a previous version. P. R. acknowledges 
the hospitality of Instituto de F\'{\i}sica at UNAM, where part of this work was done.

\section*{Appendix A}\label{Appendix_A}

We include here the expressions for the reduced amplitudes in the Standard Model case for unpolarized tau decays, as defined in Eq. (\ref{reducedamplitudes}). For simplicity in what follows we will use the following notation, $Q_i = Q\cdot p_i$ and $p_{ij} = p_i\cdot p_j$ ($i\ne j$). All results shown in the Appendices were obtained using FeynCalc \cite{Mertig:1990an}.
\begin{eqnarray}
\widehat{t}_{11} & = & -2 \big(m_1^2 + p_{12} + p_{13} + p_{14} - Q_1\big) \big(p_{23} p_{24} Q_2 + p_{34} Q_2^2 + p_{23} p_{34} Q_3 - p_{24} Q_2 Q_3 \\ & & - p_{34} Q_2 Q_3 + p_{24} Q_3^2 + M^2 \big(m^2 + p_{23}\big) \big(p_{24} + p_{34} - Q_4\big) - m^4 Q_4 - \big(-2 Q_2 Q_3 + p_{23} \nonumber\\ & & \times \big(Q_2 + Q_3\big)\big) Q_4 + 
  m^2 \big(p_{24} \big(2 Q_2 + Q_3\big) + p_{34} \big(Q_2 + 2 Q_3\big) - \big(p_{23} + Q_2 + Q_3\big) Q_4\big)\big) \nonumber\,,
\end{eqnarray}
\begin{eqnarray}
\widehat{t}_{22} & = & 2 \big(m_1^4 \big(m^2 + p_{23}\big) + m^4 \big(-2 p_{12} - 2 p_{13} + p_{14} - Q_1\big) - p_{13} \big(p_{13} \big(p_{23} - p_{24} + Q_2\big) \nonumber\\ & & + p_{23} \big(p_{14} + p_{23} + p_{34} - Q_1 - Q_3\big)\big) - p_{12} \big(2 p_{13}^2 + p_{23} \big(p_{14} + p_{23} + p_{24} - Q_1 - Q_2\big) \nonumber\\ & & + p_{13} \big(2 p_{14} + 4 p_{23} + p_{24} + p_{34} - 2 Q_1 - Q_2 - Q_3\big)\big) + m_1^2 \big(3 m^4 + p_{12} \big(-2 p_{13} + p_{23}\big) \nonumber\\ & & + p_{23} \big(p_{13} + p_{14} + 2 p_{23} + p_{24} + p_{34} - Q_1 - Q_2 - Q_3\big) + m^2 \big(p_{12} + p_{13} + p_{14} + 5 p_{23} \nonumber\\ & & + p_{24} + p_{34} - Q_1 - Q_2 - Q_3\big)\big) - p_{12}^2 \big(2 p_{13} + p_{23} - p_{34} + Q_3\big) + m^2 \big(-2 p_{12}^2 - 2 p_{13}^2 \nonumber\\ & & + p_{23} \big(p_{14} - Q_1\big) - p_{12} \big(6 p_{13} + p_{14} + 3 p_{23} + 2 p_{24} + p_{34} - Q_1 - 2 Q_2 - Q_3\big) \nonumber\\ & & + p_{13} \big(-p_{14} - 3 p_{23} - p_{24} - 2 p_{34} + Q_1 + Q_2 + 2 Q_3\big)\big)\
\big) Q_4 \,,
\end{eqnarray}
\begin{eqnarray}
\widehat{t}_{1221} & = & -2 \big(p_{13}^2 \big(-p_{34} Q_2 + p_{24} Q_3 + 2 Q_2 Q_4\big) + p_{12}^2 \big(p_{34} Q_2 - p_{24} Q_3 + 2 Q_3 Q_4\big) + m_1^2 \big(\big(p_{12} \nonumber\\ 
& & - p_{13}\big) \big(p_{34} Q_2 - p_{24} Q_3\big) + m^2 \big(-p_{14} \big(Q_2 + Q_3\big) + Q_1 \big(p_{24} + p_{34} - 2 Q_4\big)\big) + p_{23} \nonumber\\ 
& &\times \big(Q_1 \big(p_{24} + p_{34} - 2 Q_4\big) - p_{14} \big(Q_2 + Q_3\big) \big) + 2 \big(p_{13} Q_2 + p_{12} Q_3\big) Q_4\big) + m^4 \big(Q_1 \big(p_{24} \nonumber\\ 
& & + p_{34} - 4 Q_4\big) - p_{14} \big(Q_2 + Q_3 + 2 Q_4\big)\big) + M^2 \big(2 m^4 p_{14} + p_{23} \big(p_{13} p_{24} + p_{12} p_{34} - \big(p_{12} \nonumber\\ 
& & + p_{13}\big) Q_4\big) + m^2 \big(2 p_{14} p_{23} - p_{12} \big(p_{24} + Q_4\big) - p_{13} \big(p_{34} + Q_4\big)\big)\big) + p_{13} \big(\big(p_{14} + 2 p_{24} \nonumber\\ 
& & - Q_1 - 2 Q_2\big) \big(-p_{34} Q_2 + p_{24} Q_3\big) + \big(\big(2 p_{14} + p_{24} - 2 Q_1 - Q_2\big) Q_2 + \big(-p_{24} + Q_2\big) \nonumber\\
& & \times Q_3\big) Q_4 + p_{23} \big(p_{34} \big(Q_1 - Q_2\big) - p_{14} \big(Q_2 + Q_3\big) + p_{24} \big(Q_1 + Q_3 - Q_4\big) + Q_4 \big(-2 Q_1 \nonumber\\
& & + Q_2 + Q_4\big)\big)\big) + p_{12} \big(\big(p_{14} + 2 p_{34} - Q_1 - 2 Q_3\big) \big(p_{34} Q_2 - p_{24} Q_3\big) + \big(\big(2 p_{14} - 2 Q_1 \nonumber
\end{eqnarray}
\begin{eqnarray}
\hspace{7.5mm}  
& & + Q_2 - Q_3\big) Q_3 + p_{34} \big(-Q_2 + Q_3\big) + 2 p_{13} \big(Q_2 + Q_3\big)\big) Q_4 + p_{23} \big(p_{24} \big(Q_1 - Q_3\big) \nonumber\\
& & - p_{14} \big(Q_2 + Q_3\big) + p_{34} \big(Q_1 + Q_2 - Q_4\big) + Q_4 \big(-2 Q_1 + Q_3 + Q_4\big)\big)\big) - p_{23} \big(p_{14}^2 \big(Q_2 \nonumber\\
& & + Q_3\big) + p_{14} \big(-Q_1 Q_2 + p_{34} \big(-Q_1 + Q_2\big) - Q_1 Q_3 - 2 Q_2 Q_3 + p_{24} \big(-Q_1 + Q_3\big) \nonumber\\ 
& & + p_{23} \big(Q_2 + Q_3\big) + 2 Q_1 Q_4\big) + Q_1 \big(p_{34} \big(Q_1 + Q_2\big) - p_{23} \big(p_{24} + p_{34} - 2 Q_4\big) + \big(p_{34} \nonumber\\
& & - 2 Q_1 - Q_2 - Q_3\big) Q_4 + p_{24} \big(-2 p_{34} + Q_1 + Q_3 + Q_4\big)\big)\big) + m^2 \big(p_{13} p_{24} Q_1 \nonumber\\ & & + p_{14} p_{24} Q_1 + 2 p_{23} p_{24} Q_1 - p_{24}^2 Q_1 + p_{13} p_{34} Q_1 + p_{14} p_{34} Q_1 + 2 p_{23} p_{34} Q_1 - p_{34}^2 Q_1 \nonumber\\ & & - p_{24} Q_1^2 - p_{34} Q_1^2 - p_{13} p_{14} Q_2 - p_{14}^2 Q_2 - 2 p_{14} p_{23} Q_2 + p_{14} p_{24} Q_2 - p_{13} p_{34} Q_2 \nonumber\\ & & + p_{14} Q_1 Q_2 + p_{24} Q_1 Q_2 - p_{14} Q_2^2 - p_{13} p_{14} Q_3 - p_{14}^2 Q_3 - 2 p_{14} p_{23} Q_3 + p_{13} p_{24} Q_3 \nonumber\\ & & + p_{14} p_{34} Q_3 + p_{14} Q_1 Q_3 + p_{34} Q_1 Q_3 - p_{14} Q_3^2 + \big(-2 p_{14} \big(p_{23} + Q_1\big) + Q_1 \big(-6 p_{23} \nonumber\\ & & - p_{24} - p_{34} + 2 Q_1 + Q_2 + Q_3\big) + p_{13} \big(p_{34} - 2 Q_1 + 2 Q_2 + Q_3\big)\big) Q_4 + p_{13} Q_4^2 \nonumber\\ & & + p_{12} \big(p_{34} \big(Q_1 + Q_2\big) - p_{14} \big(Q_2 + Q_3\big) + p_{24} \big(Q_1 - Q_3 + Q_4\big) + Q_4 \big(-2 Q_1 + Q_2 \nonumber\\ & & + 2 Q_3 + Q_4\big)\big)\big)\big) \,,
\end{eqnarray}
\begin{eqnarray}
\widehat{t}_{1331} & = & 3 m^6 Q_4 - m^4 \big(p_{14} \big(2 Q_1 + Q_2 + 2 Q_3\big) + 2 p_{34} \big(Q_1 + Q_2 + 4 Q_3\big) + p_{24} \big(Q_1 + 2 \big(Q_2 \nonumber\\ & & + Q_3\big)\big) - \big(p_{12} + 3 p_{13} + 3 p_{23} - p_{34} + 5 Q_3\big) Q_4 - 2 Q_4^2\big) + 2 Q_3 \big(p_{12} p_{34}^2 + 2 p_{23} p_{34} Q_1 \nonumber\\ & & + p_{24} p_{34} Q_1 - 2 p_{34} Q_1 Q_2 - p_{12} p_{34} Q_3 + p_{24} Q_1 Q_3 - p_{13} \big(p_{34} \big(2 p_{23} + p_{24} - 2 Q_2\big) \nonumber\\ & & + p_{24} Q_3\big) + p_{14} \big(-2 p_{24} Q_3 - p_{23} \big(p_{34} + Q_3 - Q_4\big) + Q_2 \big(p_{34} + Q_3 - Q_4\big)\big) + p_{13} \nonumber\\ & & \times \big(2 p_{23} + p_{24} - 2 Q_2\big) Q_4 + \big(-Q_1 \big(2 p_{23} + p_{24} - 2 Q_2\big) + p_{12} \big(-2 p_{34} + Q_3\big)\big) Q_4 \nonumber\\ & & + p_{12} Q_4^2\big) - m^2 \big(p_{23} p_{34} Q_1 - p_{24} p_{34} Q_1 - p_{34}^2 Q_1 + p_{24} Q_1^2 + p_{34} Q_1^2 + 2 p_{13} p_{24} Q_2 \nonumber\\ & & + p_{13} p_{34} Q_2 - p_{34}^2 Q_2 - p_{24} Q_1 Q_2 + p_{34} Q_2^2 + p_{13} p_{24} Q_3 + 2 p_{12} p_{34} Q_3 + 6 p_{13} p_{34} Q_3 \nonumber\\ & & + 6 p_{23} p_{34} Q_3 + p_{24} p_{34} Q_3 + p_{24} Q_1 Q_3 - 4 p_{34} Q_1 Q_3 - p_{24} Q_2 Q_3 - 4 p_{34} Q_2 Q_3 + 4 p_{24} Q_3^2 \nonumber\\ & & + \big(p_{23} \big(p_{34} - Q_2 - 6 Q_3\big) + p_{12} \big(p_{34} - Q_1 - Q_2 - 3 Q_3\big) - p_{13} \big(2 p_{23} - p_{34} + Q_1 \nonumber\\ & & + 6 Q_3\big) + 2 \big(p_{24} Q_1 + p_{34} \big(Q_1 + Q_2\big) + Q_3 \big(3 \big(Q_1 + Q_2\big) + Q_3\big)\big)\big) Q_4 - 2 \big(p_{12} + p_{13} \nonumber\\ & & + p_{23} - Q_3\big) Q_4^2 + p_{14} \big(-p_{34} Q_2 + Q_2^2 + p_{34} Q_3 + Q_2 Q_3 + 4 Q_3^2 + p_{23} \big(2 Q_1 + Q_3\big) \nonumber\\ & & - Q_1 \big(Q_2 + Q_3\big) + 2 Q_2 Q_4\big)\big) + M^2 \big(2 p_{34} \big(-p_{14} p_{23} + p_{12} p_{34} + p_{23} Q_1 + p_{13} \big(-2 p_{23} \nonumber\\ & & - p_{24} + Q_2\big)\big) + \big(p_{14} p_{23} - p_{23} Q_1 + p_{13} \big(2 p_{23} + p_{24} - Q_2\big) - p_{12} \big(p_{34} + Q_3\big)\big) Q_4 \nonumber\\ & & + m^4 \big(p_{14} + p_{24} - 2 p_{34} + 3 Q_4\big) + m^2 \big(p_{14} p_{23} + 2 p_{14} p_{24} - p_{23} p_{24} - 4 p_{23} p_{34} + 2 p_{34} Q_1 \nonumber\\ & & + 2 p_{34} Q_2 - p_{12} \big(p_{14} + p_{24} + 2 p_{34} - Q_4\big) + \big(p_{14} + 3 p_{23} + p_{24} + p_{34} - Q_1 - Q_2 \nonumber\\ & & + Q_3\big) Q_4 + p_{13} \big(-p_{14} + p_{24} - 4 p_{34} + 3 Q_4\big)\big)\big) \,,
\end{eqnarray}
\begin{eqnarray}
\widehat{t}_{1441} & = & M^2 \big(m^4 \big(2 p_{14} - p_{24} + p_{34}\big) + m^2 \big(p_{23} \big(2 p_{14} - p_{24} + p_{34}\big) - p_{13} \big(p_{24} + p_{34}\big) + p_{12} \nonumber\\ & & \times \big( -p_{24} + p_{34} - 2 Q_4\big)\big) + 2 p_{12} p_{23} \big(p_{34} - Q_4\big)\big) + m^4 \big(3 p_{34} Q_1 - p_{14} Q_2 - 8 p_{34} Q_2 \nonumber\\ & & - 3 p_{14} Q_3 - \big(2 p_{14} + p_{34} + 6 Q_1 - 9 Q_2 + 9 Q_3\big) Q_4 + p_{24} \big(Q_1 + 8 Q_3 + Q_4\big)\big) + m^2 \nonumber\\ & & \times \big( 2 p_{12} p_{34} Q_1 + 2 p_{13} p_{34} Q_1 + 2 p_{14} p_{34} Q_1 - p_{34}^2 Q_1 - 2 p_{34} Q_1^2 - 5 p_{12} p_{34} Q_2 - 9 p_{13} p_{34} \nonumber\\ & & \times Q_2 - 3 p_{14} p_{34} Q_2 - p_{34}^2 Q_2 + 3 p_{34} Q_1 Q_2 - p_{14} Q_2^2 + p_{34} Q_2^2 - 2 p_{12} p_{14} Q_3 - 2 p_{13} p_{14} Q_3 \nonumber
\end{eqnarray}
\begin{eqnarray}
\hspace{10mm} & & - 2 p_{14}^2 Q_3 + p_{14} p_{34} Q_3 + 2 p_{14} Q_1 Q_3 + p_{34} Q_1 Q_3 + p_{34} Q_2 Q_3 - p_{14} Q_3^2 + p_{24}^2 \big(-Q_1 \nonumber\\ & & + Q_3\big) - \big(-p_{13} \big(p_{34} - 2 Q_1 + 11 Q_2 - 5 Q_3\big) + p_{12} \big(p_{34} + 2 Q_1 - 7 Q_2 + 3 Q_3\big) \nonumber\\ & & + 2 \big(p_{14} Q_1 - Q_1^2 - 2 p_{14} Q_2 - p_{34} Q_2 + Q_1 Q_2 + Q_2^2 + \big(2 p_{14} + p_{34} - 2 Q_1\big) Q_3 - Q_3^2\big)\big) \nonumber\\ & & \times Q_4 + 2 p_{12} Q_4^2 + p_{24} \big(-p_{34} Q_2 + Q_1 Q_2 + 5 p_{12} Q_3 + 9 p_{13} Q_3 + p_{34} Q_3 - 3 Q_1 Q_3 \nonumber\\ & & - Q_2 Q_3 - Q_3^2 + p_{14} \big(Q_2 + 3 Q_3\big) + \big(p_{12} + p_{13} - 2 \big(Q_1 - Q_2 + Q_3\big)\big) Q_4\big) + p_{23} \big( 5 p_{34} \nonumber\\ & & \times Q_1 - p_{14} Q_2 - 4 p_{34} Q_2 - 5 p_{14} Q_3 - \big(2 p_{14} + p_{34} + 8 Q_1 - 5 Q_2 + 5 Q_3\big) Q_4 + p_{24} \big(Q_1 \nonumber\\ & & + 4 Q_3 + Q_4\big)\big)\big) + 2 \big(p_{12}^2 Q_3 Q_4 + \big(p_{13} + p_{14} + p_{23} + p_{24} - Q_1 - Q_2\big) \big(p_{23} \big(p_{34} Q_1 \nonumber\\ & & - p_{14} Q_3 - Q_1 Q_4\big) + p_{13} \big(-p_{34} Q_2 + p_{24} Q_3 + Q_2 Q_4\big)\big) + p_{12} \big(-\big(p_{13} - p_{34} + Q_3\big) \nonumber\\ & & \times \big(p_{34} Q_2 - p_{24} Q_3\big) + \big(\big(p_{13} - p_{34}\big) Q_2 + \big(p_{13} + p_{14} - Q_1 + Q_2\big) Q_3\big) Q_4 + p_{23} \big(p_{34} Q_1 \nonumber\\ & & - p_{14} Q_3 - \big(p_{34} + Q_1\big) Q_4 + Q_4^2\big)\big)\big) \,,
\end{eqnarray}
\begin{eqnarray}
\widehat{t}_{2442} & = & -4 \big(2 m^6 + m^4 \big(-4 p_{12} + 3 p_{13} + p_{14} + 3 p_{23} + p_{24} - Q_1 - Q_2\big) + m^2 \big(-3 p_{12}^2 \nonumber\\ & & + \big(p_{13} + p_{23}\big) \big(p_{13} + p_{14} + p_{23} + p_{24} - Q_1 - Q_2\big) - 2 p_{12} \big(2 p_{13} + p_{14} + 2 p_{23} + p_{24} \nonumber\\ & & + p_{34} - Q_1 - Q_2 - Q_3\big)\big) - p_{12} \big(\big(p_{13} + p_{23}\big) \big(p_{13} + p_{14} + p_{23} + p_{24} - Q_1 - Q_2\big) \nonumber\\ & & + p_{12} \big(p_{13} + p_{23} - p_{34} + Q_3\big)\big)\big) Q_4 \,,
\end{eqnarray}
\begin{equation}
\widehat{t}_{33} = \widehat{t}_{11}\big|_{p_1\leftrightarrow p_2,m_1\to m}\,, \qquad
\widehat{t}_{44} = \widehat{t}_{22}\big|_{p_1\leftrightarrow p_2, m_1\to m}\,, \nonumber
\end{equation}
\begin{equation}
\widehat{t}_{2332} = \widehat{t}_{1441}\big|_{p_1\leftrightarrow p_2}\,, \qquad
\widehat{t}_{3443} = \widehat{t}_{1221}\big|_{p_1\leftrightarrow p_2,m_1\to m}\,.
\end{equation}

\section*{Appendix B}\label{Appendix_B}

Here, we collect  the expressions for the reduced spin-(in)dependent form factors in the Standard Model case for polarized taus (we recall that the SM contribution, $T_{SM}$, equals the $T^V_{LL}$ contribution of the most general low-energy effective field theory). The origin of the identities below can be traced back to the exchange of indistinguishable fermions as well as to the symmetric character of the electromagnetic tensor $p_i^\mu p_j^\mu+p_j^\mu p_i^\nu-g^{\mu\nu}(p_i\cdot p_j+m^2)$, which comes from summing over the polarizations of the $\ell'^+ \ell'^-$ pair.
\begin{eqnarray}
f^{11}_{SM} & = & 2 \big(2 \big(m_1^2 + p_{12} + p_{13} - Q_1\big) \big(\big(m^2 + p_{23} - Q_2\big) \big(M^2 p_{23} + m^2 \big(M^2 + Q_2\big)\big) \nonumber\\ & & + \big(\big(m^2 + p_{23}\big)^2 - 2 \big(m^2 + p_{23}\big) Q_2 + 2 Q_2^2\big) Q_3 + m^2 Q_3^2\big) + 2 \big(m_1^2 + p_{12} + p_{13} - Q_1\big) \nonumber\\ & & \times \big(M^4 \big(m^2 + p_{23}\big) + M^2 \big(m^4 + m^2 p_{23} - 2 Q_2 Q_3\big) - p_{23} \big(Q_2^2 + Q_3^2\big) - 2 m^2 \big(Q_2^2 \nonumber\\ & & + Q_2 Q_3 + Q_3^2\big)\big) + 2 \big(m_1^2 + p_{12} + p_{13} - Q_1\big) \big(M^2 \big(m^2 + p_{23}\big) \big(m^2 + p_{23} - Q_3\big) \nonumber\\ & & + m^4 \big(Q_2 + Q_3\big) + Q_2 \big(p_{23}^2 - 2 p_{23} Q_3 + 2 Q_3^2\big) + m^2 \big(Q_2^2 - 2 Q_2 Q_3 - Q_3^2 + p_{23} \big(2 Q_2 \nonumber\\ & & + Q_3\big)\big)\big) + \big(M^2 + 3 m_1^2 + 2 \big(m^2 + 2 p_{12} + 2 p_{13} + p_{23} - 2 Q_1 - Q_2 - Q_3\big)\big) \big(M^2 \big(m^2 \nonumber\\ & & + p_{23}\big) \big(p_{12} + p_{13} - Q_1\big) - m^4 Q_1 + p_{12} p_{23} Q_2 - p_{23} Q_1 Q_2 + p_{13} Q_2^2 + p_{13} p_{23} Q_3 \nonumber\\ & & - p_{23} Q_1 Q_3 - p_{12} Q_2 Q_3 - p_{13} Q_2 Q_3 + 2 Q_1 Q_2 Q_3 + p_{12} Q_3^2 + m^2 \big(-Q_1 \big(p_{23} + Q_2 \nonumber\\ & & + Q_3\big) + p_{12} \big(2 Q_2 + Q_3\big) + p_{13} \big(Q_2 + 2 Q_3\big)\big)\big)\big) \,,
\end{eqnarray}
\begin{eqnarray}
l^{11}_{SM} & = & 2 M \big(2 m^2 + M^2 + 3 m_1^2 + 4 p_{12} + 4 p_{13} + 2 p_{23} - 4 Q_1 - 2 Q_2 - 2 Q_3\big) \big(\big(m^2 + M^2\big) \nonumber\\ & & \times \big(m^2 + p_{23}\big) - 2 Q_2 Q_3\big) \,,
\end{eqnarray}
\begin{eqnarray}
{g_1}^{11}_{SM} & = & -2 M \big(m^4 \big(8 p_{12} + 6 p_{13} - 6 Q_1\big) + M^2 \big(p_{12} p_{23} + m^2 \big(2 p_{12} + p_{13} - Q_1\big) - p_{23} Q_1 \nonumber\\ & & + p_{13} \big(Q_2 - Q_3\big)\big) + m_1^2 \big(4 m^4 + 3 p_{12} p_{23} + 2 p_{23}^2 + m^2 \big(6 p_{12} + 3 p_{13} + 6 p_{23} - 3 Q_1 \nonumber\\ & & - 4 Q_2 - 6 Q_3\big) + 3 p_{13} \big(Q_2 - Q_3\big) + 2 Q_3 \big(Q_2 + Q_3\big) - p_{23} \big(3 Q_1 + 2 Q_2 + 4 Q_3\big)\big) \nonumber\\ & & + 2 m^2 \big(4 p_{12}^2 + 2 p_{13}^2 + p_{12} \big(6 p_{13} + 6 p_{23} - 6 Q_1 - 4 Q_2 - 5 Q_3\big) + p_{13} \big(4 p_{23} - 4 Q_1 \nonumber\\ & & - 2 Q_2 - 5 Q_3\big) + Q_1 \big(-5 p_{23} + 2 Q_1 + 3 Q_2 + 4 Q_3\big)\big) + 2 \big(2 p_{12}^2 p_{23} + 2 p_{13}^2 \big(Q_2 - Q_3\big) \nonumber\\ & & + p_{13} \big(\big(p_{23} - 2 Q_1 - Q_2\big) \big(p_{23} + Q_2\big) + \big(-3 p_{23} + 2 Q_1 + Q_2\big) Q_3 + 2 Q_3^2\big) + Q_1 \nonumber\\ & & \times \big(-2 p_{23}^2 + 2 p_{23} \big(Q_1 + Q_2\big) + 3 p_{23} Q_3 - Q_3 \big(Q_2 + Q_3\big)\big) + p_{12} \big(2 p_{23}^2 + 2 p_{13} \big(p_{23} \nonumber\\ & & + Q_2 - Q_3\big) + Q_3 \big(Q_2 + Q_3\big) - p_{23} \big(4 Q_1 + 2 Q_2 + 3 Q_3\big)\big)\big)\big) \,,
\end{eqnarray}
\begin{eqnarray}
f^{22}_{SM} & = & -2 \big(2 m^6 Q_1 + m_1^4 \big(3 m^2 \big(Q_1 + Q_2 + Q_3\big) + 3 p_{23} \big(Q_1 + Q_2 + Q_3\big)\big) - 2 m^4 \big(2 Q_1 \nonumber\\ & & \times \big(-p_{23} + Q_1 + Q_2 + Q_3\big) + p_{12} \big(2 Q_1 + 4 Q_2 + 3 Q_3\big) + p_{13} \big(2 Q_1 + 3 Q_2 + 4 Q_3\big)\big) \nonumber\\ & & - 2 m^2 \big(p_{12}^2 \big(3 Q_1 + 4 Q_2 + 2 Q_3\big) + p_{13}^2 \big(3 Q_1 + 2 Q_2 + 4 Q_3\big) + p_{23} Q_1 \big(-p_{23} + 2 \big(Q_1 \nonumber\\ & & + Q_2 + Q_3\big)\big) + 2 p_{13} \big(2 p_{23} \big(Q_1 + Q_2\big) + 3 p_{23} Q_3 - \big(Q_1 + Q_2 + Q_3\big) \big(Q_1 + Q_2 \nonumber\\ & & + 2 Q_3\big)\big) + 2 p_{12} \big(5 p_{13} \big(Q_1 + Q_2 + Q_3\big) - \big(Q_1 + Q_2 + Q_3\big) \big(Q_1 + 2 Q_2 + Q_3\big) + p_{23} \nonumber\\ & & \times \big(2 Q_1 + 3 Q_2 + 2 Q_3\big)\big)\big) + m_1^2 \big(p_{13}^2 Q_2 + p_{12}^2 Q_3 + p_{13} p_{23} \big(3 Q_1 + 4 Q_2 + 3 Q_3\big) \nonumber\\ & & + m^4 \big(9 Q_1 + 8 \big(Q_2 + Q_3\big)\big) + 2 p_{23} \big(Q_1 + Q_2 + Q_3\big) \big(3 p_{23} - 2 \big(Q_1 + Q_2 + Q_3\big)\big) \nonumber\\ & & + m^2 \big(3 p_{13} Q_1 + 15 p_{23} Q_1 - 4 Q_1^2 + 3 p_{13} Q_2 + 14 p_{23} Q_2 - 8 Q_1 Q_2 - 4 Q_2^2 + 2 \big(p_{13} \nonumber\\ & & + 7 p_{23} - 4 \big(Q_1 + Q_2\big)\big) Q_3 - 4 Q_3^2 + p_{12} \big(3 Q_1 + 2 Q_2 + 3 Q_3\big)\big) + p_{12} \big(3 p_{23} \big(Q_1 + Q_2\big) \nonumber\\ & & + 4 p_{23} Q_3 - p_{13} \big(6 Q_1 + 5 \big(Q_2 + Q_3\big)\big)\big)\big) + M^2 \big(m_1^4 \big(-2 m^2 - 2 p_{23}\big) + m^4 \big(4 \big(p_{12} \nonumber\\ & & + p_{13}\big) + 3 Q_1\big) + p_{12}^2 \big(4 p_{13} + 2 p_{23} + 3 Q_3\big) + p_{13} \big(p_{13} \big(2 p_{23} + 3 Q_2\big) + p_{23} \big(2 p_{23} \nonumber\\ & & - 3 \big(Q_1 + Q_3\big)\big)\big) + m_1^2 \big(-6 m^4 + p_{12} \big(4 p_{13} - 2 p_{23}\big) + m^2 \big(-2 p_{12} - 2 p_{13} - 10 p_{23} \nonumber\\ & & + 3 \big(Q_1 + Q_2 + Q_3\big)\big) + p_{23} \big(-2 p_{13} - 4 p_{23} + 3 \big(Q_1 + Q_2 + Q_3\big)\big)\big) + p_{12} \big(4 p_{13}^2 \nonumber\\ & & + p_{23} \big(2 p_{23} - 3 \big(Q_1 + Q_2\big)\big) + p_{13} \big(8 p_{23} - 3 \big(2 Q_1 + Q_2 + Q_3\big)\big)\big) + m^2 \big(4 p_{12}^2 + 4 p_{13}^2 \nonumber\\ & & + 3 p_{23} Q_1 + 3 p_{12} \big(4 p_{13} + 2 p_{23} - Q_1 - 2 Q_2 - Q_3\big) + p_{13} \big(6 p_{23} - 3 \big(Q_1 + Q_2 + 2 \nonumber\\ & & \times Q_3\big)\big)\big)\big) + 2 \big(p_{12}^3 Q_3 - p_{12}^2 \big(p_{13} \big(4 Q_1 + 3 Q_2 + 2 Q_3\big) + 2 \big(p_{23} \big(Q_1 + Q_2\big) + Q_3 \big(Q_1 \nonumber\\ & & + Q_2 + Q_3\big)\big)\big) + p_{13} \big(p_{13}^2 Q_2 - 2 p_{13} \big(p_{23} \big(Q_1 + Q_3\big) + Q_2 \big(Q_1 + Q_2 + Q_3\big)\big) + p_{23} \big(2 \nonumber\\ & & \times \big(Q_1 + Q_3\big) \big(Q_1 + Q_2 + Q_3\big) - p_{23} \big(2 Q_1 + Q_2 + 2 Q_3\big)\big)\big) + p_{12} \big(-p_{13}^2 \big(4 Q_1 + 2 Q_2 \nonumber\\ & & + 3 Q_3\big) + p_{23} \big(2 \big(Q_1 + Q_2\big) \big(Q_1 + Q_2 + Q_3\big) - p_{23} \big(2 \big(Q_1 + Q_2\big) + Q_3\big)\big) + 2 p_{13} \big(\big(Q_1 \nonumber\\ & & + Q_2 + Q_3\big) \big(2 Q_1 + Q_2 + Q_3\big) - p_{23} \big(4 Q_1 + 3 \big(Q_2 + Q_3\big)\big)\big)\big)\big)\big) \,,
\end{eqnarray}
\begin{eqnarray}
l^{22}_{SM} & = & -2 M \big(-2 m^6 - 3 m_1^4 \big(m^2 + p_{23}\big) + M^2 \big(-m^4 + 2 p_{12} p_{13} - m_1^2 \big(m^2 + p_{23}\big) \nonumber\\ & & + m^2 \big(p_{12} + p_{13} - p_{23}\big) + \big(p_{12} + p_{13}\big) p_{23}\big) + 2 m^4 \big(2 p_{12} + 2 p_{13} - 2 p_{23} + 2 Q_1 + Q_2 \nonumber\\ & & + Q_3\big) + 2 m^2 \big(3 p_{12}^2 + 3 p_{13}^2 + p_{13} \big(4 p_{23} - 2 \big(Q_1 + Q_2\big) - 3 Q_3\big) + p_{12} \big(10 p_{13} + 4 p_{23} \nonumber
\end{eqnarray}
\begin{eqnarray}
\hspace{10mm} & & - 2 Q_1 - 3 Q_2 - 2 Q_3\big) + p_{23} \big(-p_{23} + 2 Q_1 + Q_2 + Q_3\big)\big) + m_1^2 \big(-9 m^4 + p_{12} \big(6 p_{13} \nonumber\\ & & - 3 p_{23}\big) + m^2 \big(-3 p_{12} - 3 p_{13} - 15 p_{23} + 4 \big(Q_1 + Q_2 + Q_3\big)\big) + p_{23} \big(-3 p_{13} - 6 p_{23} \nonumber\\ & & + 4 \big(Q_1 + Q_2 + Q_3\big)\big)\big) + 2 \big(p_{13} \big(p_{13} \big(2 p_{23} + Q_2\big) + p_{23} \big(2 p_{23} - 2 Q_1 - Q_2 - 2 Q_3\big)\big) \nonumber\\ & & + p_{12}^2 \big(4 p_{13} + 2 p_{23} + Q_3\big) + p_{12} \big(4 p_{13}^2 + p_{23} \big(2 p_{23} - 2 \big(Q_1 + Q_2\big) - Q_3\big) + p_{13} \big(8 p_{23} \nonumber\\ & & - 4 Q_1 - 3 \big(Q_2 + Q_3\big)\big)\big)\big)\big) \,,
\end{eqnarray}
\begin{eqnarray}
{g_1}^{22}_{SM} & = & 2 M \big(m_1^4 \big(3 m^2 + 3 p_{23}\big) + M^2 \big(-m^2 \big(2 p_{12} + p_{13}\big) + p_{13}^2 + m_1^2 \big(m^2 + p_{23}\big) - p_{12} \big(p_{13} \nonumber\\ & & + p_{23}\big)\big) - 2 m^4 \big(4 p_{12} + 3 p_{13} + Q_1\big) + m^2 \big(-4 \big(2 p_{12}^2 + 5 p_{12} p_{13} + p_{13}^2 + 3 p_{12} p_{23} \nonumber\\ & & + 2 p_{13} p_{23}\big) + 2 \big(3 p_{12} + 2 p_{13} - p_{23}\big) Q_1 + 4 \big(2 p_{12} + p_{13}\big) Q_2 + 6 \big(p_{12} + p_{13}\big) Q_3\big) \nonumber\\ & & + m_1^2 \big(8 m^4 + p_{13}^2 + 4 p_{13} p_{23} + p_{12} \big(-5 p_{13} + 3 p_{23}\big) + m^2 \big(2 p_{12} + 3 p_{13} + 14 p_{23} \nonumber\\ & & - 4 \big(Q_1 + Q_2 + Q_3\big)\big) + 2 p_{23} \big(3 p_{23} - 2 \big(Q_1 + Q_2 + Q_3\big)\big)\big) + 2 \big(-p_{12}^2 \big(3 p_{13} + 2 p_{23} \nonumber\\ & & + Q_3\big) + p_{13} \big(p_{13}^2 + p_{23} \big(-p_{23} + Q_1 + Q_3\big) - p_{13} \big(Q_1 + 2 Q_2 + Q_3\big)\big) + p_{12} \big(-2 p_{13}^2 \nonumber\\ & & + p_{23} \big(2 \big(Q_1 + Q_2\big) + Q_3 - 2 p_{23}\big) + p_{13} \big(-6 p_{23} + 3 Q_1 + 2 \big(Q_2 + Q_3\big)\big)\big)\big)\big) \,,
\end{eqnarray}
\begin{eqnarray}
f^{1221}_{SM} & = & 2 \big(M^4 \big(3 m^2 \big(p_{12} + p_{13}\big) + 3 \big(p_{12} + p_{13}\big) p_{23}\big) + 12 m^6 Q_1 + m_1^4 \big(-3 m^2 \big(Q_2 + Q_3\big) \nonumber\\ & & - 3 p_{23} \big(Q_2 + Q_3\big)\big) + 4 m^4 \big(p_{13} \big(4 Q_1 - Q_2\big) + p_{12} \big(4 Q_1 - Q_3\big) - 2 Q_1 \big(-4 p_{23} + 3 Q_1 \nonumber\\ & & + 2 \big(Q_2 + Q_3\big)\big)\big) + 4 m^2 \big(p_{13}^2 \big(Q_1 - 2 Q_2\big) + p_{12}^2 \big(Q_1 - 2 Q_3\big) + p_{13} \big(7 p_{23} Q_1 - 3 Q_1^2 \nonumber\\ & & - 2 p_{23} Q_2 + Q_1 Q_2 + 2 Q_2^2 - Q_1 Q_3\big) + p_{12} \big(-Q_1 \big(-2 p_{13} - 7 p_{23} + 3 Q_1 + Q_2\big) \nonumber\\ & & + \big( Q_1 - 2 p_{23}\big) Q_3 + 2 Q_3^2\big) + Q_1 \big(7 p_{23}^2 + \big(Q_1 + Q_2 + Q_3\big) \big(2 Q_1 + Q_2 + Q_3\big) - p_{23} \nonumber\\ & & \times \big(10 Q_1 + 7 \big(Q_2 + Q_3\big)\big)\big)\big) + 4 \big(-p_{13}^3 Q_2 + p_{23} Q_1 \big(2 p_{23} - 2 Q_1 - Q_2 - Q_3\big) \big(p_{23} \nonumber\\ & & - Q_1 - Q_2 - Q_3\big) - p_{12}^3 Q_3 +  p_{13}^2 \big(p_{23} \big(Q_1 - 2 Q_2\big) + Q_2 \big(3 \big(Q_1 + Q_2\big) + Q_3\big)\big) + p_{12}^2 \nonumber\\ & & \times \big(p_{13} Q_2 + p_{23} \big(Q_1 - 2 Q_3\big) + Q_3 \big(3 Q_1 + Q_2 + 3 Q_3\big)\big) + p_{13} \big(p_{23}^2 \big(3 Q_1 - Q_2\big) - Q_2 \nonumber\\ & & \times \big(2 Q_1 + Q_2 - Q_3\big) \big(Q_1 + Q_2 + Q_3\big) + p_{23} \big(2 Q_2^2 - 3 Q_1^2 - 2 Q_1 Q_3\big)\big) + p_{12} \big(p_{23}^2 \big(3 Q_1 \nonumber\\ & & - Q_3\big) + p_{13}^2 Q_3 - Q_3 \big(2 Q_1 - Q_2 + Q_3\big) \big(Q_1 + Q_2 + Q_3\big) + p_{23} \big(-3 Q_1^2 - 2 Q_1 Q_2 \nonumber\\ & & + 2 Q_3^2\big) + p_{13} \big(2 p_{23} Q_1 + \big(Q_2 - Q_3\big)^2 + Q_1 \big(Q_2 + Q_3\big)\big)\big)\big) + M^2 \big(4 m^4 \big(p_{12} + p_{13} \nonumber\\ & & + 2 Q_1\big) + p_{13}^2 \big(2 p_{23} - 5 Q_2\big) + p_{12}^2 \big(2 p_{23} - 5 Q_3\big) + 3 p_{23} Q_1 \big(2 p_{23} - 2 Q_1 - Q_2 - Q_3\big) \nonumber\\ & & + m_1^2 \big(4 m^4 + p_{12} p_{23} + p_{13} p_{23} + 4 p_{23} Q_1 - 4 p_{13} Q_2 - p_{23} Q_2 + m^2 \big(p_{12} + p_{13} + 4 p_{23} \nonumber\\ & & + 4 Q_1 - Q_2 - Q_3\big) - \big(4 p_{12} + p_{23}\big) Q_3\big) + p_{12} \big(4 p_{23}^2 + p_{23} \big(Q_1 - 4 Q_2 - 8 Q_3\big) + 3 Q_3 \nonumber\\ & & \times \big(2 Q_1 - Q_2 + Q_3\big) + p_{13} \big(8 p_{23} - 3 \big(Q_2 + Q_3\big)\big)\big) + m^2 \big(p_{12} \big(4 p_{13} + 8 p_{23} + Q_1 - 4 Q_2 \nonumber\\ & & - 8 Q_3\big) + p_{13} \big(8 p_{23} + Q_1 - 8 Q_2 - 4 Q_3\big) - Q_1 \big(-14 p_{23} + 3 \big(2 Q_1 + Q_2 + Q_3\big)\big)\big) \nonumber\\ & & + p_{13} \big(4 p_{23}^2 + 3 Q_2 \big(2 Q_1 + Q_2 - Q_3\big) + p_{23} \big(Q_1 - 4 \big(2 Q_2 + Q_3\big)\big)\big)\big) + m_1^2 \big(-3 p_{13}^2 Q_2 \nonumber\\ & & - 3 p_{12}^2 Q_3 + m^2 \big(3 p_{12} Q_1 + 3 p_{13} Q_1 + 14 p_{23} Q_1 - 6 Q_1^2 - 4 p_{12} Q_2 - 8 p_{13} Q_2 - 8 p_{23} Q_2 \nonumber\\ & & - Q_1 Q_2 - \big(8 p_{12} + 4 p_{13} + 8 p_{23} + Q_1 - 4 Q_2\big) Q_3\big) + m^4 \big(8 Q_1 - 4 \big(Q_2 + Q_3\big)\big) + p_{23} \nonumber\\ & & \times \big(-6 Q_1^2 - Q_1 \big(Q_2 + Q_3\big) + 2 \big(Q_2^2 + 4 Q_2 Q_3 + Q_3^2\big) + p_{23} \big(6 Q_1 - 4 \big(Q_2 + Q_3\big)\big)\big) \nonumber
\end{eqnarray}
\begin{eqnarray}
\hspace{7mm}  & & + p_{13} \big(Q_2 \big( 6 Q_1 + 5 Q_2 + 3 Q_3\big) + p_{23} \big(3 Q_1 - 4 \big(2 Q_2 + Q_3\big)\big)\big) + p_{12} \big(3 p_{13} \big(Q_2 + Q_3\big) \nonumber\\ & & + Q_3 \big(6 Q_1 + 3 Q_2 + 5 Q_3\big) + p_{23} \big(3 Q_1 - 4 \big(Q_2 + 2 Q_3\big)\big)\big)\big)\big) \,,
\end{eqnarray}
\begin{eqnarray}
l^{1221}_{SM} & = & -2 M \big(12 m^6 + M^2 \big(8 m^4 + m^2 \big(-p_{12} - p_{13} + 10 p_{23} - 2 Q_1\big) - p_{23} \big(p_{13} - 2 p_{23} \nonumber\\ & & + 2 Q_1\big) + 2 p_{13} Q_2 - p_{12} \big(p_{23} - 2 Q_3\big)\big) + 4 m^4 \big(4 p_{12} + 4 p_{13} + 8 p_{23} - 3 \big(2 Q_1 + Q_2 \nonumber\\ & & + Q_3\big)\big) + 2 m^2 \big(2 p_{12}^2 + 2 p_{13}^2 + 14 p_{23}^2 - 20 p_{23} Q_1 + 4 Q_1^2 - 10 p_{23} Q_2 + 4 Q_1 Q_2 - Q_2^2 \nonumber\\ & & + p_{13} \big(14 p_{23} - 6 Q_1 + 2 Q_2 - Q_3\big) - 10 p_{23} Q_3 + 4 Q_1 Q_3 - Q_3^2 + p_{12} \big(4 p_{13} + 14 p_{23} \nonumber\\ & & - 6 Q_1 - Q_2 + 2 Q_3\big)\big) + m_1^2 \big(8 m^4 + 3 p_{13} \big(p_{23} + 2 Q_2\big) + 2 p_{23} \big(3 p_{23} - 3 Q_1 - Q_2 \nonumber\\ & & - Q_3\big) + 3 p_{12} \big(p_{23} + 2 Q_3\big) + m^2 \big(3 p_{12} + 3 p_{13} + 14 p_{23} - 2 \big(3 Q_1 + Q_2 + Q_3\big)\big)\big) \nonumber\\ & & + 2 \big(2 p_{13}^2 \big(p_{23} + 2 Q_2\big) + 2 p_{12}^2 \big(p_{23} + 2 Q_3\big) + 2 p_{23} \big(2 p_{23}^2 + 2 Q_1 \big(Q_1 + Q_2\big) + \big(2 Q_1 \nonumber\\ & & + Q_2\big) Q_3 - 2 p_{23} \big(2 Q_1 + Q_2 + Q_3\big)\big) + p_{13} \big(6 p_{23}^2 + p_{23} \big(-6 Q_1 + Q_2 - 2 Q_3\big) \nonumber\\ & & - Q_2 \big(4 Q_1 + 3 Q_2 + Q_3\big)\big) + p_{12} \big(6 p_{23}^2 + p_{23} \big(-6 Q_1 - 2 Q_2 + Q_3\big) + 4 p_{13} \big(p_{23} + Q_2 \nonumber\\ & & + Q_3\big) - Q_3 \big(4 Q_1 + Q_2 + 3 Q_3\big)\big)\big)\big) \,,
\end{eqnarray}
\begin{eqnarray}
{g_1}^{1221}_{SM} & = & 2 M \big(m_1^4 \big(3 m^2 + 3 p_{23}\big) + 4 m^4 \big(p_{13} + Q_1\big) + m^2 \big(8 p_{13}^2 + 8 p_{13} \big(p_{23} - Q_2\big) + 2 Q_1 \big(p_{12} \nonumber\\ & & + 4 p_{23} - 2 Q_1 - 3 Q_2 - 2 Q_3\big)\big) + m_1^2 \big(4 m^4 - 3 p_{12} p_{13} + 3 p_{13}^2 + 4 p_{12} p_{23} + 8 p_{13} p_{23} \nonumber\\ & & + 4 p_{23}^2 - p_{23} Q_1 - 5 p_{13} Q_2 - 2 p_{23} Q_2 + m^2 \big(4 p_{12} + 8 p_{13} + 8 p_{23} - Q_1 - 2 Q_3\big) \nonumber\\ & & - \big(2 p_{12} + p_{13} + 4 p_{23}\big) Q_3\big) + M^2 \big(-p_{12} \big(p_{13} - 2 p_{23}\big) + m_1^2 \big(m^2 + p_{23}\big) + p_{23} Q_1 \nonumber\\ & & + m^2 \big(4 p_{12} + 2 p_{13} + Q_1\big) + p_{13} \big(p_{13} - Q_2 + Q_3\big)\big) + 2 \big(2 p_{13}^3 + p_{13}^2 \big(4 p_{23} - 2 Q_1 \nonumber\\ & & - 6 Q_2 - Q_3\big) + p_{23} Q_1 \big(2 p_{23} - 2 \big(Q_1 + Q_2\big) - Q_3\big) - p_{12}^2 \big(2 p_{13} + Q_3\big) + p_{12} \big(2 p_{23} Q_1 \nonumber\\ & & + \big(Q_1 - Q_2 - Q_3\big) Q_3 + 2 p_{13} \big(Q_1 - Q_2 + Q_3\big)\big) + p_{13} \big(2 p_{23}^2 + p_{23} \big(Q_1 - 4 Q_2\big) \nonumber\\ & & + 3 Q_1 Q_2 + \big(2 Q_2 - Q_3\big)
 \big(Q_2 + Q_3\big)\big)\big)\big) \,,
\end{eqnarray}
\begin{equation}
{g_2}^{11}_{SM} = {g_1}^{11}_{SM} \big|_{p_2\leftrightarrow p_3} \,, \qquad
{g_2}^{22}_{SM} = {g_1}^{22}_{SM} \big|_{p_2\leftrightarrow p_3} \,, \qquad
{g_2}^{1221}_{SM} = {g_1}^{1221}_{SM} \big|_{p_2\leftrightarrow p_3} \,,
\end{equation}

\begin{equation}
f^{33}_{SM} = f^{11}_{SM} \big|_{p_1\leftrightarrow p_2, m_1\to m} \,, \qquad
l^{33}_{SM} = {g_1}^{11}_{SM} \big|_{p_1\leftrightarrow p_2, m_1\to m} \,,
\end{equation}
\begin{equation}
{g_1}^{33}_{SM} = l^{11}_{SM} \big|_{p_1\leftrightarrow p_2, m_1\to m} \,, \qquad
{g_2}^{33}_{SM} = {g_2}^{11}_{SM} \big|_{p_1\leftrightarrow p_2, m_1\to m} \,, \nonumber
\end{equation}

\begin{equation}
f^{44}_{SM} = f^{22}_{SM} \big|_{p_1\leftrightarrow p_2, m_1\to m} \,, \qquad
l^{44}_{SM} = {g_1}^{22}_{SM} \big|_{p_1\leftrightarrow p_2, m_1\to m} \,,
\end{equation}
\begin{equation}
{g_1}^{44}_{SM} = l^{22}_{SM} \big|_{p_1\leftrightarrow p_2, m_1\to m} \,, \qquad
{g_2}^{44}_{SM} = {g_2}^{22}_{SM} \big|_{p_1\leftrightarrow p_2, m_1\to m} \,, \nonumber
\end{equation}
\begin{eqnarray}
f^{1331}_{SM} & = & m^6 \big(13 \big(Q_1 + Q_2\big) + 45 Q_3\big) + m^4 \big(10 p_{12} Q_1 + 8 p_{13} Q_1 + 21 p_{23} Q_1 - 6 Q_1^2 + 10 p_{12} Q_2 \nonumber\\ & & + 21 p_{13} Q_2 + 8 p_{23} Q_2 - 32 Q_1 Q_2 - 6 Q_2^2 + \big(25 p_{12} + 64 p_{13} + 64 p_{23} - 68 \big(Q_1 + Q_2\big)\big) \nonumber\\ & & \times Q_3 - 28 Q_3^2\big) - 8 Q_3 \big(-2 \big(p_{13} - Q_1\big) \big(p_{23} - Q_2\big) \big(p_{13} + p_{23} - Q_1 - Q_2 - Q_3\big) \nonumber\\ & & + p_{12} \big(-\big(p_{13} - Q_1\big) \big(p_{23} - Q_2\big) + \big(-p_{13} - p_{23} + Q_1 + Q_2\big) Q_3 + Q_3^2\big)\big) + M^4 \nonumber\\ & & \times \big(12 m^4 + 8 p_{13} p_{23} - 3 p_{23} Q_1 - 3 p_{13} Q_2 - 3 p_{12} Q_3 + m^2 \big(4 p_{12} + 3 \big(4 p_{13} + 4 p_{23} - Q_1 \nonumber\\ & & - Q_2 + Q_3\big)\big)\big) + 2 m^2 \big(Q_1 \big(4 p_{23}^2 + 4 Q_2 \big(Q_1 + Q_2\big) - 3 p_{23} \big(Q_1 + 3 Q_2\big)\big) - p_{12}^2 Q_3 \nonumber
\end{eqnarray}
\begin{eqnarray}
\hspace{12mm} & & + 12 \big(p_{23}^2 - 3 p_{23} Q_1 + Q_1^2 - 2 p_{23} Q_2 + 3 Q_1 Q_2 + Q_2^2\big) Q_3 + 12 \big(-p_{23} + Q_1 + Q_2\big) Q_3^2 \nonumber\\ & & + 2 Q_3^3 + 4 p_{13}^2 \big(Q_2 + 3 Q_3\big) + p_{12} \big(5 p_{23} Q_1 + 5 p_{13} Q_2 - 8 Q_1 Q_2 + \big(8 p_{13} + 8 p_{23} - 9 \big(Q_1 \nonumber\\ & & + Q_2\big)\big) Q_3 + 6 Q_3^2\big) + p_{13} \big(4 p_{23} \big(Q_1 + Q_2\big) + 34 p_{23} Q_3 - 3 \big(3 Q_1 Q_2 + Q_2^2 + 8 Q_1 Q_3 \nonumber\\ & & + 12 Q_2 Q_3 + 4 Q_3^2\big)\big)\big) + M^2 \big(24 m^6 + m^4 \big(24 p_{12} + 46 p_{13} + 46 p_{23} - 20 \big(Q_1 + Q_2\big)\big) \nonumber\\ & & + 2 m^2 \big(4 p_{12}^2 + 11 p_{13}^2 + 30 p_{13} p_{23} + 11 p_{23}^2 - 12 p_{13} Q_1 - 13 p_{23} Q_1 + 2 Q_1^2 - 13 p_{13} Q_2 \nonumber\\ & & - 12 p_{23} Q_2 + 2 Q_1 Q_2 + 2 Q_2^2 + p_{12} \big(11 p_{13} + 11 p_{23} - 4 \big(Q_1 + Q_2\big) - 3 Q_3\big) - \big(3 p_{13} \nonumber\\ & & + 3 p_{23} + 7 \big(Q_1 + Q_2\big)\big) Q_3 - 4 Q_3^2\big) - 2 \big(p_{13}^2 \big(-8 p_{23} + 3 Q_2\big) + p_{12}^2 Q_3 + p_{13} \big(-8 p_{23}^2 \nonumber\\ & & + 9 p_{23} \big(Q_1 + Q_2\big) - 2 Q_2 \big(Q_1 + Q_2 - Q_3\big) + 4 p_{23} Q_3\big) + Q_1 \big(3 p_{23}^2 - 2 p_{23} \big(Q_1 + Q_2 \nonumber\\ & & - Q_3\big) - 6 Q_2 Q_3\big) + p_{12} \big(p_{23} \big(Q_1 + Q_3\big) + p_{13} \big(-4 p_{23} + Q_2 + Q_3\big) - 2 Q_3 \big(Q_1 + Q_2 \nonumber\\ & & + 2 Q_3\big)\big)\big)\big) \,,
\end{eqnarray}
\begin{eqnarray}
l^{1331}_{SM} & = & M \big(-13 m^6 + M^4 \big(m^2 + p_{23}\big) + m^4 \big(-10 p_{12} - 8 p_{13} - 21 p_{23} + 6 Q_1 + 16 Q_2 \nonumber\\ & & + 18 Q_3\big) + 4 Q_3 \big(p_{23} \big(p_{12} + 2 \big(p_{13} + p_{23} - Q_1\big)\big) - \big(p_{12} + 3 \big(p_{13} + p_{23} - Q_1\big)\big) Q_2 \nonumber\\ & & + Q_2^2 + \big(p_{12} - 2 p_{23} + 2 Q_2\big) Q_3\big) - 2 m^2 \big(5 p_{12} p_{23} + 4 p_{13} p_{23} + 4 p_{23}^2 - 3 p_{23} Q_1 \nonumber\\ & & - 4 p_{12} Q_2 - 2 p_{13} Q_2 - 7 p_{23} Q_2 + Q_1 Q_2 + 3 Q_2^2 + \big(-3 p_{12} - 5 p_{13} - 13 p_{23} + 5 Q_1 \nonumber\\ & & + 12 Q_2\big) Q_3 + 6 Q_3^2\big) + M^2 \big(4 m^4 + 2 p_{23} \big(p_{12} + 3 p_{13} + p_{23} - 2 Q_1\big) - 2 \big(p_{13} + p_{23}\big) Q_2 \nonumber\\ & & - 2 \big(p_{12} + Q_2\big) Q_3 + 2 m^2 \big(p_{12} + 3 p_{13} + 3 p_{23} - 2 Q_1 - Q_2 + Q_3\big)\big)\big) \,,
\end{eqnarray}
\begin{eqnarray}
{g_2}^{1331}_{SM} & = & M \big(-45 m^6 - M^4 p_{12} - 2 M^2 \big(p_{12}^2 + 4 p_{13} p_{23} - 2 p_{23} Q_1 + p_{12} \big(p_{13} + p_{23} - Q_1 - Q_2\big) \nonumber\\ & & - 2 p_{13} Q_2\big) + m^4 \big(-12 M^2 - 25 p_{12} - 64 p_{13} - 64 p_{23} + 50 \big(Q_1 + Q_2\big) + 28 Q_3\big) \nonumber\\ & & + m^2 \big(M^4 - 2 M^2 \big(p_{12} + 5 p_{13} + 5 p_{23} - 2 \big(Q_1 + Q_2\big)\big) + 2 \big(p_{12}^2 - 8 p_{12} p_{13} - 12 p_{13}^2 \nonumber\\ & & - 8 p_{12} p_{23} - 34 p_{13} p_{23} - 12 p_{23}^2 + 6 p_{12} Q_1 + 19 p_{13} Q_1 + 23 p_{23} Q_1 - 7 Q_1^2 + 6 p_{12} Q_2 \nonumber\\ & & + 23 p_{13} Q_2 + 19 p_{23} Q_2 - 12 Q_1 Q_2 - 7 Q_2^2 - 6 \big(p_{12} - 2 p_{13} - 2 p_{23} + Q_1 + Q_2\big) Q_3 \nonumber\\ & & - 2 Q_3^2\big)\big) - 4 \big(2 \big(2 p_{13} p_{23} - p_{23} Q_1 - p_{13} Q_2\big) \big(p_{13} + p_{23} - Q_1 - Q_2 - Q_3\big) - p_{12} \big(p_{23} \nonumber\\ & & \times \big(Q_1 - 2 Q_3\big) + p_{13} \big(-2 p_{23} + Q_2 - 2 Q_3\big) + Q_3 \big(Q_1 + Q_2 + 2 Q_3\big)\big)\big)\big) \,,
\end{eqnarray}
\begin{eqnarray}
f^{1441}_{SM} & = & -2 \big(M^4 \big(3 m^2 p_{12} + 3 p_{12} p_{23}\big) + m^6 \big(10 Q_1 - 15 Q_2 + 8 Q_3\big) + m^4 \big(7 p_{12} Q_1 + 12 p_{13} Q_1 \nonumber\\ & & + 23 p_{23} Q_1 - 15 Q_1^2 - 10 p_{12} Q_2 - 31 p_{13} Q_2 - 23 p_{23} Q_2 + 10 Q_1 Q_2 + 16 Q_2^2 + \big(19 p_{12} \nonumber\\ & & + 6 p_{13} + 12 p_{23} - 27 Q_1 + 2 Q_2\big) Q_3 - 16 Q_3^2\big) + m^2 \big(17 p_{23}^2 Q_1 - 23 p_{23} Q_1^2 + 4 Q_1^3 \nonumber\\ & & + 4 p_{13}^2 \big(Q_1 - 5 Q_2\big) - 8 p_{23}^2 Q_2 - 4 p_{23} Q_1 Q_2 + 12 p_{23} Q_2^2 - 8 Q_1 Q_2^2 - 4 Q_2^3 + 9 p_{12}^2 Q_3 \nonumber\\ & & + \big(4 p_{23}^2 - 25 p_{23} Q_1 + 12 Q_1^2 + 4 \big(p_{23} + Q_1\big) Q_2 - 4 Q_2^2\big) Q_3 + 2 \big(-5 p_{23} + 6 Q_1 + 2 Q_2\big) \nonumber\\ & & \times Q_3^2 + 4 Q_3^3 + p_{12} \big(4 p_{13} Q_1 + 11 p_{23} Q_1 - 4 Q_1^2 - 13 p_{13} Q_2 - 10 p_{23} Q_2 + 8 Q_1 Q_2 \nonumber\\ & & + 8 Q_2^2 + \big(16 p_{13} + 13 p_{23} - 13 Q_1 - 8 Q_2\big) Q_3 - 11 Q_3^2\big) + p_{13} \big(20 p_{23} Q_1 - 8 Q_1^2 \nonumber\\ & & - 29 p_{23} Q_2 + 23 Q_1 Q_2 + 24 Q_2^2 + 6 p_{23} Q_3 - 12 Q_1 Q_3 + 11 Q_2 Q_3 - 8 Q_3^2\big)\big) + M^2 \big(2 m^6 \nonumber\\ & & + 3 \big(p_{13} + p_{23} - Q_1 - Q_2\big) \big(p_{23} Q_1 - p_{13} Q_2\big) + 2 p_{12}^2 \big(p_{23} - Q_3\big) + m^4 \big(4 p_{12} + p_{13} \nonumber\\ & & + 2 p_{23} + 6 Q_1 - 10 Q_2 + 9 Q_3\big) + p_{12} \big(4 p_{23}^2 + p_{13} \big(4 p_{23} - 2 Q_2 - Q_3\big) + 3 \big(Q_1 - Q_2\big) \nonumber
\end{eqnarray}
\begin{eqnarray}
\hspace{12mm} & & \times Q_3 - p_{23} \big(2 Q_1 + 4 Q_2 + 5 Q_3\big)\big) + m^2 \big(p_{12}^2 - p_{13}^2 + 9 p_{23} Q_1 - 3 Q_1^2 - 6 p_{23} Q_2 \nonumber\\ & & + 3 Q_1 Q_2 + 3 Q_2^2 + 5 p_{23} Q_3 - 6 Q_1 Q_3 - 3 Q_3^2 + p_{12} \big(2 p_{13} + 8 p_{23} - 2 \big(Q_1 + 5 Q_2\big) \nonumber\\ & & + Q_3\big) + p_{13} \big(p_{23} + 3 Q_1 - 13 Q_2 + 6 Q_3\big)\big)\big) + 4 \big(-\big(p_{13} + p_{23} - Q_1 - Q_2\big) \big(-p_{23} Q_1 \nonumber\\ & & + p_{13} Q_2\big) \big(p_{13} + p_{23} - Q_1 - Q_2 - Q_3\big) + p_{12}^2 Q_3 \big(p_{13} + Q_1 + Q_3\big) + p_{12} \big(p_{13}^2 \big(-Q_2 \nonumber\\ & & + Q_3\big) + \big(p_{23} - Q_1 - Q_2 - Q_3\big) \big(p_{23} Q_1 + \big(Q_1 - Q_2\big) Q_3\big) +  p_{13} \big(Q_2 \big(Q_1 + Q_2 - Q_3\big) \nonumber\\ & & + p_{23} \big(Q_1 - Q_2 + Q_3\big)\big)\big)\big)\big) \,,
\end{eqnarray}
\begin{eqnarray}
l^{1441}_{SM} & = & 2 M \big(10 m^6 + 4 \big(p_{12} + p_{13} + p_{23} - Q_1 - Q_2\big) \big(p_{23} \big(p_{13} + p_{23} - Q_1\big) + p_{13} Q_2\big) + m^4 \nonumber\\ & & \times \big(4 M^2 + 7 p_{12} + 12 p_{13} + 23 p_{23} - 15 Q_1 + 7 Q_2 - 21 Q_3\big) + 2 \big(2 p_{12}^2 + p_{12} \big(2 p_{13} + p_{23} \nonumber\\ & & - 2 Q_1\big) - 2 p_{23} \big(p_{13} + p_{23} - Q_1\big) - \big(p_{13} - p_{23}\big) Q_2\big) Q_3 - 2 p_{12} Q_3^2 + m^2 \big(4 p_{13}^2 + 20 p_{13} \nonumber\\ & & \times p_{23} + 17 p_{23}^2 - 8 p_{13} Q_1 - 23 p_{23} Q_1 + 4 Q_1^2 + 14 p_{13} Q_2 - 3 p_{23} Q_2 - 4 Q_1 Q_2 - 7 Q_2^2 \nonumber\\ & & + p_{12} \big(4 p_{13} + 11 p_{23} - 4 Q_1 + 11 Q_2 - 4 Q_3\big) + M^2 \big(-2 p_{12} + p_{13} + 5 p_{23} - Q_1 + 2 Q_2 \nonumber\\ & & - 2 Q_3\big) - 13 p_{13} Q_3 - 19 p_{23} Q_3 + 12 Q_1 Q_3 + 5 Q_3^2\big) + M^2 \big(p_{23} \big(p_{13} + p_{23} - Q_1\big) \nonumber\\ & & + p_{13} Q_2 + p_{12} \big(-2 p_{23} + Q_3\big)\big)\big) \,,
\end{eqnarray}
\begin{eqnarray}
{g_1}^{1441}_{SM} & = & -2 M \big(15 m^6 + 4 \big(p_{13}^2 + p_{23} Q_1 + p_{13} \big(p_{23} - Q_2\big)\big) \big(p_{12} + p_{13} + p_{23} - Q_1 - Q_2\big) \nonumber\\ & & - 2 \big(p_{12}^2 + 2 p_{13}^2 + p_{23} Q_1 + p_{13} \big(2 p_{23} - Q_1 - 2 Q_2\big) + p_{12} \big(p_{13} - Q_1 + Q_2\big)\big) Q_3 \nonumber\\ & & - 2 p_{12} Q_3^2 + M^2 \big(3 m^4 + p_{13}^2 + p_{23} \big(2 p_{12} + Q_1\big) + p_{13} \big(p_{23} - Q_2\big) + m^2 \big(3 p_{12} + 4 p_{13} \nonumber\\ & & + 3 p_{23} + Q_1 - Q_2 + Q_3\big)\big) + m^4 \big(10 p_{12} + 31 p_{13} + 23 p_{23} - 3 Q_1 - 2 \big(8 Q_2 + Q_3\big)\big) \nonumber\\ & & + m^2 \big(20 p_{13}^2 + 8 p_{23}^2 + p_{23} Q_1 - 4 Q_1^2 - 12 p_{23} Q_2 + Q_1 Q_2 + 4 Q_2^2 - 4 p_{23} Q_3 - 5 Q_1 Q_3 \nonumber\\ & & + Q_2 Q_3 - 3 Q_3^2 + p_{12} \big(13 p_{13} + 10 p_{23} + 3 Q_1 - 8 Q_2 + Q_3\big) + p_{13} \big(29 p_{23} - 9 Q_1 \nonumber\\ & & - 8 \big(3 Q_2 + Q_3\big)\big)\big)\big) \,,
\end{eqnarray}
\begin{eqnarray}
{g_2}^{1441}_{SM} & = & 2 M \big(8 m^6 + m^4 \big(2 M^2 + 19 p_{12} + 6 p_{13} + 12 p_{23} - 6 Q_1 - 16 Q_3\big) + m^2 \big(9 p_{12}^2 + 6 p_{13} p_{23} \nonumber\\ & & + 4 p_{23}^2 + p_{13} Q_1 - 6 p_{23} Q_1 + 3 p_{13} Q_2 - Q_1 Q_2 - 3 Q_2^2 + p_{12} \big(16 p_{13} + 13 p_{23} - 9 Q_1 \nonumber\\ & & - 7 Q_2 - 11 Q_3\big) + M^2 \big(2 p_{12} - p_{13} + 2 p_{23} + Q_2 - Q_3\big) - 8 p_{13} Q_3 - 10 p_{23} Q_3 + 7 Q_1 \nonumber\\ & & \times Q_3 + Q_2 Q_3 + 4 Q_3^2\big) + p_{12} \big(M^2 \big(p_{13} + p_{23} - Q_2\big) + 2 \big(2 p_{12} p_{13} + 2 p_{13}^2 + 2 p_{13} p_{23} \nonumber\\ & & - 2 p_{13} Q_1 - p_{23} Q_1 - p_{12} Q_2 - 3 p_{13} Q_2 - 2 p_{23} Q_2 + Q_1 Q_2 + Q_2^2 + \big(2 p_{12} - Q_1 + Q_2\big) \nonumber\\ & & \times Q_3\big)\big)\big) \,,
\end{eqnarray}
\begin{eqnarray}
f^{2442}_{SM} & = & 4 \big(m^6 \big(7 \big(Q_1 + Q_2\big) + 4 Q_3\big) + m^4 \big(-\big(Q_1 + Q_2\big) \big(12 p_{12} - 11 p_{13} - 11 p_{23} + 4 \big(Q_1 \nonumber\\ & & + Q_2\big)\big) - 2 \big(7 p_{12} - 3 p_{13} - 3 p_{23} + 2 \big(Q_1 + Q_2\big)\big) Q_3\big) + 2 p_{12} \big(-2 \big(p_{13} + p_{23}\big) \big(p_{12} \nonumber\\ & & + p_{13} + p_{23} - Q_1 - Q_2\big) \big(Q_1 + Q_2\big) + \big(p_{12} - p_{13} - p_{23}\big) \big(p_{12} + p_{13} + p_{23} - 2 \big(Q_1 \nonumber\\ & & + Q_2\big)\big) Q_3 - 2 p_{12} Q_3^2\big) + m^2 \big(-p_{12}^2 \big(10 \big(Q_1 + Q_2\big) + 7 Q_3\big) + 2 \big(p_{13} + p_{23}\big) \big(2 \big(p_{13} \nonumber\\ & & + p_{23} - Q_1 - Q_2\big) \big(Q_1 + Q_2\big) + \big(p_{13} + p_{23} - 2 \big(Q_1 + Q_2\big)\big) Q_3\big) + p_{12} \big(-13 p_{13} \big(Q_1 \nonumber\\ & & + Q_2\big) - 13 p_{23} \big(Q_1 + Q_2\big) - 12 \big(p_{13}+p_{23}\big) Q_3 + 8 \big(Q_1 + Q_2 + Q_3\big)^2\big)\big) + M^2 \nonumber\\ & & \times \big(-4 m^6 + m^4 \big(8 p_{12} + 3 \big(-2 p_{13} - 2 p_{23} + Q_1 + Q_2\big)\big) + p_{12} \big(\big(p_{13} + p_{23}\big) \big(2 p_{12} \nonumber
\end{eqnarray}
\begin{eqnarray}
\hspace{7mm} & & + 2 p_{13} + 2 p_{23} - 3 \big(Q_1 + Q_2\big)\big) + 3 p_{12} Q_3\big) + m^2 \big(6 p_{12}^2 - \big(p_{13} + p_{23}\big) \big(2 p_{13} + 2 p_{23} \nonumber\\ & & - 3 \big(Q_1 + Q_2\big)\big) + p_{12} \big(8 p_{13} + 8 p_{23} - 6 \big(Q_1 + Q_2 + Q_3\big)\big)\big)\big)\big) \,,
\end{eqnarray}
\begin{eqnarray}
l^{2442}_{SM} & = & 4 M \big(-7 m^6 + p_{12} \big(p_{13} + p_{23}\big) \big(M^2 + 4 \big(p_{12} + p_{13} + p_{23} - Q_1 - Q_2\big)\big) + 2 p_{12} \big(p_{12} \nonumber\\ & & - p_{13} - p_{23}\big) Q_3 + m^4 \big(-M^2 + 12 p_{12} - 11 p_{13} - 11 p_{23} + 4 \big(Q_1 + Q_2\big) + 2 Q_3\big) \nonumber\\ & & + m^2 \big(10 p_{12}^2 + M^2 \big(2 p_{12} - p_{13} - p_{23}\big) + 2 \big(p_{13} + p_{23}\big) \big(-2 p_{13} - 2 p_{23} + 2 \big(Q_1 + Q_2\big) \nonumber\\ & & + Q_3\big) + p_{12} \big(13 p_{13} + 13 p_{23} - 8 \big(Q_1 + Q_2 + Q_3\big)\big)\big)\big) \,,
\end{eqnarray}
\begin{eqnarray}
{g_2}^{2442}_{SM} & = & 4 M \big(-4 m^6 + 2 m^4 \big(7 p_{12} - 3 p_{13} - 3 p_{23} + Q_1 + Q_2\big) +  p_{12} \big(-M^2 p_{12} - 2 \big(p_{12} - p_{13} \nonumber\\ & & - p_{23}\big) \big(p_{12} + p_{13} + p_{23} - Q_1 - Q_2\big) + 4 p_{12} Q_3\big) + m^2 \big(2 M^2 p_{12} + 7 p_{12}^2 - 2 \big(p_{13} + p_{23}\big) \nonumber\\ & & \times \big(p_{13} + p_{23} - Q_1 - Q_2\big) + 4 p_{12} \big(3 p_{13} + 3 p_{23} - 2 \big(Q_1 + Q_2 + Q_3\big)\big)\big)\big) \,,
\end{eqnarray}
\begin{equation}
{g_1}^{1331}_{SM} = l^{1331}_{SM} \big{|}_{p_1\leftrightarrow p_2} \,, \qquad
{g_1}^{2442}_{SM} = l^{2442}_{SM} \,,
\end{equation}

\begin{equation}
f^{2332}_{SM} = f^{1441}_{SM} \big|_{p_1\leftrightarrow p_2} \,, \qquad
l^{2332}_{SM} = {g_1}^{1441}_{SM} \big|_{p_1\leftrightarrow p_2} \,,
\end{equation}
\begin{equation}
{g_1}^{2332}_{SM} = l^{1441}_{SM} \big|_{p_1\leftrightarrow p_2} \,, \qquad
{g_2}^{2332}_{SM} = {g_2}^{1441}_{SM} \big|_{p_1\leftrightarrow p_2} \,, \nonumber
\end{equation}

\begin{equation}
f^{3443}_{SM} = f^{1221}_{SM} \big|_{p_1\leftrightarrow p_2, m_1\to m} \,, \qquad
l^{3443}_{SM} = {g_1}^{1221}_{SM} \big|_{p_1\leftrightarrow p_2, m_1\to m} \,,
\end{equation}
\begin{equation}
{g_1}^{3443}_{SM} = l^{1221}_{SM} \big|_{p_1\leftrightarrow p_2, m_1\to m} \,, \qquad
{g_2}^{3443}_{SM} = {g_2}^{1221}_{SM} \big|_{p_1\leftrightarrow p_2, m_1\to m} \,. \nonumber
\end{equation}

\section*{Appendix C}\label{Appendix_C}

For decays with five or more particles in the final state, not all of the scalar products of their momenta can be written as linear combinations of the invariant variables. In these cases, symmetry considerations allow to derive their expressions \cite{Kumar, Kumar2}. Although the procedure is quite transparent, results are not always compact. In our case this is so for a particular product, $p_3\cdot p_4$, whose explicit expression we quote in this Appendix for the reader's convenience.

\begin{eqnarray}
p_3\cdot p_4 & = & \Big[\big(m^2 + M^2 - u_2\big) \big(s_3 \big(s_3 + t_2 - m_1^2\big) \big(u_1 + u_2 -2 m^2\big) - 2 s_3 \big(t_2 - t_3\big) \big(u_1 + u_2 \nonumber\\
& & -2 m^2\big) - 2 s_3 t_2 \big(s_1 + u_1 + u_2 -2 m^2 - M^2 - m_1^2\big) +  \big(s_3 + t_2 - m_1^2\big) \big(t_2 - t_3\big) \big(s_1 \nonumber\\
& & + u_1 + u_2 -2 m^2 - M^2 - m_1^2\big) + 4 s_3 t_2 \big(M^2 - u_3\big) - \big(s_3 + t_2 - m_1^2\big)^2 \big(M^2 - u_3\big)\big) \nonumber\\
& & + \big(s_2 - s_3 -m^2\big) \big(4 M^2 s_3 t_2 - 2 M^2 \big(s_3 + t_2 - m_1^2\big) \big(t_2 - t_3\big) - s_3 \big(u_1 + u_2 - 2 m^2\big)^2 \nonumber\\
& & + \big(t_2 - t_3\big) \big(u_1 + u_2 - 2 m^2\big) \big(s_1 + u_1 + u_2 -2 m^2 - M^2 - m_1^2\big) + \big(s_3 + t_2 - m_1^2\big) \nonumber\\
& & \times \big(u_1 + u_2 - 2 m^2\big) \big(M^2 - u_3\big) - 2 t_2 \big(s_1 + u_1 + u_2 -2 m^2 - M^2 - m_1^2\big) \big(M^2 - u_3\big)\big) \nonumber\\
& & + \big(u_1 - t_2 - m^2\big) \big(2 M^2 s_3 \big(m_1^2 - s_3 - t_2\big) + 4 M^2 s_3 \big(t_2 - t_3\big) + s_3 \big(u_1 + u_2 - 2 m^2\big) \nonumber\\
& & \times \big(s_1 + u_1 + u_2 -2 m^2 - M^2 - m_1^2\big) - \big(t_2 - t_3\big) \big(s_1 + u_1 + u_2 -2 m^2 - M^2 \nonumber\\
& & - m_1^2\big)^2 - 2 s_3 \big(u_1 + u_2 -2 m^2\big) \big(M^2 - u_3\big) + \big(s_3 + t_2 - m_1^2\big) \big(s_1 + u_1 + u_2 -2 m^2 \nonumber
\end{eqnarray}
\begin{eqnarray}
\hspace{5mm}
& & - M^2 - m_1^2\big) \big(M^2 - u_3\big)\big)\Big] \Big[4 \big(4 M^2 s_3 t_2 - M^2 \big(s_3 + t_2 - m_1^2\big)^2 - s_3 \big(u_1 + u_2 \nonumber\\
& & - 2 m^2\big)^2 + \big(s_3 + t_2 - m_1^2\big) \big(u_1 + u_2 - 2 m^2\big) \big(s_1 + u_1 + u_2 -2 m^2 - M^2 - m_1^2\big) \nonumber\\
& & - t_2 \big(s_1 + u_1 + u_2 -2 m^2 - M^2 - m_1^2\big)^2\big)\Big]^{-1} \,.
\end{eqnarray}

\section*{Appendix D}\label{Appendix_D}

Here we give the expressions for $T^S_{RL}$, $T^V_{RL}$, $T^{SV}_{LRRL}$ and $T^{SV}_{LLRR}$ contributions introduced in Section \ref{EFT} in terms of the reduced form factors $f$, $g_1$, $g_2$ and $l$, which are defined in the same way as those of the SM contribution ($T_{SM}$).\\

$(i)$ For the $T^S_{RL}$ contribution, in terms of the reduced form factors (we use the generic expression $x=f,\, l,\, g_1,\, g_2$), we have
\begin{equation}
x^{S,11}_{RL} = \frac{x^{11}_{SM}}{4} \,,\qquad
x^{S,22}_{RL} = \frac{x^{22}_{SM}}{4} \,, \qquad
x^{S,33}_{RL} = \frac{x^{33}_{SM}}{4} \,, \nonumber
\end{equation}
\begin{equation}
x^{S,44}_{RL} = \frac{x^{44}_{SM}}{4} \,, \qquad
x^{S,2442}_{RL} = \frac{x^{2442}_{SM}}{4} \,,
\end{equation}
\begin{eqnarray}
f^{S,1221}_{RL} & = & \frac{1}{2} \bigg[M^4 \big(3 m^2 \big(p_{12} + p_{13}\big) + 3 \big(p_{12} + p_{13}\big) p_{23}\big) + m_1^4 \big(-3 m^2 \big(Q_2 + Q_3\big) - 3 p_{23} \nonumber\\
& & \times \big(Q_2 + Q_3\big)\big) + 4 m^4 \big(p_{12} \big(Q_1 + 2 Q_2\big) - Q_1 \big(-p_{23} + Q_1 + Q_2 + Q_3\big) + p_{13} \nonumber\\
& & \times \big(Q_1 + 2 Q_3\big)\big) + 4 m^2 \big(p_{12}^2 \big(Q_1 + 2 Q_2 - Q_3\big) + Q_1 \big(-p_{23} + Q_1 + Q_2 + Q_3\big) \nonumber\\
& & \times \big(-2 p_{23} + 2 Q_1 + Q_2 + Q_3\big) + p_{13}^2 \big(Q_1 - Q_2 + 2 Q_3\big) - p_{13} \big(3 Q_1^2 + Q_1 Q_2 \nonumber\\
& & - Q_2^2 + p_{23} \big(-3 Q_1 + Q_2 - 3 Q_3\big) + 6 Q_1 Q_3 + 3 Q_2 Q_3 + 2 Q_3^2\big) + p_{12} \big(3 p_{23} Q_1 \nonumber\\
& & - 3 Q_1^2 + 3 p_{23} Q_2 - 6 Q_1 Q_2 - 2 Q_2^2 - \big(p_{23} + Q_1 + 3 Q_2\big) Q_3 + Q_3^2 + p_{13} \big(2 Q_1 \nonumber\\
& & + 3 \big(Q_2 + Q_3\big)\big)\big)\big) + 4 \big(-p_{13}^3 Q_2 + p_{23} Q_1 \big(p_{23} - 2 Q_1 - Q_2 - Q_3\big) \big(p_{23} - Q_1 \nonumber\\
& & - Q_2 - Q_3\big) - p_{12}^3 Q_3 + p_{13}^2 \big(p_{23} \big(Q_1 - 2 Q_2 + Q_3\big) + Q_2 \big(3 Q_1 + 2 Q_2 + Q_3\big)\big) \nonumber\\
& & + p_{13} \big(p_{23}^2 \big(2 Q_1 - Q_2 + Q_3\big) - Q_2 \big(2 Q_1 + Q_2 - Q_3\big) \big(Q_1 + Q_2 + Q_3\big) - p_{23} \nonumber\\
& & \times \big(Q_1 - Q_2 + Q_3\big) \big(3 Q_1 + 2 Q_2 + Q_3\big)\big) + p_{12}^2 \big(p_{13} Q_2 + p_{23} \big(Q_1 + Q_2 - 2 Q_3\big) \nonumber\\
& & + Q_3 \big(3 Q_1 + Q_2 + 2 Q_3\big)\big) + p_{12} \big(p_{23}^2 \big(2 Q_1 + Q_2 - Q_3\big) + p_{13}^2 Q_3 - Q_3 \big(2 Q_1 \nonumber\\
& & - Q_2 + Q_3\big) \big(Q_1 + Q_2 + Q_3\big) - p_{23} \big(Q_1 + Q_2 - Q_3\big) \big(3 Q_1 + Q_2 + 2 Q_3\big) + p_{13} \nonumber\\ 
& & \times \big(Q_2 \big(Q_1 + Q_2\big) + Q_1 Q_3 + Q_3^2 + p_{23} \big(2 Q_1 + Q_2 + Q_3\big)\big)\big)\big) + M^2 \big(m^4 \big(4 \big(p_{12} \nonumber\\ 
& & + p_{13}\big) + 2 Q_1\big) + p_{13}^2 \big(4 p_{23} - 5 Q_2\big) + p_{12}^2 \big(4 p_{23} - 5 Q_3\big) + m_1^2 \big(-4 m^4 + p_{12} p_{23} \nonumber\\ 
& & + p_{13} p_{23} + 4 p_{23} Q_1 - 4 p_{13} Q_2 - p_{23} Q_2 + m^2 \big(p_{12} + p_{13} - 4 p_{23} + 4 Q_1 - Q_2 \nonumber\\ 
& & - Q_3\big) - \big(4 p_{12} + p_{23}\big) Q_3\big) + p_{23} Q_1 \big(4 p_{23} - 3 \big(2 Q_1 + Q_2 + Q_3\big)\big) + m^2 \big(6 p_{12}^2 \nonumber\\ 
& & + 6 p_{13}^2 + p_{13} \big(8 p_{23} + Q_1 - 6 Q_2\big) + p_{12} \big(8 p_{13} + 8 p_{23} + Q_1 - 6 Q_3\big) + 3 Q_1 \big(2 p_{23} \nonumber\\ 
& & - 2 Q_1 - Q_2 - Q_3\big)\big) + p_{12} \big(4 p_{23}^2 + p_{23} \big(Q_1 - 2 Q_2 - 8 Q_3\big) + 3 Q_3 \big(2 Q_1 - Q_2 \nonumber\\ 
& & + Q_3\big) + p_{13} \big(4 p_{23} - 3 \big(Q_2 + Q_3\big)\big)\big) + p_{13} \big(4 p_{23}^2 + 3 Q_2 \big(2 Q_1 + Q_2 - Q_3\big) \nonumber
\end{eqnarray}
\begin{eqnarray}
\hspace{11mm} 
& & + p_{23} \big(Q_1 - 2 \big(4 Q_2 + Q_3\big)\big)\big)\big) + m_1^2 \big(-3 p_{13}^2 Q_2 - 3 p_{12}^2 Q_3 + m^2 \big(3 p_{12} Q_1 \nonumber\\
& & + 3 p_{13} Q_1 + 6 p_{23} Q_1 - 6 Q_1^2 - 6 p_{13} Q_2 - 8 p_{23} Q_2 - Q_1 Q_2 + 6 Q_2^2 - \big(6 p_{12} \nonumber\\ 
& & + 8 p_{23} + Q_1 - 8 Q_2\big) Q_3 + 6 Q_3^2\big) + 2 m^4 \big(Q_1 - 2 \big(Q_2 + Q_3\big)\big) + p_{23} \big(-6 Q_1^2 \nonumber\\ 
& & + 4 p_{23} \big(Q_1 - Q_2 - Q_3\big) - Q_1 \big(Q_2 + Q_3\big) + 4 \big(Q_2^2 + Q_2 Q_3 + Q_3^2\big)\big) + p_{13} \big(Q_2 \nonumber\\ 
& & \times \big(6 Q_1 + 5 Q_2 + 3 Q_3\big) + p_{23} \big(3 Q_1 - 2 \big(4 Q_2 + Q_3\big)\big)\big) + p_{12} \big(3 p_{13} \big(Q_2 + Q_3\big) \nonumber\\ 
& & + Q_3 \big(6 Q_1 + 3 Q_2 + 5 Q_3\big) + p_{23} \big(3 Q_1 - 2 \big(Q_2 + 4 Q_3\big)\big)\big)\big)\bigg] \,,
\end{eqnarray}
\begin{eqnarray}
l^{S,1221}_{RL} & = & \frac{M}{2} \bigg[-4 m^4 \big(p_{12} + p_{13} + p_{23} - Q_1\big) + M^2 \big(6 m^4 + 2 p_{23} Q_1 + m^2 \big(p_{12} + p_{13} \nonumber\\
& & + 6 p_{23} + 2 Q_1\big) + p_{13} \big(p_{23} - 2 Q_2\big) + p_{12} \big(p_{23} - 2 Q_3\big)\big) + m_1^2 \big(-2 m^4 - 3 p_{13} \nonumber\\ 
& & \times \big(p_{23} + 2 Q_2\big) + 2 p_{23} \big(-2 p_{23} + 3 Q_1 + Q_2 + Q_3\big) - 3 p_{12} \big(p_{23} + 2 Q_3\big) + m^2 \nonumber\\ 
& & \times \big(-3 p_{12} - 3 p_{13} + 2 \big(-3 p_{23} + 3 Q_1 + Q_2 + Q_3\big)\big)\big) - 2 m^2 \big(2 p_{12}^2 + 2 p_{13}^2 \nonumber\\ 
& & + p_{13} \big(6 p_{23} - 6 Q_1 + Q_2 - 4 Q_3\big) + p_{12} \big(4 p_{13} + 6 p_{23} - 6 Q_1 - 4 Q_2 + Q_3\big) \nonumber\\ 
& & + 2 \big(2 p_{23}^2 + 2 Q_1^2 + Q_2^2 + Q_2 Q_3 + Q_3^2 + 2 Q_1 \big(Q_2 + Q_3\big) - p_{23} \big(4 Q_1 + Q_2 \nonumber\\ 
& & + Q_3\big)\big)\big) + 2 \big(-2 p_{13}^2 \big(p_{23} + 2 Q_2\big) - 2 p_{12}^2 \big(p_{23} + 2 Q_3\big) - p_{23} \big(2 p_{23}^2 + \big(2 Q_1 \nonumber\\ 
& & + Q_2\big)^2 + 4 Q_1 Q_3 + Q_3^2 - 2 p_{23} \big(3 Q_1 + Q_2 + Q_3\big)\big) + p_{13} \big(-4 p_{23}^2 + Q_2 \big(4 Q_1 \nonumber\\ 
& & + 3 Q_2 + Q_3\big) + p_{23} \big(6 Q_1 - 2 Q_2 + 3 Q_3\big)\big) + p_{12} \big(-4 p_{23}^2 + p_{23} \big(6 Q_1 + 3 Q_2 \nonumber\\ 
& & - 2 Q_3\big) - 4 p_{13} \big(p_{23} + Q_2 + Q_3\big) + Q_3 \big(4 Q_1 + Q_2 + 3 Q_3\big)\big)\big)\bigg] \,,
\end{eqnarray}
\begin{eqnarray}
{g_1}^{S,1221}_{RL} & = & \frac{M}{2} \bigg[4 p_{13}^3 + m_1^4 \big(3 m^2 + 3 p_{23}\big) + m^4 \big(-8 p_{12} + 4 Q_1\big) + 2 p_{23} Q_1 \big(2 p_{23} - 2 Q_1 \nonumber\\ 
& & - Q_2 - 2 Q_3\big) - 2 p_{12}^2 \big(2 \big(p_{13} + p_{23}\big) + Q_3\big) + m_1^2 \big(4 m^4 - 3 p_{12} p_{13} + 3 p_{13}^2 \nonumber\\ 
& & + 2 p_{12} p_{23} + 8 p_{13} p_{23} + 4 p_{23}^2 - p_{23} Q_1 - 5 p_{13} Q_2 - 4 p_{23} Q_2 + m^2 \big(6 p_{13} + 8 p_{23} \nonumber\\
& & - Q_1 - 6 Q_2 - 4 Q_3\big) - \big(2 p_{12} + p_{13} + 2 p_{23}\big) Q_3\big) + M^2 \big(p_{13}^2 + m_1^2 \big(m^2 + p_{23}\big) \nonumber\\ 
& & - p_{12} \big(p_{13} + 2 p_{23}\big) + p_{23} Q_1 + m^2 \big(-6 p_{12} - 2 p_{13} + Q_1\big) + p_{13} \big(2 p_{23} - Q_2 \nonumber\\ 
& & + Q_3\big)\big) + p_{13} \big(4 p_{23}^2 - 8 p_{23} Q_2 + 6 Q_1 Q_2 + 2 \big(Q_2 - Q_3\big) \big(2 Q_2 + Q_3\big)\big) + p_{13}^2 \nonumber\\ 
& & \times \big(8 p_{23} - 2 \big(2 Q_1 + 4 Q_2 + Q_3\big)\big) + 2 p_{12} \big(-2 p_{23}^2 - 2 p_{13} \big(p_{23} - Q_1 + Q_2\big) \nonumber\\ 
& & + Q_3 \big(Q_1 - Q_2 + Q_3\big) + p_{23} \big(5 Q_1 + 2 \big(Q_2 + Q_3\big)\big)\big) + 2 m^2 \big(-4 p_{12}^2 + 2 p_{13}^2 \nonumber\\ 
& & - Q_1 \big(-4 p_{23} + 2 Q_1 + Q_3\big) + p_{13} \big(2 p_{23} + 3 Q_1 - 2 Q_2 + 2 Q_3\big) + p_{12} \big(-6 p_{13} \nonumber\\ 
& & - 6 p_{23} + 4 \big(2 Q_1 + Q_2 + Q_3\big)\big)\big)\bigg] \,,
\end{eqnarray}
\begin{eqnarray}
f^{S,1331}_{RL} & = & \frac{1}{4} \bigg[M^4 \big(4 p_{13} p_{23} - 3 p_{23} Q_1 - 3 p_{13} Q_2 - 3 p_{12} Q_3\big) + m^4 \big(4 M^4 + 10 p_{13} Q_1 + 23 p_{23} Q_1 \nonumber\\ 
& & - 4 Q_1^2 + 23 p_{13} Q_2 + 10 p_{23} Q_2 - 48 Q_1 Q_2 - 4 Q_2^2 + 2 p_{12} \big(Q_1 + Q_2\big) + 2 M^2 \big(6 p_{12} \nonumber\\ 
& & + 11 p_{13} + 11 p_{23} - 6 \big(Q_1 + Q_2\big)\big) - 11 p_{12} Q_3 + 2 \big(5 p_{13} + 5 p_{23} - 11 \big(Q_1 + Q_2\big)\big) Q_3 \nonumber
\end{eqnarray}
\begin{eqnarray}
\hspace{16mm} 
& & - 8 Q_3^2\big) + m^6 \big(16 M^2 + 13 \big(Q_1 + Q_2 + Q_3\big)\big) + m^2 \big(- 2 Q_1 \big(2 p_{12}^2 - 7 p_{13} p_{23} - 5 p_{23}^2 \nonumber\\ 
& & + p_{12} \big(2 p_{13} + p_{23} - 4 Q_1\big) - 2 p_{13} Q_1 + 2 Q_1^2\big) - 2 \big(2 p_{12}^2 - p_{13} \big(5 p_{13} + 7 p_{23}\big) + 20 \big(p_{13} \nonumber\\
& & + p_{23}\big) Q_1 - 8 Q_1^2 + p_{12} \big(p_{13} + 2 p_{23} + 4 Q_1\big)\big) Q_2 + 4 \big(2 p_{12} + p_{23} + 4 Q_1\big) Q_2^2 - 4 Q_2^3 \nonumber\\ 
& & + M^4 \big(4 p_{13} + 4 p_{23} - 3 \big(Q_1 + Q_2 - Q_3\big)\big) - 2 \big(3 p_{12}^2 - 2 \big(p_{13} p_{23} - 5 p_{23} Q_1 - 5 p_{13} Q_2 \nonumber\\ 
& & + 13 Q_1 Q_2\big) + p_{12} \big(7 p_{13} + 7 p_{23} - 4 \big(Q_1 + Q_2\big)\big)\big) Q_3 + 4 \big(5 p_{12} + Q_1 + Q_2\big) Q_3^2 \nonumber\\ 
& & + 2 M^2 \big(3 p_{13}^2 + 12 p_{13} p_{23} + 3 p_{23}^2 - 5 p_{13} Q_1 - 8 p_{23} Q_1 + 3 Q_1^2 - 8 p_{13} Q_2 - 5 p_{23} Q_2 \nonumber\\ 
& & + 2 Q_1 Q_2 + 3 Q_2^2 + p_{12} \big(5 p_{13} + 5 p_{23} - 4 \big(Q_1 + Q_2\big) - 5 Q_3\big) - 2 \big(p_{13} + p_{23} + Q_1 \nonumber\\ 
& & + Q_2\big) Q_3 - 2 Q_3^2\big)\big) + 2 M^2 \big(2 p_{13}^2 \big(p_{23} - Q_2\big) + p_{23} Q_1 \big(-2 p_{23} + 3 Q_1 + Q_2\big) - p_{12}^2 \nonumber\\ 
& & \times Q_3 + 6 Q_1 Q_2 Q_3 + p_{13} \big(2 p_{23}^2 + Q_2 \big(Q_1 + 3 Q_2\big) - 2 p_{23} \big(2 \big(Q_1 + Q_2\big) + Q_3\big)\big) + p_{12} \nonumber\\ 
& & \times \big(p_{13} \big(4 p_{23} - 3 Q_2 - 2 Q_3\big) - p_{23} \big(3 Q_1 + 2 Q_3\big) + Q_3 \big(Q_1 + Q_2 + 4 Q_3\big)\big)\big) + 4 \big(p_{13}^2 \nonumber\\ 
& & \times Q_2 \big(p_{23} - Q_1 + Q_2 - Q_3\big) + p_{12}^3 Q_3 + p_{12}^2 \big(-p_{23} Q_1 - p_{13} Q_2 + \big(p_{13} + p_{23} - 2 \big(Q_1 \nonumber\\ 
& & + Q_2\big)\big) Q_3\big) + p_{12} \big(-p_{23} \big(p_{13} + p_{23} - 2 Q_1\big) Q_1 - p_{13} \big(p_{13} + p_{23}\big) Q_2 + 2 p_{13} Q_2^2 \nonumber\\ 
& & - \big(\big(p_{23} - Q_1\big) Q_1 + \big(p_{23} - 5 Q_1\big) Q_2 - Q_2^2 + p_{13} \big(p_{23} + Q_1 + Q_2\big)\big) Q_3 + \big(p_{13} + p_{23}\big) \nonumber\\ 
& & \times Q_3^2 - Q_3^3\big) + p_{13} \big(p_{23}^2 Q_1 + p_{23} \big(Q_1^2 - 6 Q_1 Q_2 + Q_2^2 - \big(Q_1 + Q_2\big) Q_3\big) + Q_2 \big(Q_1^2 \nonumber\\ 
& & + Q_1 Q_2 - Q_2^2 + 6 Q_1 Q_3 + Q_3^2\big)\big) + Q_1 \big(p_{23}^2 \big(Q_1 - Q_2 - Q_3\big) - 4 Q_2 Q_3 \big(Q_1 + Q_2 \nonumber\\ 
& & + Q_3\big) + p_{23} \big(-Q_1^2 + Q_1 Q_2 + Q_2^2 + 6 Q_2 Q_3 + Q_3^2\big)\big)\big)\bigg] \,,
\end{eqnarray}
\begin{eqnarray}
l^{S,1331}_{RL} & = & \frac{M}{4} \bigg[-13 m^6 + M^4 p_{23} + m^4 \big(4 M^2 - 2 p_{12} - 10 p_{13} - 23 p_{23} + 4 Q_1 + 24 Q_2 \nonumber\\ 
& & + 2 Q_3\big) + 2 M^2 \big(-3 p_{23} Q_1 + p_{13} \big(2 p_{23} + Q_2\big) - \big(p_{23} + Q_2\big) Q_3 + p_{12} \big(p_{23} \nonumber\\ 
& & + Q_3\big)\big) + m^2 \big(M^4 + 2 M^2 \big(3 p_{12} + 4 p_{13} + 2 p_{23} - 3 Q_1 + Q_2 - 3 Q_3\big) + 2 \big(2 p_{12}^2 \nonumber\\ 
& & - 5 p_{23}^2 + 2 Q_1^2 + 10 p_{23} Q_2 - 3 Q_1 Q_2 - 5 Q_2^2 + p_{12} \big(2 p_{13} + p_{23} - 4 Q_1 + 2 Q_2 \nonumber\\
& & - Q_3\big) + 4 p_{23} Q_3 + 3 Q_1 Q_3 - 8 Q_2 Q_3 + 2 Q_3^2 - p_{13} \big(7 p_{23} + 2 Q_1 - 10 Q_2 \nonumber\\ 
& & + 3 Q_3\big)\big)\big) + 4 \big(p_{13}^2 Q_2 + p_{23} Q_1 \big(-p_{23} + Q_1 + Q_2\big) + \big(p_{23}^2 + p_{23} \big(Q_1 - 2 Q_2\big) \nonumber\\ 
& & + Q_2 \big(2 Q_1 + Q_2\big)\big) Q_3 + Q_2 Q_3^2 + p_{12}^2 \big(p_{23} + Q_3\big) + p_{12} \big(p_{23}^2 - p_{23} \big(2 Q_1 + Q_2 \nonumber\\ 
& & - Q_3\big) + p_{13} \big(p_{23} + Q_2 + Q_3\big) - Q_3 \big(Q_1 + 2 Q_2 + Q_3\big)\big) - p_{13} \big(p_{23}^2 + p_{23} \big(Q_1 \nonumber\\ 
& & - 3 Q_2\big) + Q_2 \big(Q_1 + 2 Q_2 + 3 Q_3\big)\big)\big)\bigg] \,,
\end{eqnarray}
\begin{eqnarray}
{g_2}^{S,1331}_{RL} & = & \frac{M}{4} \bigg[-13 m^6 + M^4 \big(m^2 - p_{12}\big) + m^4 \big(11 p_{12} - 10 \big(p_{13} + p_{23} - 2 \big(Q_1 + Q_2\big)\big) \nonumber\\ 
& & + 8 Q_3\big) + 2 m^2 \big(3 p_{12}^2 + 6 p_{23} Q_1 - 3 Q_1^2 + 3 p_{23} Q_2 - 10 Q_1 Q_2 - 3 Q_2^2 - p_{13} \big(2 p_{23} \nonumber\\ 
& & - 3 Q_1 - 6 Q_2\big) + p_{12} \big(7 p_{13} + 7 p_{23} - 3 \big(Q_1 + Q_2\big) - 10 Q_3\big) - 4 \big(Q_1 + Q_2\big) Q_3\big) \nonumber\\ 
& & + 4 \big(-p_{12}^3 + \big(p_{13} - Q_1\big) \big(p_{23} - Q_2\big) \big(Q_1 + Q_2\big) + p_{12}^2 \big(-p_{13} - p_{23} + Q_1 + Q_2\big) \nonumber\\ 
& & - \big(p_{23} Q_1 + \big(p_{13} - 2 Q_1\big) Q_2\big) Q_3 + p_{12} \big(-Q_1 Q_2 + p_{13} \big(p_{23} - Q_3\big) + Q_3 \big(  Q_1 \nonumber
\end{eqnarray}
\begin{eqnarray}
\hspace{9mm} 
& & + Q_2 + Q_3 -p_{23} \big)\big)\big) + M^2 \big(4 m^4 - 2 \big(p_{12}^2 + p_{23} Q_1 - p_{13} \big(2 p_{23} - Q_2\big)\big) + m^2 \nonumber\\ 
& & \times \big(10 p_{12} + 8 p_{13} + 8 p_{23} - 2 \big(Q_1 + Q_2 + 2 Q_3\big)\big)\big)\bigg] \,,\,
\end{eqnarray}
\begin{eqnarray}
f^{S,1441}_{RL} & = & \frac{1}{2} \bigg[-3 M^4 p_{12} p_{23} + m^6 \big(2 M^2 - Q_1 + 14 Q_2 - 7 Q_3\big) - m^4 \big(6 p_{12} Q_1 + p_{13} Q_1 \nonumber\\ 
& & + 5 p_{23} Q_1 - 5 Q_1^2 - p_{12} Q_2 - 25 p_{13} Q_2 - 22 p_{23} Q_2 + 14 Q_1 Q_2 + 18 Q_2^2 + \big(19 p_{12} \nonumber\\ 
& & + 9 p_{13} + 11 p_{23} - 19 Q_1 + 4 Q_2\big) Q_3 - 12 Q_3^2 + M^2 \big(6 p_{12} - p_{13} - 2 p_{23} + 3 Q_1 \nonumber\\ 
& & - 9 Q_2 + 8 Q_3\big)\big) - m^2 \big(3 M^4 p_{12} + 3 p_{13} p_{23} Q_1 + 6 p_{23}^2 Q_1 - 4 p_{13} Q_1^2 - 11 p_{23} Q_1^2 \nonumber\\ 
& & + 4 Q_1^3 - 11 p_{13}^2 Q_2 - 22 p_{13} p_{23} Q_2 - 8 p_{23}^2 Q_2 + 19 p_{13} Q_1 Q_2 + 2 p_{23} Q_1 Q_2 + 22 \nonumber\\ 
& & \times p_{13} Q_2^2 + 12 p_{23} Q_2^2 - 8 Q_1 Q_2^2 - 4 Q_2^3 + \big(2 p_{13}^2 + 4 p_{23}^2 - 15 p_{23} Q_1 + 12 Q_1^2 + 2 \nonumber\\ 
& & \times p_{23} Q_2 + 4 Q_1 Q_2 - 4 Q_2^2 + p_{13} \big(8 p_{23} - 16 Q_1 + 7 Q_2\big)\big) Q_3 + 2 \big(-5 p_{13} - 4 p_{23} \nonumber\\ 
& & + 6 Q_1 + 2 Q_2\big) Q_3^2 + 4 Q_3^3 + p_{12}^2 \big(4 Q_1 + 6 Q_2 + 4 Q_3\big) + p_{12} \big(2 \big(2 \big( p_{13} + 3 p_{23} \nonumber\\ 
& & - 2 Q_1\big) Q_1 + \big( p_{13} + p_{23} - 4 Q_1\big) Q_2 + Q_2^2\big) + \big(13 p_{13} + 10 p_{23} - 17 Q_1 - 16 Q_2\big) \nonumber\\ 
& & \times Q_3 - 13 Q_3^2\big) + M^2 \big(5 p_{12}^2 + p_{13}^2 + 5 p_{23} Q_1 - 3 Q_1^2 - 5 p_{23} Q_2 + 3 Q_1 Q_2 + 3 Q_2^2 \nonumber\\ 
& & - p_{13} \big( p_{23} - 2 Q_1 + 11 Q_2 - 7 Q_3\big) + 4 p_{23} Q_3 - 6 Q_1 Q_3 - 3 Q_3^2 + p_{12} \big(4 p_{13} + 10 \nonumber\\ 
& & \times p_{23} - Q_1 - 7 Q_2 + Q_3\big)\big)\big) + M^2 \big(\big(- p_{23} Q_1 + p_{13} Q_2\big) \big(2 p_{13} + 2 p_{23} - 3 \big( Q_1 \nonumber\\ 
& & + Q_2\big)\big) + p_{12}^2 \big(-4 p_{23} + 3 Q_3\big) + p_{12} \big(-4 p_{23}^2 + 3 \big(- Q_1 + Q_2\big) Q_3 - p_{13} \big(2 p_{23} \nonumber\\ 
& & - Q_2 - 2 Q_3\big) + p_{23} \big( Q_1 + 2 Q_2 + 6 Q_3\big)\big)\big) + 2 \big(2 p_{12}^3 Q_3 + \big(- p_{23} Q_1 + p_{13} Q_2\big) \nonumber\\ 
& & \times \big(\big( p_{13} + p_{23} - Q_1 - Q_2\big) \big( p_{23} - 2 \big( Q_1 + Q_2\big)\big) - \big( p_{13} + p_{23} - 2 \big( Q_1 + Q_2\big)\big) \nonumber\\ 
& & \times Q_3\big) - p_{12}^2 \big(2 p_{23} \big( Q_1 + Q_2\big) + 2 p_{13} \big( Q_2 - Q_3\big) - 3 p_{23} Q_3 + Q_3 \big(4 Q_1 + Q_2 \nonumber\\ 
& & + 2 Q_3\big)\big) + p_{12} \big(-2 p_{13}^2 Q_2 + p_{23}^2 \big(-3 Q_1 - 2 Q_2 + Q_3\big) + 2 \big( Q_1 - Q_2\big) Q_3 \big( Q_1 \nonumber\\
& & + Q_2 + Q_3\big) + p_{23} \big(4 Q_1^2 + 7 Q_1 Q_2 + 2 Q_2^2 - Q_1 Q_3 + 3 Q_2 Q_3 - Q_3^2\big) - p_{13} \big( Q_2^2 \nonumber\\ 
& & + p_{23} \big(2 Q_1 + 3 Q_2 - Q_3\big) + 2 Q_1 Q_3 + Q_3^2\big)\big)\big)\bigg] \,,
\end{eqnarray}
\begin{eqnarray}
l^{S,1441}_{RL} & = & \frac{M}{2} \bigg[m^6 + m^4 \big(-3 M^2 + 6 p_{12} + p_{13} + 5 p_{23} - 5 Q_1 + 11 Q_2 - 13 Q_3\big) + M^2 \nonumber\\ 
& & \times \big(-p_{23} \big( p_{12} + Q_1\big) + p_{13} Q_2 + p_{12} Q_3\big) + m^2 \big(4 p_{12}^2 + 3 p_{13} p_{23} + 6 p_{23}^2 - 4 p_{13} Q_1 \nonumber\\ 
& & - 11 p_{23} Q_1 + 4 Q_1^2 + 14 p_{13} Q_2 + p_{23} Q_2 - 2 Q_1 Q_2 - 3 Q_2^2 + p_{12} \big(4 p_{13} + 12 p_{23} \nonumber\\ 
& & - 8 Q_1 + 3 Q_2 - 6 Q_3\big) - 11 p_{13} Q_3 - 7 p_{23} Q_3 + 10 Q_1 Q_3 + 2 Q_2 Q_3 + 7 Q_3^2 - M^2 \nonumber\\ 
& & \times \big( p_{12} + 3 p_{23} + Q_1 - 2 Q_2 + 2 Q_3\big)\big) + 2 \big(2 p_{13}^2 Q_2 + 2 p_{12}^2 \big( p_{23} + Q_3\big) + p_{23} \big( p_{23}^2 \nonumber\\ 
& & - 3 p_{23} Q_1 + 2 Q_1^2 - 2 p_{23} Q_2 + 3 Q_1 Q_2 + Q_2^2 + Q_1 Q_3\big) + p_{12} \big(3 p_{23}^2 - Q_3 \big(2 Q_1 \nonumber\\ 
& & + Q_3\big) + p_{23} \big(-4 Q_1 - 3 Q_2 + Q_3\big) + 2 p_{13} \big( p_{23} + Q_2 + Q_3\big)\big) + p_{13} \big( p_{23}^2 + p_{23} \nonumber\\ 
& & \times \big(-2 Q_1 + Q_2\big) - Q_2 \big(2 \big( Q_1 + Q_2\big) + Q_3\big)\big)\big)\bigg] \,,
\end{eqnarray}
\begin{eqnarray}
{g_1}^{S,1441}_{RL} & = & \frac{M}{2} \bigg[-14 m^6 + M^2 \big(p_{12} \big(p_{13} + 2 p_{23}\big) - p_{23} Q_1 + p_{13} Q_2\big) - m^4 \big(p_{12} + 25 p_{13} \nonumber\\
& & + 22 p_{23} - 3 \big(Q_1 + 6 Q_2\big) - 2 Q_3\big) + m^2 \big(6 p_{12}^2 - 11 p_{13}^2 - 22 p_{13} p_{23} - 8 p_{23}^2 \nonumber\\ 
& & + 5 p_{13} Q_1 + p_{23} Q_1 + 2 Q_1^2 + 22 p_{13} Q_2 + 12 p_{23} Q_2 - 5 Q_1 Q_2 - 4 Q_2^2 + p_{12} \nonumber\\ 
& & \times \big(2 p_{13} + 2 p_{23} - 11 Q_1 + 2 Q_2 - 9 Q_3\big) + M^2 \big(4 p_{12} - Q_1 + Q_2 - Q_3\big) \nonumber\\ 
& & + \big(4 p_{13} + 2 p_{23} + 3 Q_1 - 3 Q_2\big) Q_3 + Q_3^2\big) + 2 \big(2 p_{12}^2 \big(p_{13} + p_{23}\big) - p_{13}^2 \big(p_{23} \nonumber\\ 
& & - 2 Q_2\big) - p_{23} Q_1 \big(p_{23} - Q_1 - Q_2 - Q_3\big) - p_{13} \big(p_{23}^2 - 3 p_{23} Q_2 + Q_2 \big(2 \big(Q_1 \nonumber\\ 
& & + Q_2\big) + Q_3\big)\big) + p_{12} \big(2 p_{13}^2 + 2 p_{23}^2 + p_{13} \big(3 p_{23} - 2 Q_1 + Q_2 - Q_3\big) + Q_2 Q_3 \nonumber\\ 
& & - p_{23} \big(4 Q_1 + 2 Q_2 + 3 Q_3\big)\big)\big)\bigg] \,,
\end{eqnarray}
\begin{eqnarray}
{g_2}^{S,1441}_{RL} & = & -\frac{M}{2} \bigg[-7 m^6 + M^2 p_{12} \big(p_{12} + 2 p_{23} + Q_2\big) + m^4 \big(M^2 - 19 p_{12} - 9 p_{13} - 11 p_{23} \nonumber\\ 
& & + 6 Q_1 - 2 Q_2 + 12 Q_3\big) + 2 \big(2 p_{12}^3 + p_{12} \big(p_{23} - Q_2\big) \big(p_{13} + p_{23} + Q_2\big) - \big(p_{13} \nonumber\\ 
& & + p_{23} - Q_1 - Q_2\big) \big(p_{13} Q_2 - p_{23} Q_1\big) + p_{12}^2 \big(2 p_{13} + 3 p_{23} - 2 Q_1 - Q_2 - 2 Q_3\big) \nonumber\\ 
& & - p_{12} \big(p_{13} + p_{23} - Q_1 + 2 Q_2\big) Q_3\big) + m^2 \big(-4 p_{12}^2 - 2 p_{13}^2 - 8 p_{13} p_{23} - 4 p_{23}^2 \nonumber\\ 
& & + 5 p_{13} Q_1 + 8 p_{23} Q_1 - 2 Q_1^2 - 3 p_{13} Q_2 + Q_1 Q_2 + Q_2^2 + \big(10 p_{13} + 8 p_{23} - 5 Q_1 \nonumber\\ 
& & - 3 Q_2\big) Q_3 - 4 Q_3^2 + M^2 \big(-2 p_{12} - 2 p_{13} + p_{23} - Q_2 + Q_3\big) + p_{12} \big(-13 p_{13} \nonumber\\ 
& & - 10 p_{23} + 11 Q_1 + 7 Q_2 + 13 Q_3\big)\big)\big)\bigg] \,,
\end{eqnarray}
\begin{equation}
{g_2}^{S,1221}_{RL} = {g_1}^{S,1221}_{RL} \big|_{p_2 \leftrightarrow p_3} \,,
\qquad
{g_1}^{S,1331}_{RL} = l^{S,1331}_{RL} \big|_{p_1 \leftrightarrow p_2} \,,
\end{equation}

\begin{equation}
f^{S,2332}_{RL} = f^{S,1441}_{RL} \big|_{p_1 \leftrightarrow p_2} \,, \qquad
l^{S,2332}_{RL} = {g_1}^{S,1441}_{RL} \big|_{p_1 \leftrightarrow p_2} \,,
\end{equation}
\begin{equation}
{g_1}^{S,2332}_{RL} = l^{S,1441}_{RL} \big|_{p_1 \leftrightarrow p_2} \,, \qquad
{g_2}^{S,2332}_{RL} = {g_2}^{S,1441}_{RL} \big|_{p_1 \leftrightarrow p_2} \,, \nonumber
\end{equation}

\begin{equation}
f^{S,3443}_{RL} = f^{S,1221}_{RL} \big|_{p_1 \leftrightarrow p_2, m_1 \to m} \,, \qquad
l^{S,3443}_{RL} = {g_1}^{S,1221}_{RL} \big|_{p_1 \leftrightarrow p_2, m_1 \to m} \,,
\end{equation}
\begin{equation}
{g_1}^{S,3443}_{RL} = l^{S,1221}_{RL} \big|_{p_1 \leftrightarrow p_2, m_1 \to m} \,, \qquad
{g_2}^{S,3443}_{RL} = {g_2}^{S,1221}_{RL} \big|_{p_1 \leftrightarrow p_2, m_1 \to m} \,. \nonumber
\end{equation}

$(ii)$ Regarding the expressions for the $T^V_{RL}$ contribution, the reduced form factors shown below must be multiplied by the common factor $(M^2 + m_1^2 + 2(m^2 + p_{12} + p_{13} + p_{23}$ $- Q_{1} - Q_{2} - Q_{3}))$ ($m_1=m$ if there are identical particles):
\begin{eqnarray}
f^{V,11}_{RL} & = & 12\left[M^2(m^2 + p_{23}) (p_{12} + p_{13} - Q_{1}) - m^4 Q_{1} + p_{12} p_{23} Q_{2} - p_{23} Q_{1} Q_{2} + p_{13} Q_{2}^2 \right. \nonumber\\
& & \left. + p_{13} p_{23} Q_{3} - p_{23} Q_{1} Q_{3} - p_{12} Q_{2} Q_{3} - p_{13} Q_{2} Q_{3} + 2 Q_{1} Q_{2} Q_{3} +  p_{12} Q_{3}^2 + m^2  \right. \nonumber\\
& & \left. \times (-Q_{1}(p_{23} + Q_{2} + Q_{3}) + p_{12} (2 Q_{2} + Q_{3}) + p_{13} (Q_{2} + 2 Q_{3}))\right]\,, \\ 
\nonumber\\
l^{V,11}_{RL} & = & 12 M \left[(m^2 + M^2) (m^2 + p_{23}) - 2 Q_2 Q_3\right]\,,
\end{eqnarray}
\begin{eqnarray}
{g_1}^{V,11}_{RL} & = & 12 M  \left[p_{23} Q_1 - p_{12} p_{23} + m^2 (Q_1- 2 p_{12} - p_{13}) - p_{13} Q_2 + p_{13} Q_3\right]\,, \\ 
\nonumber\\
f^{V,22}_{RL} & = & 12  \left[2 p_{12} p_{13} Q_1 + p_{12} p_{23} Q_1 + p_{13} p_{23} Q_1 + p_{12} p_{13} Q_2 - m^4 Q_1 - p_{13}^2 Q_2 + p_{12} p_{23} Q_2 \right.\nonumber\\
&& \left. - p_{12}^2 Q_3 + p_{12} p_{13} Q_3 + p_{13} p_{23} Q_3 - m_1^2 (m^2 + p_{23}) (Q_1 + Q_2 + Q_3) + m^2 (-p_{23} Q_1 \right.\nonumber\\
&& \left. + p_{12} (Q_1 + 2 Q_2 + Q_3) + p_{13} (Q_1 + Q_2 + 2 Q_3))\right]\,, \\
\nonumber\\
l^{V,22}_{RL} & = & 12 M \left[M^2 + m_1^2 + 2(m^2 + p_{12} + p_{13} + p_{23} - Q_1 - Q_2 - Q_3)\right]\,, \\
\nonumber\\
f^{V,1221}_{RL} & = & 12 \left[M^2(p_{12} + p_{13})p_{23} - 2m^4Q_1 + p_{12}p_{23}Q_1 + p_{13}p_{23}Q_1 - 2p_{23}Q_1^2 + p_{12}p_{13}Q_2 \right.\nonumber\\
&& \left. - p_{13}^2Q_2 - m_1^2p_{23}Q_2 + 2p_{12}p_{23}Q_2 + 2p_{13}Q_1Q_2 - p_{23}Q_1Q_2 + p_{13}Q_2^2 - (p_{12}^2 + p_{23} \right.\nonumber\\
&& \left. \times (m_1^2 - 2p_{13} + Q_1) + p_{13}Q_2 + p_{12}(Q_2 - 2Q_1 -p_{13}))Q_3 + p_{12}Q_3^2 + m^2(M^2 \right.\nonumber\\
&& \left. \times (p_{12} + p_{13}) + p_{13}Q_1 - 2p_{23}Q_1 - 2Q_1^2 + 2p_{13}Q_2 - Q_1Q_2 + 4p_{13}Q_3 - Q_1Q_3 \right.\nonumber\\
&& \left. - m_1^2(Q_2 + Q_3)+ p_{12}(Q_1 + 4Q_2 + 2Q_3))\right]\,,\\
\nonumber\\
l^{V,1221}_{RL} & = & 12 M \left[2 m^4 - p_{13} p_{23} + 2 p_{23} Q_1 - m^2 (p_{12} + p_{13} - 2 (p_{23} + Q_1)) - 2 p_{13} Q_2 \right. \nonumber\\
& & \left. - p_{12} (p_{23} + 2 Q_3)\right]\,,\\
\nonumber\\
{g_1}^{V,1221}_{RL} & = & 12 M \left[p_{23} (m_1^2 + Q_1) - p_{12} (p_{13} + 2 p_{23}) +  m^2 (m_1^2 - 4 p_{12} - 2 p_{13} + Q_1) \right.\nonumber\\
&& \left. + p_{13} (p_{13} - Q_2 + Q_3)\right]\,,
\end{eqnarray}
\begin{equation}
{g_2}^{V,11}_{RL} = {g_1}^{V,11}_{RL}|_{p_2\leftrightarrow p_3} \, , \qquad {g_1}^{V,22}_{RL} = {l}^{V,22}_{RL} \, ,\nonumber
\end{equation}
\begin{equation}
{g_2}^{V,22}_{RL} = {g_1}^{V,22}_{RL}|_{p_2\leftrightarrow p_3} \, , \qquad {g_2}^{V,1221}_{RL} = {g_1}^{V,1221}_{RL}|_{p_2\leftrightarrow p_3}\,,
\end{equation}

\begin{equation}
f^{V,33}_{RL} = f^{V,11}_{RL}|_{p_1\leftrightarrow p_2, m_1\to m}\,,\qquad l^{V,33}_{RL} \, = \,{g_1}^{V,11}_{RL}|_{p_1\leftrightarrow p_2, m_1\to m} \,,\nonumber
\end{equation}
\begin{equation}
{g_1}^{V,33}_{RL} = \, l^{V,11}_{RL}|_{p_1\leftrightarrow p_2, m_1\to m},\qquad {g_2}^{V,33}_{RL} \, = \, {g_2}^{V,11}_{RL}|_{p_1\leftrightarrow p_2, m_1\to m} \, ,
\end{equation}

\begin{equation}
f^{V,44}_{RL} = f^{V,22}_{RL}|_{p_1\leftrightarrow p_2, m_1\to m}\,,\qquad l^{V,44}_{RL} \, = \, {g_1}^{V,22}_{RL}|_{p_1\leftrightarrow p_2, m_1\to m}\,,\nonumber
\end{equation}
\begin{equation}
{g_1}^{V,44}_{RL} = l^{V,22}_{RL}|_{p_1\leftrightarrow p_2, m_1\to m}\, ,\qquad {g_2}^{V,44}_{RL} \, = \, {g_2}^{V,22}_{RL}|_{p_1\leftrightarrow p_2, m_1\to m}\, ,
\end{equation}
\begin{eqnarray}
f^{V,1331}_{RL} & = & -6 (m^4 (2 M^2 - Q_1 - Q_2 + Q_3) + M^2 (p_{23} Q_1 + p_{13} Q_2 + p_{12} Q_3) + m^2 (4 Q_1 Q_2 \nonumber\\
&& - p_{23} Q_1 - 2 Q_1^2 - p_{13} Q_2 - 2 Q_2^2 + 2 p_{12} (Q_1 + Q_2) + M^2 (2 p_{12} + 2 p_{13} + 2 p_{23} \nonumber\\
&& + Q_1 + Q_2 - Q_3) +  5 p_{12} Q_3 + 2 (p_{13} + p_{23} - 2 (Q_1 + Q_2)) Q_3 - 4 Q_3^2) + 2 \nonumber\\
&& \times ((Q_2 - Q_1) (p_{23} Q_1 - p_{13} Q_2) - p_{12}^2 Q_3 - 2 Q_1 Q_2 Q_3 + p_{12} (p_{23} Q_1 + p_{13} Q_2 \nonumber\\
& & + (Q_1 + Q_2) Q_3))) \,,
\end{eqnarray}
\begin{eqnarray}
l^{V,1331}_{RL} & = & -6 M (m^4 - M^2 p_{23} - m^2 (M^2 + 2 p_{12} - p_{23} - 2 Q_1 + 2 Q_2 - 4 Q_3) \nonumber\\
&& + 2 (p_{23} Q_1 - Q_2 (p_{13} - Q_3) - p_{12} (p_{23} + Q_3))) \,,\\
\nonumber\\
{g_2}^{V,1331}_{RL} & = & 6 M (m^4 - p_{12} (M^2 + 2 p_{12}) + m^2 (M^2 + 5 p_{12} + 2 (p_{13} + p_{23} - 2 Q_3))) \, ,\\
\nonumber\\
f^{V,1441}_{RL} & = & -12(M^2 p_{12} p_{23} + p_{12} p_{23} Q_1 - p_{23} Q_1^2 + p_{12} p_{13} Q_2 + 2 p_{12} p_{23} Q_2 + p_{13} Q_1 Q_2 \nonumber\\ 
&& - p_{23} Q_1 Q_2 + p_{13} Q_2^2 - m^4 (Q_1 + Q_2) - p_{12} (p_{12} - Q_1 + Q_2) Q_3 + m^2 (M^2 p_{12}  \nonumber\\
&& - p_{23} Q_1 - Q_1^2 - p_{13} Q_2 - p_{23} Q_2 + Q_1 Q_2 + Q_2^2 + p_{13} Q_3 - 2 Q_1 Q_3 - Q_3^2 \nonumber\\
&& + p_{12} (Q_1 + 3 (Q_2 + Q_3))))\,,\\
\nonumber\\
l^{V,1441}_{RL} & = & -12 M (m^4 + p_{23} Q_1 - p_{13} Q_2 - p_{12} (p_{23} + Q_3) + m^2 (Q_1 - 2 Q_2 + 2 Q_3 \nonumber\\ & & - p_{12} + p_{23})) \,,\\
\nonumber\\
{g_1}^{V,1441}_{RL} & = & -12 M (m^4 - p_{12} (p_{13} + 2 p_{23}) + p_{23} Q_1 - p_{13} Q_2 + m^2 (Q_1 - Q_2 + Q_3 \nonumber\\ & & - 3 p_{12} + p_{13} + p_{23})) \, ,\\
\nonumber\\
{g_2}^{V,1441}_{RL} & = & -12 M (p_{12} (p_{12} + Q_2) - m^2 (3 p_{12} + p_{13} + Q_2 - Q_3)) \,,\\
\nonumber\\
f^{V,2442}_{RL} & = & 24 ((m^4 - p_{12} (p_{13} + p_{23}) + m^2 (-2 p_{12} + p_{13} + p_{23})) (Q_1 + Q_2) \nonumber\\ & & - (2 m^2 - p_{12}) p_{12} Q_3)\,, \\
\nonumber\\
l^{V,2442}_{RL} & = & -24 M (m^4 - p_{12} (p_{13} + p_{23}) + m^2 (p_{13} + p_{23} - 2 p_{12}))\,,\\
\nonumber\\
{g_2}^{V,2442}_{RL} & = & -24 M p_{12} (p_{12} - 2 m^2) \, ,
\end{eqnarray}
\begin{equation}
{g_1}^{V,1331}_{RL} = l^{V,1331}_{RL}|_{p_1\leftrightarrow p_2}\, , \qquad
{g_1}^{V,2442}_{RL} = l^{V,2442}_{RL}|_{p_1\leftrightarrow p_2}\, ,  
\end{equation}

\begin{equation}
f^{V,2332}_{RL} = f^{V,1441}_{RL}|_{p_1\leftrightarrow p_2} \,,\qquad l^{V,2332}_{RL} \, = \, {g_1}^{V,1441}_{RL}|_{p_1\leftrightarrow p_2}\,, \nonumber
\end{equation}
\begin{equation}
{g_1}^{V,2332}_{RL} = l^{V,1441}_{RL}|_{p_1\leftrightarrow p_2}\, , \qquad {g_2}^{V,2332}_{RL} \, = \, {g_2}^{V,1441}_{RL}|_{p_1\leftrightarrow p_2}\,,
\end{equation}

\begin{equation}
f^{V,3443}_{RL} = f^{V,1221}_{RL}|_{p_1\leftrightarrow p_2}\,,\qquad l^{V,3443}_{RL} \, = \, {g_1}^{V,1221}_{RL}|_{p_1\leftrightarrow p_2}\,,\nonumber
\end{equation}
\begin{equation}
{g_1}^{V,3443}_{RL} = l^{V,1221}_{RL}|_{p_1\leftrightarrow p_2}\, ,\qquad {g_2}^{V,3443}_{RL} \, = \, {g_2}^{V,1221}_{RL}|_{p_1\leftrightarrow p_2}\, .  
\end{equation}

$(iii)$ For the $T^{SV}_{LRRL}$ contribution all reduced spin-dependent form factors are zero. The results of all reduced spin-independent form factors, which are shown below, must be multiplied by the common factor $M m_1(M^2 + m_1^2 + 2 (m^2 + p_{12} + p_{13} + p_{23} - Q_1 - Q_2 - Q_3))$ ($m_1=m$ if there are identical particles):
\begin{eqnarray}
f^{SV,11}_{LRRL} & = & -6144 (2 m^4 + 3 m^2 p_{23} + p_{23}^2 + M^2 (m^2 + p_{23}) - 2 Q_2 Q_3)\,,\\
\nonumber\\
f^{SV,22}_{LRRL} & = & -6144 (2 m^4 - 2 p_{12} p_{13} + 3 m^2 p_{23} + p_{23}^2 + m_1^2 (m^2 + p_{23})) \,,\\
\nonumber\\
f^{SV,1221}_{LRRL} & = & -12288 (2 m^4 + p_{23}^2 + p_{23} Q_1 + m^2 (3 p_{23} + Q_1) - p_{13} Q_2 - p_{12} Q_3)\,,
\end{eqnarray} 

\begin{equation}
f^{SV,33}_{LRRL} = f^{SV,11}_{LRRL}|_{p_1\leftrightarrow p_2, m_1\to m} \,, \qquad f^{SV,44}_{LRRL} = f^{SV,22}_{LRRL}|_{p_1\leftrightarrow p_2, m_1\to m}\,,
\end{equation}
\begin{eqnarray}
f^{SV,1331}_{LRRL} & = & -3072 ( M^2 (p_{12} + p_{13} + p_{23}) - 3 m^4 + m^2 (M^2 + p_{12} - 3 p_{13} - 3 p_{23} \nonumber\\ && + 2 Q_1 + 2 Q_2 - 6 Q_3) - 2 ((-p_{12} + Q_1 + Q_2) Q_3 + p_{23} (-Q_1 + Q_3) \nonumber\\ & & + p_{13} (2 p_{23} - Q_2 + Q_3)))\,,\\
\nonumber\\
f^{SV,1441}_{LRRL} & = & 6144 (2 m^4 + 2 p_{13} p_{23} + p_{23}^2 + p_{23} Q_1 - 2 p_{13} Q_2 - p_{12} (2 p_{23} + Q_2) \nonumber\\ & & + p_{13} Q_3 + m^2 (Q_1 - 2 Q_2 + 2 Q_3 - 2 p_{12} + 2 p_{13} + 3 p_{23}))\,,\\
\\
f^{SV,2442}_{LRRL} & = & 6144 (3 m^4 - p_{12}^2 - 2 p_{12} (p_{13} + p_{23}) + (p_{13} + p_{23})^2 \nonumber\\ & & + 4 m^2 (p_{13} + p_{23} - p_{12}))\, ,
\end{eqnarray}

\begin{equation}
f^{SV,2332}_{LRRL} = f^{SV,1441}_{LRRL}|_{p_1\leftrightarrow p_2}\,, \qquad f^{SV,3443}_{LRRL} = f^{SV,1221}_{LRRL}|_{p_1\leftrightarrow p_2, m_1\to m}\, .
\end{equation}

$(iv)$ As in the previous case, for the $T^{SV}_{LLRR}$ contribution only the reduced spin-independent form factors are different from zero. The expressions of the reduced form factors $f^{SV,ii}_{LLRR}$ ($i=1,\,2,\,3,\,4$) and $f^{SV,2442}_{LLRR}$ given below have to be multiplied by the common factor $M m_1 (M^2 + m_1^2 + 2 (m^2 + p_{12} + p_{13} + p_{23} - Q_1 - Q_2 - Q_3))$ ($m_1=m$ if there are identical particles):
\begin{eqnarray}
f^{SV,11}_{LLRR} & = & - 6144 (2 m^4 + 3 m^2 p_{23} + p_{23}^2 + M^2 (m^2 + p_{23}) - 2 Q_2 Q_3)\,,\\
\nonumber\\
f^{SV,22}_{LLRR} &= & -6144 (2 m^4 - 2 p_{12} p_{13} + 3 m^2 p_{23} + p_{23}^2 + m_1^2 (m^2 + p_{23}))\,,\\
\nonumber\\
f^{SV,1221}_{LLRR} & = & 4096 M m_1 (-4 m^6 - 4 m^4 p_{12} - 2 m^2 p_{12}^2 - 4 m^4 p_{13} - 2 m^2 p_{12} p_{13} - 2 m^2 p_{13}^2 \nonumber\\ && - 10 m^4 p_{23} - 6 m^2 p_{12} p_{23} - p_{12}^2 p_{23} - 6 m^2 p_{13} p_{23} - p_{13}^2 p_{23} - 8 m^2 p_{23}^2 - 2 p_{12} p_{23}^2  \nonumber\\ && - 2 p_{13} p_{23}^2 - 2 p_{23}^3 - 8 m^4 Q_1 - 6 m^2 p_{12} Q_1 - 6 m^2 p_{13} Q_1 - 14 m^2 p_{23} Q_1 - 6 p_{12} p_{23} Q_1 \nonumber\\ && - 6 p_{13} p_{23} Q_1 - 6 p_{23}^2 Q_1 + 6 m^2 Q_1^2 + 6 p_{23} Q_1^2 + 4 m^4 Q_2 + 4 m^2 p_{12} Q_2 + 8 m^2 p_{13} Q_2 \nonumber\\ && + 6 p_{12} p_{13} Q_2 + 6 p_{13}^2 Q_2 + 6 m^2 p_{23} Q_2 + 2 p_{12} p_{23} Q_2 + 6 p_{13} p_{23} Q_2 + 2 p_{23}^2 Q_2 \nonumber\\ && + 6 m^2 Q_1 Q_2 - 6 p_{13} Q_1 Q_2 + 6 p_{23} Q_1 Q_2 - 2 m^2 Q_2^2 - 6 p_{13} Q_2^2 - p_{23} Q_2^2 + 4 m^4 Q_3 \nonumber
\end{eqnarray}
\begin{eqnarray}
\hspace{6.5mm} && + 8 m^2 p_{12} Q_3 + 6 p_{12}^2 Q_3 + 4 m^2 p_{13} Q_3 + 6 p_{12} p_{13} Q_3 + 6 m^2 p_{23} Q_3 + 2 p_{13} p_{23} Q_3 \nonumber\\ && + 2 p_{23}^2 Q_3  + 6 m^2 Q_1 Q_3 - 6 p_{12} Q_1 Q_3 + 6 p_{23} Q_1 Q_3 - 2 m^2 Q_2 Q_3 - 6 p_{12} Q_2 Q_3 \nonumber\\ && - 6 p_{13} Q_2 Q_3 - 2 m^2 Q_3^2 - 6 p_{12} Q_3^2 - p_{23} Q_3^2 + M^2 (m^4 + m^2 (p_{23} - 3 Q_1) \nonumber\\ && + 3 (p_{13} Q_2 + p_{12} Q_3 - p_{23} Q_1)) + m_1^2 (m^4 + m^2 (p_{23} - 3 Q_1) + 3 (p_{13} Q_2 \nonumber\\ && + p_{12} Q_3 -p_{23} Q_1))) \,,
\end{eqnarray} 

\begin{equation}
f^{SV,33}_{LLRR} = f^{SV,11}_{LLRR}|_{p_1\leftrightarrow p_2, m_1\to m}\,, \qquad f^{SV,44}_{LLRR} = f^{SV,22}_{LLRR}|_{p_1\leftrightarrow p_2, m_1\to m} \,,
\end{equation}
\begin{eqnarray}
f^{SV,1331}_{LLRR} & = & -1024 m M (11 m^6 + 3 M^4 (p_{12} + p_{13} + p_{23}) + m^4 (22 M^2 + 43 p_{12} + 29 p_{13} + 29 p_{23} \nonumber\\ && - 8 Q_1 - 8 Q_2 - 4 Q_3) + 2 M^2 (6 p_{12}^2 + 6 p_{13}^2 + 6 p_{23}^2 - 5 p_{23} Q_1 - 4 p_{23} Q_2 + p_{13} (12 p_{23} \nonumber\\ && - 4 Q_1 - 5 Q_2 - 4 Q_3) + 4 p_{12} (3 p_{13} + 3 p_{23} - Q_1 - Q_2 - Q_3) - 4 p_{23} Q_3 - 3 Q_1 Q_3  \nonumber\\ && - 3 Q_2 Q_3) + 2 (4 p_{12}^3 + 4 p_{13}^3 + 4 p_{23}^3 - 3 p_{23}^2 Q_1 - p_{23} Q_1^2 - 4 p_{23}^2 Q_2 - p_{23} Q_1 Q_2 + 2 Q_1^2 \nonumber\\ && \times Q_2 + 2 Q_1 Q_2^2 + 2 p_{12}^2 (6 p_{13} + 6 p_{23} - 2 Q_1 - 2 Q_2 - 3 Q_3) - 6 p_{23}^2 Q_3 - 12 p_{23} Q_1 Q_3 \nonumber\\ && + 6 Q_1^2 Q_3 - 4 p_{23} Q_2 Q_3 + 20 Q_1 Q_2 Q_3 + 6 Q_2^2 Q_3 + 8 Q_1 Q_3^2 + 8 Q_2 Q_3^2 + p_{13}^2 (6 p_{23} \nonumber\\ && - 4 Q_1 - 3 (Q_2 + 2 Q_3)) + p_{12} (12 p_{13}^2 + 12 p_{23}^2 - 2 Q_1 Q_2 + p_{13} (18 p_{23} - 8 Q_1 - 6 Q_2 \nonumber\\ && - 9 Q_3) - 7 Q_1 Q_3 - 7 Q_2 Q_3 - p_{23} (6 Q_1 + 8 Q_2 + 9 Q_3)) + p_{13} (6 p_{23}^2 - 3 p_{23} (Q_1 \nonumber\\ && + Q_2) - Q_1 (Q_2 + 4 Q_3) - Q_2 (Q_2 + 12 Q_3))) + m^2 (3 M^4 + 2 M^2 (17 p_{12} + 17 p_{13} \nonumber\\ && + 17 p_{23} - 5 Q_1 - 5 Q_2 - 4 Q_3) + 2 (18 p_{12}^2 + 13 p_{13}^2 + 13 p_{23}^2 - 7 p_{23} Q_1 - Q_1^2 - 8 p_{23} \nonumber\\ && \times Q_2 - 2 Q_1 Q_2 - Q_2^2 + p_{12} (31 p_{13} + 31 p_{23} - 11 Q_1 - 11 Q_2 - 12 Q_3) - 7 p_{23} Q_3  \nonumber\\ && - 14 Q_1 Q_3 - 14 Q_2 Q_3 - 2 Q_3^2 + p_{13} (18 p_{23} - 8 Q_1 - 7 (Q_2 + Q_3)))) \nonumber\\ && + 3 i \varepsilon_{\alpha\beta\gamma\delta}p_1^\alpha p_2^\beta p_3^\gamma Q^\delta (p_{13} - p_{23} - Q_1 + Q_2))\,,\\
\nonumber\\
f^{SV,1441}_{LLRR} & = & 2048 m M (3 m^6 + 2 p_{13}^2 p_{23} + 4 p_{13} p_{23}^2 + 2 p_{23}^3 + 5 p_{12} p_{23} Q_1 + 7 p_{13} p_{23} Q_1 + 6 p_{23}^2 Q_1 \nonumber\\ && - 6 p_{23} Q_1^2 - 6 p_{12}^2 Q_2 - 21 p_{12} p_{13} Q_2 - 15 p_{13}^2 Q_2 - 8 p_{12} p_{23} Q_2 - 18 p_{13} p_{23} Q_2 - 4 p_{23}^2 \nonumber\\ && \times Q_2 + 6 p_{12} Q_1 Q_2 + 12 p_{13} Q_1 Q_2 - 7 p_{23} Q_1 Q_2 + 6 p_{12} Q_2^2 + 15 p_{13} Q_2^2 + 2 p_{23} Q_2^2 \nonumber\\ && + M^2 (3 p_{23} Q_1 + p_{12} (p_{23} - 3 Q_2) - p_{13} (p_{23} + 6 Q_2 - 3 Q_3)) + 3 p_{12}^2 Q_3 + 9 p_{12} p_{13} Q_3 \nonumber\\ && + 6 p_{13}^2 Q_3 + 6 p_{12} p_{23} Q_3 + 4 p_{13} p_{23} Q_3 - 6 p_{13} Q_1 Q_3 - 5 p_{23} Q_1 Q_3 + 3 p_{12} Q_2 Q_3 + 9 p_{13} \nonumber\\ && \times Q_2 Q_3 - 3 p_{12} Q_3^2 - 6 p_{13} Q_3^2 + m^4 (-M^2 + 2 p_{12} + 6 p_{13} + 9 p_{23} + 11 Q_1 - 23 Q_2 \nonumber\\ && + 15 Q_3) + m^2 (p_{12}^2 + 3 p_{13}^2 + 10 p_{13} p_{23} + 8 p_{23}^2 + 7 p_{13} Q_1 + 17 p_{23} Q_1 - 6 Q_1^2 - 38 p_{13} \nonumber\\ && \times Q_2 - 21 p_{23} Q_2 + 5 Q_1 Q_2 + 15 Q_2^2 + 17 p_{13} Q_3 + 9 p_{23} Q_3 - 17 Q_1 Q_3 + 2 Q_2 Q_3 \nonumber\\ && - 11 Q_3^2 + M^2 (p_{12} - p_{13} - p_{23} + 3 Q_1 - 6 Q_2 + 6 Q_3) + p_{12} (2 p_{13} + 2 p_{23} + 5 Q_1 \nonumber\\ && - 25 Q_2 + 16 Q_3)) - 3 i \varepsilon_{\alpha\beta\gamma\delta}p_1^\alpha p_2^\beta p_3^\gamma Q^\delta (2 m^2 + p_{12} + p_{13} + 2 p_{23} - Q_2 - Q_3))\,, \nonumber\\
\\
f^{SV,2442}_{LLRR} & = & 6144 (3 m^4 - p_{12}^2 - 2 p_{12} (p_{13} + p_{23}) + (p_{13} + p_{23})^2 + 4 m^2 (-p_{12} + p_{13} + p_{23})) \,, \nonumber\\
\end{eqnarray} 
\begin{equation}
f^{SV,2332}_{LLRR} = f^{SV,1441}_{LLRR}|_{p_1\leftrightarrow p_2}\,, \qquad f^{SV,3443}_{LLRR} = f^{SV,1221}_{LLRR}|_{p_1\leftrightarrow p_2, m_1\to m}\,.
\end{equation}

\end{document}